\documentclass[11pt]{article}
\global\arraycolsep=1pt
\usepackage{amsmath,amssymb,graphicx}
\usepackage{hyperref}
\usepackage{multicol,color,longtable}
\definecolor{darkred}{rgb}{0.65,0.15,0}
\hypersetup{pdfborder={0 0 0},colorlinks=true,urlcolor=darkred,citecolor=blue,linkcolor=darkred,linktocpage=true}

\usepackage{cite}

\usepackage{tikz}
\usetikzlibrary{arrows}
\usepackage{pgfplots}

\usepackage{amsmath}
\usepackage{amsthm}
\usepackage{amssymb}
\usepackage{mathrsfs}
\usepackage[Symbol]{upgreek}
\usepackage{dsfont}
\usepackage[vcentermath]{youngtab}
\usepackage{textcomp}

\usepackage{latexsym}
\usepackage{graphicx}

\usepackage{calc}
\makeatletter
\newlength\@SizeOfCirc%
\newcommand{\CricArrowRight}[1]{%
    \setlength{\@SizeOfCirc}{\maxof{\widthof{#1}}{\heightof{#1}}}%
    \tikz [x=1.0ex,y=1.0ex,line width=.15ex, draw=black]%
        \draw [->,anchor=center]%
            node (0,0) {#1}%
            (0,1.2\@SizeOfCirc) arc (-240:85:1.2\@SizeOfCirc);%
}%
\makeatother

\newcommand{\eprint}[1]{{\href{http://arxiv.org/abs/#1}{\texttt{[#1}]}}}
\newcommand{\eprintN}[1]{{\href{http://arxiv.org/abs/#1}{\texttt{#1 [hep-th]}}}}

\newcommand{\allowDbreak}[1]{\begingroup \allowdisplaybreaks #1 \endgroup}

\newcommand{\baa}{/ \hspace{-1.4ex}}

\newcommand{\baS}{/ \hspace{-1.0ex}} 

\newcommand{\mf}[1]{{\mathfrak{#1}}}
\newcommand{\lp}{\left(}
\newcommand{\rp}{\right)}
\def\taubis{{u}}

\def\DJo{$\;$\kern-.4em \hbox{D\kern-.8em\raise.15ex\hbox{--}\kern.35em okovi\'c}}

\def\DEVII#1#2#3#4#5#6#7{{\tiny $ { \left[ \begin{array}{ccccccc}  & & \mathfrak{#2} \hspace{-0.7mm}&&&& \vspace{ -1.5mm} \\ \mathfrak{#1}\hspace{-0.7mm} &  \mathfrak{#3} \hspace{-0.7mm}& \mathfrak{#4} \hspace{-0.7mm} & \mathfrak{#5}\hspace{-0.7mm}&\mathfrak{#6}\hspace{-0.7mm}& #7 \hspace{-0.8mm} \end{array}\right] }$}}
\def\DEVIIC#1#2#3#4#5#6#7{{\tiny $ { \left[ \begin{array}{ccccccc}  & & {#2} \hspace{-0.7mm}&&&& \vspace{ -1mm} \\ {#1}\hspace{-0.7mm} &  {#3} \hspace{-0.7mm}& {#4} \hspace{-0.7mm} & {#5}\hspace{-0.7mm}&\mathfrak{#6}\hspace{-0.7mm}& #7 \hspace{-0.8mm} \end{array}\right] }$}}
\def\DEVIII#1#2#3#4#5#6#7#8{{\tiny $ { \left[ \begin{array}{ccccccc}  & & \mathfrak{#2} \hspace{-0.7mm}&&&& \vspace{ -1.5mm} \\ \mathfrak{#1}\hspace{-0.7mm} &  \mathfrak{#3} \hspace{-0.7mm}& \mathfrak{#4} \hspace{-0.7mm} & \mathfrak{#5}\hspace{-0.7mm}&\mathfrak{#6}\hspace{-0.7mm}&\mathfrak{#7}\hspace{-0.7mm}&\mathfrak{#8} \end{array}\right] }$}}

\newfont{\bbbold}{msbm10 scaled \magstep1}

\def\cB{{\cal B}}

\def\cD{{\cal D}}
\def\cE{{\cal E}}
\def\cF{{\cal F}}

\def\cM{{\cal M}}
\def\cN{{\cal N}}

\def\cV{{\cal V}}

\newfont{\goth}{eufm10 scaled \magstep1}

\def\c{\gamma}

\def\be{\begin{equation}}\def\ee{\end{equation}}
\def\bea{\begin{eqnarray}}\def\eea{\end{eqnarray}}
\def\barr{\begin{array}}\def\earr{\end{array}}

\def\nn{\nonumber}
\def\bea{\begin{eqnarray}}
\def\eea{\end{eqnarray}}

\def\DEVI#1#2#3#4#5#6{{\tiny $ { \left[ \begin{array}{cccccc}  & & \mathfrak{#2} \hspace{-0.7mm}&&& \vspace{ -1.5mm} \\ \mathfrak{#1}\hspace{-0.7mm} &  \mathfrak{#3} \hspace{-0.7mm}& \mathfrak{#4} \hspace{-0.7mm} & \mathfrak{#5}\hspace{-0.7mm}& \mathfrak{#6} \end{array}\right] }$}}

\def\DSOX#1#2#3#4#5{{\tiny $ {   \biggl[ \begin{array}{ccc}  &&\mathfrak{#3}  \vspace{ -1.5mm} \\  \mathfrak{#1}\hspace{0.2mm}\mathfrak{#2}\hspace{-0.6mm} &\mathfrak{#4} \hspace{-0.9mm}&\vspace{-1.5mm}\\ && \mathfrak{#5}  \end{array}\biggr] }$}}

\def\EiEVIII#1{{E_{\fontsize{6.35pt}{6pt}\selectfont   \left[ \begin{array}{ccccccc}  & & \mathfrak{0} \hspace{-0.6mm}&&& \vspace{ -1.0mm} \\ #1 \hspace{-0.6mm} &  \mathfrak{0} \hspace{-0.6mm}& \mathfrak{0} \hspace{-0.6mm} & \mathfrak{0}\hspace{-0.6mm} & \mathfrak{0}\hspace{-0.6mm}&\mathfrak{0}\hspace{-0.6mm}&\mathfrak{0} \end{array}\right] \fontsize{12.35pt}{12pt}\selectfont }}}

\def\EiEVII#1{{E_{\fontsize{6.35pt}{6pt}\selectfont   \left[ \begin{array}{cccccc}  & & \mathfrak{0} \hspace{-0.6mm}&&& \vspace{ -1.0mm} \\ #1 \hspace{-0.6mm} &  \mathfrak{0} \hspace{-0.6mm}& \mathfrak{0} \hspace{-0.6mm} & \mathfrak{0}\hspace{-0.6mm}&\mathfrak{0}\hspace{-0.6mm}&\mathfrak{0} \end{array}\right] \fontsize{12.35pt}{12pt}\selectfont }}}

\def\EiEVI#1{{E_{\fontsize{6.35pt}{6pt}\selectfont   \left[ \begin{array}{cccccc}  & & \mathfrak{0} \hspace{-0.6mm}&& \vspace{ -1.0mm} \\ #1 \hspace{-0.6mm} &  \mathfrak{0} \hspace{-0.6mm}& \mathfrak{0} \hspace{-0.6mm} & \mathfrak{0}\hspace{-0.6mm}&\mathfrak{0} \end{array}\right] \fontsize{12.35pt}{12pt}\selectfont }}}

\def\det{{\rm det\,}}

\makeatletter

\@addtoreset{equation}{section} \makeatother

\newcommand{\Scal}[1]{\Bigl ({#1} \Bigr )}
\newcommand{\scal}[1]{\bigl ({#1} \bigr )}

\newcommand{\CR}{\nonumber \\*}

\DeclareMathAlphabet{\mathpzc}{OT1}{pzc}{m}{it}

\newcommand{\gra}[2]{{\scriptscriptstyle (#1 , #2 )}}
\newcommand{\ord}[1]{{\scriptscriptstyle (#1)}}

\def\cN{\mathcal{N}}

\def\ie{{\it i.e.}\ }
\def\eg{{\it e.g.}\ }

\newcommand{\stfrac}[2]{{\scriptscriptstyle \frac{#1}{#2}}}

\def\Td{{\fontsize{7.35pt}{7pt}\selectfont \mbox{$T^d$} \fontsize{12.35pt}{12pt}\selectfont }}

\begin{document}

\thispagestyle{empty}

{\flushright {CPHT-RR041.1015}\\[15mm]}

\begin{center}
{\LARGE \bf Loops in exceptional field theory}\\[10mm]

\vspace{8mm}
\normalsize
{\large  Guillaume Bossard${}^{1}$ and Axel Kleinschmidt${}^{2,3}$}

\vspace{10mm}
${}^1${\it Centre de Physique Th\'eorique, Ecole Polytechnique, CNRS\\
Universit\'e Paris-Saclay 91128 Palaiseau cedex, France}
\vskip 1 em
${}^2${\it Max-Planck-Institut f\"{u}r Gravitationsphysik (Albert-Einstein-Institut)\\
Am M\"{u}hlenberg 1, DE-14476 Potsdam, Germany}
\vskip 1 em
${}^3${\it International Solvay Institutes\\
ULB-Campus Plaine CP231, BE-1050 Brussels, Belgium}

\vspace{20mm}

\hrule

\vspace{10mm}

\begin{tabular}{p{12cm}}
{\small
We study certain four-graviton amplitudes in exceptional field theory in dimensions $D\geq 4$ up to two loops. As the formulation is manifestly invariant under the U-duality group $E_{11-D}(\mathds{Z})$, our resulting expressions can be expressed in terms of automorphic forms. In the low energy expansion, we find terms in the M-theory effective action of type $R^4$, $\nabla^4R^4$ and $\nabla^6 R^4$ with automorphic coefficient functions in agreement with independent derivations from string theory. This provides in particular an explicit integral formula for the exact string theory $\nabla^6 R^4$ threshold function. We exhibit moreover that the usual supergravity logarithmic divergences cancel out in the full exceptional field theory amplitude, within an appropriately defined dimensional regularisation scheme. We also comment on terms of higher derivative order and the role of the section constraint for possible counterterms.

}
\end{tabular}
\vspace{7mm}
\hrule
\end{center}

\newpage
\setcounter{page}{1}

\setcounter{tocdepth}{2}
\tableofcontents

\vspace{5mm}
\hrule
\vspace{5mm}

\section{Introduction}

Exceptional field theory \cite{Koepsell:2000xg,Hull:2007zu,Pacheco:2008ps,Berman:2010is,Berman:2011cg,Cederwall:2013naa,Hohm:2013pua,Hohm:2013vpa,Hohm:2013uia,Aldazabal:2013via,Godazgar:2014nqa,Hohm:2014fxa} provides in principle a framework to compute manifestly U-duality invariant amplitudes in perturbation theory. However, the enforcement of the strong section constraint makes it difficult to define Feynman rules from the Lagrangian. In this paper we will argue that such Feynman rules are not needed to compute on-shell amplitudes with external states carrying no momentum along the exceptional coordinates, and that unitarity and U-duality determine uniquely the correct loop integrand.

\begin{figure}
\centering
\begin{picture}(200,50)
\thicklines
\multiput(10,10)(30,0){4}{\circle*{10}}
\multiput(97,10)(10,0){5}{\line(1,0){5}}
\put(10,10){\line(1,0){90}}
\put(150,10){\line(1,0){30}}
\put(150,10){\circle*{10}}
\put(180,10){\circle*{10}}
\put(70,40){\circle*{10}}
\put(70,10){\line(0,1){30}}
\put(7,-5){$1$}
\put(57,38){$2$}
\put(37,-5){$3$}
\put(67,-5){$4$}
\put(97,-5){$5$}
\put(142,-5){$d\!-\!1$}
\put(177,-5){$d$}
\end{picture}
\caption{\label{fig:dynk}\small Dynkin diagram of $E_{d(d)}(\mathds{R})$ with labelling of nodes used in the text.}
\end{figure}
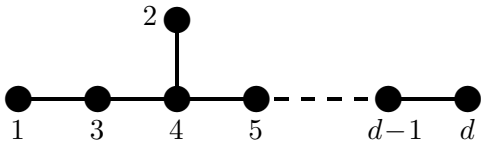

In this framework we will consider amplitudes on $\mathds{R}^{11-d}\times M^{\dim \overline{\bf R}_{\alpha_d}}$, where $\overline{\bf R}_{\alpha_d}$ denotes the  representation of the hidden symmetry group $E_{d(d)}(\mathds{R})$ associated with the last node of the $E_{d}$ Dynkin diagram of figure~\ref{fig:dynk}. The representations $\overline{\bf R}_{\alpha_d}$ for the various $d$ are listed in table~\ref{tab:reps} and we restrict ourselves mainly to $d<8$.  The whole space $\mathds{R}^{11-d}\times M^{\dim \overline{\bf R}_{\alpha_d}}$ has coordinates $(x^\mu,Y^M)$ with $\mu=1,\ldots,11-d$ and $M=1,\ldots, \dim \overline{\bf R}_{\alpha_d}$, where the dependence of fields on the $Y^M$ are restricted by the strong section constraint 
\begin{align}
\label{seccond}
\left.\frac{\partial \phi_1}{\partial Y^M}  \times \frac{\partial \phi_2}{\partial Y^N} \right|_{\overline{\bf R}_{\alpha_1}} =0
\end{align}
for any fields $\phi_1(x,Y)$ and $\phi_2(x,Y)$. The representation $\overline{\bf R}_{\alpha_1}$ is associated with the first node and contained in the (symmetric) tensor product of the momentum representation $\overline{\bf R}_{\alpha_d}$ with itself. The condition~\eqref{seccond} means that the resulting expression in $\overline{\bf R}_{\alpha_d}\otimes \overline{\bf R}_{\alpha_d}$ must vanish on the representation $\overline{\bf R}_{\alpha_1}$.\footnote{\label{fn:addSC}There are additional constraints in four and three dimensions that we do not indicate in the table. In $D=4$ dimensions the section constraint is often written to lie in ${\bf 133} \oplus {\bf 1}$ of $E_{7(7)}$ (see \eg~\cite{Coimbra:2011ky,Hohm:2013uia}). In $D=3$ dimensions, the section constraint is given as  the ${\bf 3875}\oplus {\bf 248} \oplus {\bf 1}$ of $E_{8(8)}$ in~\cite{Hohm:2014fxa}. We will see later that the additional representations are consequences of the $\overline{\bf R}_{\alpha_1}$ constraints listed in the table.} 
This condition is a non-linear but $E_{d(d)}(\mathds{R})$-covariant condition. Linear spaces of solutions to the strong section constraint permit to define truncations of exceptional field theories to standard Lagrangian theories. There are two standard solutions of the strong section condition that can be obtained by decomposing with respect to either the $GL(d,\mathds{R})$ subgroup of $E_{d(d)}(\mathds{R})$, such that the $11$ coordinates $x,Y$ on which the fields depend parametrize the eleven-dimensional space-time in 11-dimensional supergravity, or a $GL(d-1,\mathds{R})\times SL(2,\mathds{R})$ subgroup, such that the $10$ coordinates $x,Y$ on which the fields depend parametrize the ten-dimensional space-time in type IIB supergravity~\cite{Blair:2013gqa}. We will show in section~\ref{sec:sssols} that in general the linear spaces of dimension $k$ of solutions to the strong section constraint define a unique $E_{d(d)}(\mathds{R})$ orbit for $k\le d-2$, whereas they define two independent orbits for $k>d-2$, which correspond respectively to the 11-dimensional supergravity and the type IIB solutions.

\begin{table}
\centering
\begin{tabular}{c|c||c|c}
Space-time dimension& Hidden symmetry & coordinates $Y^M$ & Section constraint\\
$D=11-d$ & $E_{d(d)}(\mathds{R})$ & $\overline{\bf R}_{\alpha_d}$ & $\overline{\bf R}_{\alpha_1}$\\[2mm]\hline
$9$ & $GL(2,\mathds{R})$ & ${\bf 1}^\ord{-4}\oplus {\bf 2}^\ord{3}$ & ${\bf 2}^\ord{-1}$\\
$8$ & $SL(2,\mathds{R})\times SL(3,\mathds{R})$ & $({\bf 2}, {\bf 3})$ & $({\bf 1},\overline{\bf 3})$\\
$7$ & $SL(5,\mathds{R})$ & ${\bf 10}$ & ${\bf \overline{5}}$\\[2mm]
$6$ & $SO(5,5,\mathds{R})$ & ${\bf 16}$ & ${\bf 10}$\\[2mm]
$5$ & $E_{6(6)}(\mathds{R})$ & ${\bf 27}$ & ${\bf \overline{27}}$\\[2mm]
$4$ & $E_{7(7)}(\mathds{R})$ & ${\bf 56}$ & ${\bf 133}$\\[2mm]
$3$ &$E_{8(8)}(\mathds{R})$ & ${\bf 248}$ & ${\bf 3875}$\\[2mm]
\end{tabular}
\caption{\label{tab:reps}\small Coordinate representation $\overline{\bf R}_{\alpha_d}$ and strong section constraint representation $\overline{\bf R}_{\alpha_1}$ for hidden symmetry groups $E_{d(d)}(\mathds{R})$ in dimension $D=11-d$ for $4\leq d \leq 7$. The conditions are stated here for the coordinates although we will mainly use them in their Fourier transformed versions such that the conjugate representations ${\bf R}_{\alpha_1}$ and ${\bf R}_{\alpha_d}$ occur.}
\end{table}

We will consider the case when the `extended space'  $M^{\dim \overline{\bf R}_{\alpha_d}}$ is compact and therefore all momenta in the exceptional directions are quantised. For a square `extended' torus there are then discrete charges $\Gamma\in \mathds{Z}^{\dim {\bf R}_{\alpha_d}}$. As we will explain in more detail in section~\ref{sec:Frules}, the strong section constraint~\eqref{seccond} for the charges becomes
\begin{align}
\label{intro:sc}
\left.\Gamma_i \times \Gamma_j\right|_{{\bf R}_{\alpha_1}} =0
\end{align}
for any set of charges $\Gamma_i$ that are connected through a vertex in the Feynman diagram. This last condition is important and reflects the locality of the classical exceptional field theory Lagrangian. As we will see in more detail in section~\ref{sec:Frules}, the condition~\eqref{intro:sc} does \textit{not} imply that all charges appearing in the theory have to mutually satsify the section constraint. This is important for the construction of potential higher derivative counterterms in exceptional field theory because they can not necessarily be expressed in terms of local expressions subject to the section constraint. This might be an important lesson for understanding the role of the strong constraint and higher derivative terms purely within exceptional field theory.\footnote{We are grateful to M. Cederwall for discussions of this point, see also~\cite{Cederwall:2015jfa}.}
 
The particular process we will focus on is the four-graviton amplitude that has been evaluated at one and two loops in $D=11$ supergravity on  a torus in~\cite{Green:1997as,Russo:1997mk,Green:1999pu,deWit:1999ir,Green:2005ba,Green:2008bf}, and at 3-loop in \cite{Basu:2014hsa,Basu:2014uba}. Since the lowest order terms in the energy expansion in Mandelstam variables enjoy BPS protection, one can be sure that the field theory calculation produces correct results for the M-theory effective action if one includes the corresponding BPS states. No other M-theory states contribute to the four-graviton process at lowest order. One major result of the analysis of~\cite{Green:1999pu} was the prediction of the Eisenstein series $E_{[\stfrac52]}$ multiplying the $\nabla^4 R^4$ interaction in type IIB theory in $D=10$ space-time dimensions. This has served as an important reference point for the subsequent construction of $\cE_\gra{1}{0}^D$ for $D<10$ in terms of Eisenstein series on $E_{d(d)}(\mathds{R})$ where $D=11-d$. 

The function $\cE_\gra{1}{0}^D$ that arose from the two-loop calculation of~\cite{Green:1999pu} is part of a family of functions $\cE_\gra{p}{q}^D$ that appear in the analytic part of the $\alpha'$-expansion of the four-graviton scattering amplitude in type II string theory on a $10-D$-dimensional torus~\cite{Green:2010wi}. From the point of view of the effective action they multiply higher derivative terms of the form $\nabla^{4p+6q} R^4$ and are required to be invariant under U-duality. For the lowest values up to $4p+6q\leq 6$, the conjectured expressions for $\cE_\gra{p}{q}^D$ have been subjected to numerous consistency checks~\cite{Green:2005ba,Green:2008bf,Kiritsis:1997em,Obers:1999um,Green:2010wi,Pioline:2010kb,Green:2010kv,Green:2011vz}. Not all of them are Eisenstein series. 

In this paper, we will perform the one- and two-loop calculation of the four-graviton amplitude in exceptional field theory. This will automatically produce U-duality invariant expressions and include full multiplets of $\tfrac{1}{2}$-BPS states in the construction. We will see that the results at one and two loops reproduce in a direct manner the expressions for $\cE_\gra{0}{0}^D$ and $\cE_\gra{1}{0}^D$ that have been hitherto only obtained indirectly. Particular attention will be paid to infrared and ultraviolet divergences that arise in the exceptional field theory calculations in various dimensions. We will deal with ultra-violet divergences associated to both large $D$-dimensional momentum integration and the infinite sums over the discrete momenta in $\mathds{Z}^{\dim {\bf R}_{\alpha_d}}$ in dimensional regularisation, whereas we will introduce a small mass regularisation for the infrared divergences  that will define the sliding scale separating the analytic and the non-analytic components of the amplitudes. We will find in particular that both the amplitude divergences arising at 1-loop in eight dimensions \cite{Fradkin} and at 2-loop in seven dimensions \cite{Bern:1998ug} cancel nicely in the full exceptional field theory amplitude. There is also an additional divergence that occurs at 2-loops in six dimensions in exceptional field theory, which is associated to the 1-loop $R^4$ type form factor divergence in supergravity  (as advocated in \cite{Bossard:2014lra}). We analyse this amplitude in detail and show that the divergences cancel out in this case as well in the complete amplitude. These cancelations provide a direct physical correspondence between the Eisenstein series poles and the supergravity logarithmic divergences in supergravity, which reflects the correspondence between the logarithmic terms in the Mandelstam variables and the string coupling modulus analysed in \cite{Green:2010sp}.

Our calculations also allow an investigation of correction terms of yet higher derivative order of the form $\nabla^{2k} R^4$ for $k>2$. We will work out  in detail the 2-loop contribution to the $\nabla^6 R^4$ threshold function. We obtain a generalisation of the integral formula of the type IIB  $\nabla^6 R^4$ threshold function derived in \cite{Green:2005ba}, valid in all dimensions $D\ge 4$ (see \cite{Pioline:2015yea} for an alternative proposal in six dimensions). We prove in particular that this threshold function satisfies the Poisson equation with quadratic source derived in \cite{Green:2005ba,Green:2010wi,Green:2010kv,Basu:2007ck}. The computations works similarly as in \cite{Green:2005ba}, although the strong section constraint and special properties of the $R^4$ threshold function as an automorphic function associated to the minimal unitary representation of $E_{d(d)}$ play an important role. We also prove that this function satisfies the tensorial differential equations (of inhomogeneous type) required by supersymmetry \cite{Bossard:2015uga}. We moreover analyse the types of instantons that contribute to the various higher derivative terms; in mathematical terms this is captured by the so-called wavefront set of the 2-loop  $\nabla^{2k} R^4$ threshold functions. The wavefront set associated to the $R^4$ and $\nabla^4 R^4$ threshold functions were analysed in detail in \cite{Green:2011vz,Fleig:2013psa}, and the one associated to $\nabla^6 R^6$ in \cite{Bossard:2015uga,Bossard:2015oxa}. 
 
The structure of this article is as follows. In section~\ref{sec:Frules}, we explain the method for computing amplitudes in exceptional field theory based on the two-derivative action, including a discussion of the form and relevance of the section constraint in this set-up. In section~\ref{sec:1loop}, we apply the formalism to the one-loop four-graviton amplitude. The two-loop amplitude is studied in detail in section~\ref{sec:2loops}, with particular attention to the contributions to the $\nabla^4 R^4$ and the $\nabla^6 R^4$  threshold functions in all dimensions. Section~\ref{sec:Specs} contains additional comments on our results and possible extensions. For completeness, we include the short appendix~\ref{app:ES} that summarises some properties of Eisenstein series that we use and our notations. Appendix~\ref{MinimalOrtho} we study certain properties of $SO(n,n)$ Eisenstein series that are used in analysing the $\nabla^6 R^4$ correction and appendix~\ref{Epstein1/4} discusses the possibility of defining automorphic functions from a sum over $\tfrac{1}{4}$-BPS charges.

\section{Feynman amplitudes in exceptional field theory}
\label{sec:Frules}

In this section, we outline the calculation of scattering amplitudes in the framework of exceptional field theory and in particular the treatment of the strong section constraint~\eqref{seccond}. The final scattering amplitudes we are considering will not carry any momentum/charges in the extended space on the external legs.

\subsection{Set-up and BPS multiplets}

To begin with, we consider a constant eleven-dimensional background metric of the form
\be 
ds^2_{\rm \scriptscriptstyle 11} = e^{\frac{9-d}{3} \phi} M_{IJ} dy^I dy^J  + e^{-\frac{d}{3} \phi} \eta_{\mu\nu} dx^\mu dx^\nu \ . 
\ee 
Here, we are envisaging manipulations adapted to a split $11=D+d$ such that the $D$-dimensional space-time is Minkowskian and $d$-dimensional `internal' space is a torus $T^d$ with uni-modular metric $M_{IJ}$ and volume set by a dilaton field $\phi$. The Kaluza--Klein vectors are set to zero for simplicity. The variables $y^I$ are in the range $y^I\in [0, 2\pi\ell)$ with $I=1,\ldots, d$ and $\ell$ is our moduli-independent reference length scale. The connection to the gravitational coupling $\kappa_{11-d}$ is provided by~\cite{Green:2005ba}
\be
\label{eq:kappad}
2   \kappa_{\scriptscriptstyle 11-d}^{\; 2}  =  (2\pi)^{8-d}\ell^{9-d}\ .
\ee
(The length scale $\ell$ is related to the standard Planck length by $\ell^2=32\, \ell_{\mathrm{Pl}}^{\; 2}$.)

Due to the toroidal internal space, the conjugate momenta take the form $p_{11} = \lp p, \tfrac{n_I}{\ell}\rp$.  The norm square of a loop momentum in a Feynman diagram is associated to the kinetic term in the background metric such that it includes the measure factor, leading to 
\be 
\label{eq:loopmom}
\sqrt{-g_{ \scriptscriptstyle 11}} \ p^2_{ \scriptscriptstyle 11} = p^2 +  \ell^{-2} e^{-3\phi} M^{IJ} n_I n_J \ .
\ee 
The powers of $\ell$ are needed on dimensional grounds. 

We can Fourier expand on the toroidal space such that one sums over the $d$-dimensional lattice $\mathds{Z}^d$ of integer modes $n_I$. In the case of a scalar field this reads
\begin{align}
\label{D11Fourier}
\phi(x,y) = \sum_{n\in \mathds{Z}^d}   \int_{\mathds{R}^{11-d}} \frac{d^{11-d}  p}{(2\pi)^{11-d}} e^{i \ell^{-1} n_I y^I + i (p,x) } \phi_n(p) \ , 
\end{align}
and the kinetic term indeed becomes
\bea
&&  \frac{1}{2 \kappa_{\scriptscriptstyle 11}^{\; 2}} \int_{\mathds{R}^{11-d}\times \mathds{T}^d} \hspace{-4mm}d^{11-d}xd^dy  \sqrt{-g_{\rm \scriptscriptstyle 11}} \  \scal{   \nabla  \phi  (x,y), \nabla \phi(x,y) } \\
&=& \frac{(2\pi\ell)^d}{2 \kappa_{\scriptscriptstyle 11}^{\; 2}} \sum_{n\in \mathds{Z}^d} \int_{\mathds{R}^{11-d}} \frac{d^{11-d}  p}{(2\pi)^{11-d}}  \scal{ p^2 + \ell^{-2} e^{-3\phi} M^{IJ} n_I n_J }  \phi_n  (p) \phi_{-n}(-  p)\ .\nn
\eea
The Newton coupling constant $ \kappa_{\scriptscriptstyle 11-d}^{\; 2}$ in $11-d$ dimensions is defined by~\eqref{eq:kappad}. For simplicity we will always avoid writing the subscript, such that $\kappa \equiv \kappa_{\scriptscriptstyle 11-d}$. There is no scalar field in 11-dimensional supergravity, but the same computation can be repeated for the various fields of the theory to find the same substitution~\eqref{eq:loopmom} for the loop momenta. 

The sum over discrete momenta $n_I$ breaks by construction U-duality to $GL(d,\mathds{Z})$, because they only span part of an $E_{d(d)}$ multiplet of charges $\Gamma$. For $d\leq 7$ this $E_{d(d)}$ multiplet of charges is ${\bf R}_{\alpha_d}$ (conjugate to the coordinates of table~\ref{tab:reps}) and  branches as
\begin{align}
\Gamma = \lp n_I, n^{IJ}, n^{I_1\ldots I_5}, n^{I_1\ldots I_7, J} \rp
\label{Decompose} 
\end{align}
in $GL(d,\mathds{Z})$ decomposition. The $n_I$ only correspond to the highest degree components of $\Gamma$. The additional integral charges $n^{IJ}=n^{[IJ]}$ can be interpreted as the winding of the M2-brane along the torus, $n^{I_1\ldots I_5}=n^{[I_1\ldots I_5]}$ as the winding of the M5-brane, and $n^{I_1I_2I_3I_4I_5I_6I_7,J}=n^{[I_1I_2I_3I_4I_5I_6I_7],J}$ with $n^{[I_1\ldots I_7,J]}=0$ is the Kaluza--Klein monopole charge, see for example~\cite{Obers:1998fb} for a review. The Kaluza--Klein monopole charge is only non-zero in four dimensions, and in this case reduces to a vector that defines the $U(1)^7$ Chern class of the Taub-NUT solution over any $S^2$ surrounding the monopole. All the states carrying these quantum numbers are expected to contribute to the amplitude in M-theory. The decomposition \eqref{Decompose} includes more components for $d=8$, but they are not all associated to non-perturbative states. In $D=3$ dimensions the quantum numbers are defined by the $E_{8(8)}(\mathds{Z})$ monodromy of the corresponding soliton, rather than an integral vector in the adjoint representation. This non-linear realisation is reflected in exceptional field theory by the presence of an additional constrained $E_{8(8)}$ gauge symmetry \cite{Hohm:2014fxa}. We do not necessarily expect our construction to be well defined in three dimensions, but because the formulae can be generalised straightforwardly, we will nevertheless discuss the naive extension of our results to $d>7$. In terms of the Kac--Moody extension to $E_{11}$~\cite{West:2001as} one can summarise the charges in terms of the so-called $l_1$ representation~\cite{West:2003fc,Kleinschmidt:2003jf}, truncated decompositions of which also give the $E_{d(d)}$ multiplets of table~\ref{tab:reps}, see~\cite{West:2004kb}. We note, however, that the constructions in~\cite{Hohm:2013uia,Hohm:2014fxa} utilise additional vector fields whose relation to $E_{11}$ is not clear at present. 

For $d\le 7$ one has the natural $E_{d(d)}$ invariant norm~\cite{Pioline:1997pu,Obers:1997kk} (where we set the  background 3-form $a_{IJK}$ and 6-form $a_{IJKLPQ}$ along the torus to zero for simplicity)
\begin{multline}  
\hspace{-3mm} |Z(\Gamma)|^2 = e^{-3\phi} M^{IJ} n_I n_J + \frac{1}{2} e^{(6-d)\phi} M_{IK}  M_{JL}n^{IJ} n^{KL} +\frac{1}{5!} e^{(15-2d)\phi} \prod_{i=1}^5 M_{I_iJ_i}  n^{I_1I_2I_3I_4I_5} n^{J_1J_2J_3J_4J_5} \\ + \frac{1}{7!} e^{(24-3d) \phi} \prod_{i=1}^7 M_{I_iJ_i}  M_{KL} n^{I_1I_2I_3I_4I_5I_6I_7,K} \,    n^{J_1J_2J_3J_4J_5J_6J_7,L} \ . 
\label{Znorm}
\end{multline} 
A similar expression with more components exists also for $E_{8(8)}$ and formally also for $E_{d(d)}$ with $d>8$ but the precise interpretation of all the non-perturbative states is less well-established. Below we will sometimes formally discuss the extension to these cases. Further comments on the extension of our results to the Kac--Moody cases can be found in section~\ref{sec:KMext}.

Although the effective interactions between general non-perturbative states of arbitrary quantum numbers are unknown, the interactions of states of quantum numbers restricted to the U-duality orbit of the standard Kaluza--Klein momenta $n_I$ can be obtained from the standard eleven-dimensional supergravity interactions by symmetry.  Eleven-dimensional supergravity preserves $GL(d,\mathds{Z})$ by diffeomorphism invariance, such that it admits a unique $E_{d(d)}(\mathds{Z})$ invariant (minimal) extension, which  is obtained by summing over all charges satisfying the $\tfrac12$-BPS constraint (see \eg~\cite{Ferrara:1997ci,Obers:1998fb,Obers:1999um})
\bea 
n^{IJ} n_J &=&0 \ , \qquad 3 n^{[IJ} n^{KL]} = n^{IJKLP} n_P \ , \CR
6 n^{I[J} n^{KLPQR]} &=& - n^{I,JKLPQRS} n_S + n^{S,IJKLPQR} n_S \ , \CR
7 n^{IJK[PQ} n^{RSTUV]} &=& 2 n^{[IJ} n^{K],PQRSTUV} \ , \qquad n^{[IJKLP} n^{Q],RSTUVWX} = 0 \ . \label{DecomCons} 
\eea

\subsection{Fourier transform of the strong section constraint}
\label{sec:FT}

As already mentioned, the charges $\Gamma$ belong in general to the irreducible representation ${\bf R}_{\alpha_d}$ of $E_{d(d)}$ that is conjugate to the exceptional coordinates $Y\in\overline{\bf R}_{\alpha_d}$ and it is associated to the last node of the Dynkin diagram in the $E_{d}$ convention  \DEVIIC12345{...}{$d$} (cf. figure~\ref{fig:dynk}). The $\tfrac12$-BPS constraint~\eqref{DecomCons} can be written for any two charges such that their tensor product restricted to the first node irreducible representation vanishes, \ie 
\be 
\Gamma_{i} \times \Gamma_{j} \big|_{{\bf R}_{\alpha_1}} = 0 \ .
\label{2ChargeRk1} 
\ee
In the following we will simply write $\Gamma_{i} \times \Gamma_{j}$ as a cross product that is restricted to the representation ${\bf R}_{\alpha_1}$ by definition. We will now argue that~\eqref{2ChargeRk1}  is  the manifestation of the strong section constraint in exceptional field theory in Fourier space when the extended space is periodic. 

In exceptional field theory one promotes the dependence of the fields in the internal space coordinates $T^d$ to coordinates $Y$ (of the extended space) in the irreducible representation $\overline{\bf R}_{\alpha_d}$ of $E_{d(d)}$, such that any field $\phi_i$ admits a restricted Fourier expansion in the compact extended space generalising~\eqref{D11Fourier}:\footnote{Compared to the introduction we now suppress the coordinate superscript on the coordinates $Y$. The notation $d(\alpha_d)$ denotes the dimension of the representation ${\bf R}_{\alpha_d}$ of $E_{d(d)}$. We also reiterate that by $\times$ in equations of this type we always  mean the projection of the product to the representation ${\bf R}_{\alpha_1}$.}
\begin{align}
\phi_i(Y) = \sum_{\substack{\Gamma_i \in \mathds{Z}^{d(\alpha_d)} \\ \Gamma_i \times \Gamma_i = 0   }}  e^{i \ell^{-1} \langle Y , \Gamma_i \rangle} \phi_{i\, \Gamma_i}\ , 
\end{align}
 with the strong section constraint that for any two fields \cite{Hohm:2013vpa,Hohm:2013uia}
 \begin{align}
\label{eq:sc2}
0  = \frac{ \partial \phi_1(Y)}{\partial Y} \times  \frac{ \partial \phi_2(Y)}{\partial Y}   
= -\ell^{-2} \sum_{\substack{\Gamma_i \in \mathds{Z}^{2d(\alpha_d)} \\ \Gamma_i \times \Gamma_i = 0   }} e^{ i \ell^{-1} \langle Y , \Gamma_1 + \Gamma_2 \rangle } \scal{\Gamma_{1} \times \Gamma_{2} }  \phi_{1\,  \Gamma_1} \phi_{2\,  \Gamma_2} \ . 
\end{align}
This is satisfied by construction for all products in the Lagrangian if one considers the restricted convolution for a product of $n$ elementary fields
\be 
\label{eq:nproduct}
\phi_1(Y)  \cdots  \phi_n(Y) \longrightarrow  \sum_{\substack{\Gamma_i \in \mathds{Z}^{n\, d(\alpha_d)} \\ \Gamma_i \times \Gamma_j = 0   }} e^{ i \ell^{-1} \langle Y , \Gamma_1 + \ldots + \Gamma_n \rangle }  \phi_{1\,  \Gamma_1} \cdots \phi_{n\,  \Gamma_n}  \ ,
\ee
where the strong section constraint $\Gamma_i\times \Gamma_j=0$ is satisfied for all pairs $i$ and $j$. This definition of terms in exceptional field theory Lagrangians is compatible with the linearised gauge invariance of all the fields. However, when one varies cubic or higher order terms in the Lagrangian with non-linear contributions from the gauge transformations one does not obtain an expression that in Fourier space is not composed out of factors that all satisfy mutually the strong section constraint, \ie, it is not of the form~\eqref{eq:nproduct}.

%A general solution to the strong section constraint~\eqref{eq:sc2} therefore amounts to defining the field product as the convolution with the Kronecker delta restricted to the charges satisfying the strong section constraint, \ie 
%\be 
%\phi_1(Y) * \phi_2(Y) \equiv  \sum_{\substack{\Gamma_i \in \mathds{Z}^{2d(\alpha_d)} \\ \Gamma_i \times \Gamma_j = 0   }} e^{ i \ell^{-1} \langle Y , \Gamma_1 + \Gamma_2 \rangle }  \phi_{1\,  \Gamma_1} \phi_{2\,  \Gamma_2}  \ . 
%\ee
%
%%This product is commutative and associative and the derivative satisfies the Leibniz identity. All the exceptional field theory identities are therefore valid for this product. The section constraint~\eqref{eq:sc2} is solved if the charges $\Gamma_i$ of the fields entering in it satisfy~\eqref{2ChargeRk1}. In analogy with this condition for matrices we will call the corresponding charges `rank one charges'.
%
For example, a local cubic scalar field vertex is then given by a restricted sum over integral charges satisfying the strong section constraint \eqref{2ChargeRk1}, \ie
\begin{align}
\label{3-point}
&\quad  \frac{1}{2 \kappa_{\scriptscriptstyle 11}^{\; 2}} \int_{\mathds{R}^{11-d}\times {\bf R}_{\alpha_d}} \hspace{-4mm}d^{11-d}xd^{d(\alpha_d)}Y  \sqrt{-g_{\rm \scriptscriptstyle 11}}   \phi(x,Y) \scal{ \nabla \phi  (x,Y) , \nabla  \phi(x,Y) }  &\\
 &\longrightarrow \frac{1}{2 \kappa^{2}}\hspace{-0.5mm}\sum_{\substack{\Gamma_i \in \mathds{Z}^{2d(\alpha_d)} \\ \Gamma_i \times \Gamma_j = 0   }} \hspace{-1mm}\int_{\mathds{R}^{11-d}} \hspace{-6mm} d^{11-d} x \phi_{-\Gamma_1 - \Gamma_2}(x)  \Scal{ \partial_\mu \phi_{\Gamma_1 }(x)\partial^\mu \phi_{\Gamma_2 }(x)-\ell^{-2} \langle Z( \Gamma_1) , Z(\Gamma_2) \rangle \phi_{\Gamma_1 }(x) \phi_{\Gamma_2 }(x)}  \nn
\end{align}
and the gauge variation of this expression together with the linearised kinetic term vanishes up to quartic terms, because the fact that two factors satisfying the strong section constraints in a cubic term in the fields implies that all factors do~\cite{Hull:2009mi}. However, starting from quartic terms in the varied Lagrangian the variation of the kinetic and three-point interactions will generally be no longer of the form~\eqref{eq:nproduct}.\footnote{We thank Olaf Hohm and Henning Samtleben for bringing this subtlety to our attention.} 

To illustrate this point, we consider the toy example of a non-linear sigma on $SL(2,\mathds{R})/SO(2)$. The Lagrangian can be written as
\begin{align}
-\mathcal{L}_{SL(2)} &= \frac12 (\partial\phi )^2 + \frac12 e^{2\phi} (\partial a)^2  \nn\\
&=  \frac12 (\partial\phi )^2 +\frac12 (\partial a)^2 \left(1 + 2\phi + 2\phi^2 + \ldots\right)\nn\\
-\mathcal{L}^E_{SL(2)} &= \sum_{\Gamma \in \mathds{Z}^{ d(\alpha_d)}\atop \Gamma \times \Gamma =0} \frac12 \left( \partial \phi_\Gamma \partial \phi_{-\Gamma} +  \partial a_\Gamma \partial a_{-\Gamma} \right) 
+\sum_{\Gamma^i \in \mathds{Z}^{2 d(\alpha_d)}\atop \Gamma_i \times \Gamma_j =0} \partial a_{\Gamma_1} \partial a_{\Gamma_2} \phi_{-\Gamma_1-\Gamma_2}\nn\\
&\quad + \sum_{\Gamma^i \in \mathds{Z}^{3 d(\alpha_d)}\atop \Gamma_i \times \Gamma_j =0} \partial a_{\Gamma_1} \partial a_{\Gamma_2} \phi_{\Gamma_3}\phi_{-\Gamma_1-\Gamma_2-\Gamma_3}+\ldots,
\end{align}
where we have expanded everything up to quartic order. The role of the gauge transformations are played by the global non-linear transformation
\begin{align}
\delta a &= a^2-e^{-2\phi} \quad &\longrightarrow \quad \delta a_{\Gamma_1} &= -\delta_{\Gamma_1,0} + 2 \phi_{\Gamma_1} + \sum_{\Gamma_2\in \mathds{Z}^{d((\alpha_d)} \atop \Gamma_i\times \Gamma_j=0}  \left( a_{-\Gamma_2} a_{\Gamma_1+\Gamma_2} -2\phi_{-\Gamma_2} \phi_{\Gamma_1+\Gamma_2} 
\right)\nn\\
&&& \quad\quad\quad + \frac43 \sum_{\Gamma_2,\Gamma_3\in \mathds{Z}^{2d((\alpha_d)} \atop \Gamma_i\times \Gamma_j=0}  \phi_{-\Gamma_2} \phi_{-\Gamma_3} \phi_{\Gamma_1+\Gamma_2+\Gamma_3}+\ldots\, , \nn\\
 \delta \phi& =-2a \quad &\longrightarrow \quad \delta \phi_\Gamma & =-2 a_\Gamma\, .
\end{align}
Varying the Lagrangian under these transformations yields
\begin{align}
\label{eq:varL}
\delta \mathcal{L}^E_{SL(2)} = -4 \!\!\!\!\!\! \sum_{\substack{\Gamma_i \in \mathds{Z}^{3 d(\alpha_d)}\,,\, \Gamma_i\times \Gamma_i=0\\ \Gamma_2 \times \Gamma_3=0\,,\, \Gamma_1\times(\Gamma_2+\Gamma_3)=0\\ \Gamma_1\times \Gamma_2 \neq 0}}\!\!\!\!\!\!\! \Big( \phi_{-\Gamma_1} a_{-\Gamma_2} \partial a_{-\Gamma_3} \partial a_{\Gamma_1+\Gamma_2+\Gamma_3} -2\partial a_{-\Gamma_1} \partial\phi_{-\Gamma_2} \phi_{-\Gamma_3} \phi_{\Gamma_1+\Gamma_2+\Gamma_3}\Big)\,,
\end{align}
where the important point is that the charges $\Gamma_1$ and $\Gamma_2$ do not satisfy the strong section constraint $\Gamma_1\times \Gamma_2 =0$.  A similar discussion can be found in~\cite{Hull:2009mi}.

The terms in the variation~\eqref{eq:varL} can be compensated for at the price of introducing new  fields into the toy model that satisfy weaker constraints than the $\tfrac12$-BPS constraint $\Gamma\times \Gamma=0$. Instead they solve the $\tfrac14$-BPS constraint that $\Gamma\times \Gamma\neq 0$ but a cubic constraint of the form $(\Gamma, \Gamma, \Gamma)=0$.\footnote{This constraint is automatically satisfied for $d\leq 5$ (as there are no larger charge orbits then) and this tri-linear form is simply the $E_6$ invariant for $d=6$ and the Freudenthal triplet for $E_7$.} In the example coonsidered above, one can define an invariant Lagrangian at this order by introducing extra terms violating the strong section constraint and an additional field $X$, as\begin{align}
-\mathcal{L}^{E^\prime}_{SL(2)} &= \sum_{\Gamma \in \mathds{Z}^{ d(\alpha_d)}\atop \Gamma \times \Gamma =0} \frac12 \left( \partial \phi_\Gamma \partial \phi_{-\Gamma} +  \partial a_\Gamma \partial a_{-\Gamma} \right) 
+\sum_{\Gamma^i \in \mathds{Z}^{2 d(\alpha_d)}\atop \Gamma_i \times \Gamma_j =0} \partial a_{\Gamma_1} \partial a_{\Gamma_2} \phi_{-\Gamma_1-\Gamma_2}\nn\\
&\quad + \!\!\!\!\!\! \sum_{\substack{\Gamma_i \in \mathds{Z}^{3 d(\alpha_d)}\,,\, \Gamma_i\times \Gamma_i=0\\ \Gamma_2 \times \Gamma_3=0\,,\, \Gamma_1\times(\Gamma_2+\Gamma_3)=0}}\!\!\!\!\!\!\!\partial a_{\Gamma_1} \partial a_{\Gamma_2} \phi_{\Gamma_3}\phi_{-\Gamma_1-\Gamma_2-\Gamma_3}+\ldots\nn \\
&\hspace{30mm}+4 \sum_{\substack{\Gamma_i \in \mathds{Z}^{2 d(\alpha_d)}\,,\, \Gamma_i\times \Gamma_i=0\\ \Gamma_1\times \Gamma_2\neq 0}}
\!\!\ X_{\Gamma_1+\Gamma_2}  \partial a_{-\Gamma_1} \partial \phi_{-\Gamma_2}+ \dots  ,
\end{align}
such that the variation of this field is
\begin{align}
\label{eq:NewFields}
\delta X_{\Gamma} = \sum_{\substack{\Gamma_1 \in \mathds{Z}^{d(\alpha_d)}\,,\, \Gamma_1\times \Gamma_1=0\\ \Gamma\times \Gamma= 2\Gamma\times \Gamma_1 \ne 0} } \phi_{\Gamma_1} \phi_{\Gamma-\Gamma_1}\, . 
\end{align}
This toy model computation suggests that it is necessary to add $\tfrac14$-BPS multiplets of fields, represented here by $X_\Gamma$, to preserve exceptional diffeomorphism invariance while including all the $\tfrac12$-BPS multiplets, \ie not choosing a fixed truncation solving the strong section constraint. Returning to the general discussion of the strong section constraint, it seems plausible to assume that this process will continue at higher orders and one should include all possible non-perturbative states in order to obtain a gauge invariant effective action, including BPS and non-BPS states. Taking into account the fact that all charges $\Gamma$ for $d\leq 7$ satisfy some BPS condition it would be interesting to investigate whether a truncation to only BPS states for $d\leq 7$ is consistent.

The thus completed exceptional field theory Lagrangian defines a standard Lagrangian in $11-d$ dimensions that involves \textit{infinitely many} fields. It is important to note that this Lagrangian involves more fields than for any explicit solution to the section constraint defining a consistent supergravity truncation. The section constraint only implies that the fields associated to incompatible truncations (to \eg the M-theory or type IIB frame) do not interact \textit{directly} through local monomials in the Lagrangian. As we have seen above, in a more general setting we should only assume that they do not interact through three-point vertex. For example, the explicit solution associated to 11-dimensional supergravity corresponds by construction to considering charges such that only the Kaluza--Klein momentum $n_I$ in~\eqref{Decompose} is non-zero. A solution to the constraint associated to type IIB supergravity is obtained by T-duality~\cite{Blair:2013gqa}, by considering instead that only a rank 2 M2-brane wrapping number $n_{IJ}$ and its transverse momentum $n_I$ are non-zero, \ie 
\be 
n^{[IJ} n^{KL]} = 0 \ , \qquad n^{IJ} n_J= 0 \  .  
\ee
Without loss of generality, one can consider an $SL(2)\times SL(d-2)\subset SL(d)$ split of the indices $I$ to $\hat{r}=1,\, 2$ and $r$ ranging from $3$ to $d$, such that only $n^{12}$ and $n_r$ are non-zero, and define the $d-1$ type IIB Kaluza--Klein momenta. 
 By construction, one will have type IIB n-point interactions with only the charges $n_i^{12},\, n_{i\, r}$ being non-zero, and 11-dimensional supergravity interactions with only the charges  $n_{i\, \hat{r}},\, n_{i\, r}$ being non-zero. By the completion induced by gauge invariance as above one, interactions involving Fourier modes with both $n_i^{12}$ and $n_{j\, \hat{r}}$ non-zero for any $i$ and $j$ will be mediated by $\tfrac14$-BPS states or higher point vertices.

The problem with gauge invariance discussed above only arises when there are more than two independent charges $\Gamma$ involved in the process from which we conclude that tree-level processes with at most two non-trivial charges $\Gamma$ on the external lines can be consistently treated in our framework. Equivalently, any loop diagram with not more than two non-trivial charges propagating do not suffer from the above pathologies. Since we restrict to two-loop diagrams in this work, the subtlety will not affect our analysis. Therefore we will in the following only consider the fields arising in exceptional field theory in the standard framework. 
 
Exceptional field theory is invariant with respect to infinitesimal general coordinate transformations along the internal extended space, that act on a generalised vector as \cite{Berman:2012vc,Hohm:2013vpa,Hohm:2013uia}
\be
\delta_\Lambda V^M  = \Lambda^N \partial_N V^M - n_d ( \partial_N \Lambda^M)|_{ \mathfrak{e}_{d(d)}} V^N + \lambda\,  \partial_N \Lambda^N V^M \ , 
\ee
where the matrix $\partial_N \Lambda^M$ in the second term is projected to the adjoint representation of $E_{d(d)}$ and\footnote{$n_d$ defines the multiplicity of the fundamental representation ${\bf d}$ of $SL(d)\subset E_{d}$ in the representation $R_{\alpha_d}$, such that for $h\in \mathfrak{sl}_d\subset \mathfrak{e}_{d}$, ${\rm Tr}_{R_{\alpha_d}} h^2 = n_d {\rm Tr}_{{\bf d}} h^2$. For $d=8$ the formula above does not include all representations and one would have $n_8  = 60$.}
\be 
n_d = d-1 + \tfrac{(d-2)(d-3)(d-4)(d-5)}{24} + \tfrac{(d-2)(d-3)(d-4)(d-5)(d-6)d(4d-25)}{2520} \ .   
\ee
The variable $\lambda$ above denotes the density weight of the generalised vector under scaling transformations. For a recent reformulation of the generalised coordinate transformation in terms of a Borcherds algebra see~\cite{Palmkvist:2015dea}.

All the fields in the theory also transform covariantly with respect to rigid  transformations $g\in E_{d(d)}$,
\be 
g_{\mu\nu}(x,Y) \rightarrow  g_{\mu\nu}(x,Yg^{-1} ) \ , \qquad A_{\mu}^M(x,Y)  \rightarrow  A_{\mu}^N(x, Yg^{-1}) g_N{}^{M} \ ,\quad  \dots 
\ee
See for example\cite{Rey:2015mba} for an explicit construction in seven dimensions. On the dual lattice space we will therefore get the action $\Gamma_i \rightarrow g \Gamma_i$ that will be a symmetry of the theory for $g\in E_{d(d)}(\mathds{Z})$. The generalised diffeomorphism invariance of the exceptional field theory Lagrangian (when appropriately completed by the additional fields discussed above) should therefore ensure the theory on $\mathds{R}^{11-d} \times T^d$ to be invariant with respect to the U-duality symmetry $E_{d(d)}(\mathds{Z})$. In even dimensions one should in principle moreover consider a generalised Henneaux--Teitelboim formulation of the theory in order to define properly the $E_{d(d)}$ Ward identities \cite{Henneaux:1988gg,Hillmann:2009zf,Bossard:2010dq}, but in this paper we will assume without proof that the amplitudes satisfy the $E_{d(d)}$ symmetry. 

As alluded to in footnote~\ref{fn:addSC},  there is an additional section constraint in four dimensions related to the invariant symplectic trace~\cite{Hohm:2013uia}, but we will see in section \ref{Nondege} that the algebraic constraint \eqref{2ChargeRk1} in Fourier space is sufficient to imply that all the charge vectors must be isotropic (\ie that the symplectic products $\langle \Gamma_i, \Gamma_j\rangle $ vanish). Equivalently in three dimensions, we will see that \eqref{2ChargeRk1} implies the two $\mathfrak{e}_{8(8)}$ elements $\Gamma_i$ to commute and to be orthogonal with respect to the Cartan bilinear form such that the additional sections constraints  \cite{Hohm:2014fxa} are automatically satisfied.

\subsection{Example: section constraint and propagators for $E_{6(6)}$}

In this section, we will derive the modification of the exceptional field theory propagators due to the charges $\Gamma\in {\bf R}_{\alpha_d}$ in the exemplary case of $E_{6(6)}$. The final result is intuitive and used in the next sections, so that the reader who is not interested in these technical details can skip immediately to section~\ref{sec:EFTamps}.

For each integral charge $\Gamma$ one obtains a $\tfrac12$-BPS massive multiplet of particles that can be interpreted as a Kaluza--Klein mode of a massless multiplet in one space-time dimension higher. The various fields of exceptional field theory related by generalised gauge transformations combine in this way to carry massive degrees of freedom through a Higgs-like mechanism. We discuss this point in some more detail in five dimensions as a preparation to a more general discussion of the construction of Feynman-like rules in the next section. 

In the $D=5$ case a rank one charge $\Gamma_M$ satisfying \eqref{2ChargeRk1} breaks the $E_{6(6)}$ symmetry to the subgroup 
\be 
Spin(5,5) \ltimes \mathds{R}^{16} \subset E_{6(6)} \ ,
\ee
that represents the stabiliser of the charge. The dimension of the corresponding coset is $78-(45+16)=17$ in correspondence with (half) the size of the minimal nilpotent $E_7(\mathds{C})$ orbit of dimension $34$. In the linearised approximation all free fields then decompose in representations of $Spin(5)\times Spin(5)\cong Sp(2)\times Sp(2) \subset Sp(4)$. 

We consider the expansion around the Minkowski metric $ \eta_{\mu\nu}$ and the moduli\footnote{The moduli matrix $M^{MN}$ is the extension of the torus metric $M^{IJ}$ to include all $E_{6(6)}$ moduli.} $M^{MN} \in E_{6(6)}$
\be 
g_{\mu\nu} = \eta_{\mu\nu} + h_{\mu\nu} \ , \qquad {\cal M}^{MN} = M^{MP} \exp( M \Phi )_P{}^N =  M^{MN} + \Phi^{MN}  + \mathcal{O}(\Phi^2) \ ,  
\ee
where $M=1,\ldots,27$ and with the constraint 
\be 
t_{MPR} t^{NQR} M_{QS} \Phi^{PS} = 0 \ , 
\ee
ensuring that $M_{MP} \Phi^{PN} \in \mathfrak{e}_{6(6)}$, where the $E_{6(6)}$ invariant tensor $t_{MNP}$ defines the Jordan cross product and the cubic form through
\be 
( \Lambda \times \Lambda )_M = \frac{1}{2} t_{MNP} \Lambda^N \Lambda^P  \ , \qquad \det \Lambda = \frac{1}{6}  t_{MNP}\Lambda^M  \Lambda^N \Lambda^P\ , 
\ee
that satisfy\footnote{For convenience we define $t^{MNP} = \sqrt{10} d^{MNP}$ and rescale accordingly $B_{\mu\nu M}$ and $\Xi_{\mu\nu M}$ by $\sqrt{10}$ with respect to the conventions of \cite{Hohm:2013vpa}.}
\be  
( \Lambda \times \Lambda ) \times   ( \Lambda \times \Lambda )  =\lp\det \Lambda\rp \, \Lambda \ . 
\ee
The gauge transformations of the bosons displayed in \cite{Hohm:2013vpa} reduce in the linearised approximation to the following expressions in momentum space (where we avoid writing that all fields and gauge parameters have momenta $(p_\mu,\ell^{-1}\Gamma_M)$):
\allowDbreak{\begin{subequations}
\begin{align}
\delta h_{\mu\nu} &=2 p_{(\mu} \xi_{\nu)}  + \frac{2}{3} \ell^{-1}\eta_{\mu\nu} \Gamma_M \Lambda^M  \ , &\\
\delta A_\mu^M &= \ell^{-1} M^{MN} \Gamma_N \xi_\mu + p_\mu \Lambda^M - \ell^{-1} t^{MNP} \Gamma_N \Xi_{\mu P}\ , & \\
\delta B_{\mu\nu M} &= 2 p_{[\mu} \Xi_{\nu]M} + \mathcal{O}_{\mu\nu M} \ , &\\
\delta \Phi^{MN} &= 2\ell^{-1} \Gamma_S \Lambda^{P}  t_{PQR} t^{RS(M}  M^{N)Q}  - 2\ell^{-1} \Gamma_P \Lambda^{(M} M^{N)P} - \frac{2}{3}\ell^{-1} \Lambda^P \Gamma_P M^{MN} &\\
&= - 12\ell^{-1} ( \Gamma_P \Lambda^{(M})|_{\mathfrak{e}_{6(6)}} M^{N)P}  \ .&
\end{align}
\end{subequations}
The charge $\Gamma$ defines the following generator $H_M{}^N\in \mathfrak{e}_{6(6)}$
\be 
H_M{}^N \equiv \frac{1}{|Z(\Gamma)|^2} \Scal{ 3 \Gamma_M M^{NP} \Gamma_P - 3t_{MPR} t^{NQR} \Gamma_Q M^{PS} \Gamma_S  + \delta_M^N M^{PQ} \Gamma_P \Gamma_Q } \ , 
\ee
that defines the following graded decomposition of $ \mathfrak{e}_{6(6)}$
\be 
\mathfrak{e}_{6(6)} \cong \overline{\bf 16}^\ord{-3} \oplus ( \mathfrak{gl}_1 \oplus \mathfrak{so}(5,5) )^\ord{0} \oplus {\bf 16}^\ord{3} \ , \qquad {\bf 27} \cong {\bf 1}^\ord{-4} \oplus {\bf 16}^\ord{-1} \oplus {\bf 10}^\ord{2}\ . 
\ee
The projection of $\Gamma_M \Lambda^N$ to the adjoint representation of $Spin(5,5)$ includes the ${\bf 1}^\ord{0}$ singlet and the ${\bf 16}^\ord{3}$ spinor component. The linearised diffeomorphisms and the singlet component $\Gamma_M \Lambda^M$ mix together the metric $g_{\mu\nu}$ and the $Spin(5,5)$ singlet components of $A_\mu^M$ and ${\cal M}^{MN}$, producing the $9=5+3+1$ on-shell degrees of freedom of a massive spin $(1,1)$ particle. The ${\bf 16}$ component of $\Lambda^M$ mixes the ${\bf 16}$ components of the vector field and the scalar, producing the $4=3+1$ on-shell degrees of freedom of a massive spin $(\tfrac{1}{2},\tfrac{1}{2})$ particle in four dimensions. The ${\bf 10}$ components of the vector and the tensor field mix under the tensor field gauge symmetry to define $5$  spin $(1,0)$ and 5 spin $(0,1)$ massive particles. The gauge parameter  $\mathcal{O}_{\mu\nu}$ is a general function satisfying $ \Gamma\times \mathcal{O}_{\mu\nu}= 0$, and permits to gauge away the other components of the tensor field, \ie the ${\bf 1}^\ord{4}$ singlet and the ${\bf 16}^\ord{1}$ spinor components. }

To understand this in general one must consider all the propagators together because they potentially exchange all the bosonic fields associated to the same particles.  In the linearised approximation, one finds using \cite{Hohm:2013vpa} that the bosonic fields of the theory satisfy the following equations with external sources (where $\tilde{\varphi}_A$ is the source of the field $\varphi^A$)
\bea
\tilde{h}_{\mu\nu}&=& 2 G_{\mu\nu}(h) +  \ell^{-2} |Z(\Gamma)|^2 ( h_{\mu\nu} - \eta_{\mu\nu} h_\sigma{}^\sigma) - 2\ell^{-1} \langle p_{(\mu} A_{\nu)} - \eta_{\mu\nu} p^\sigma A_\sigma , \Gamma\rangle -  \eta_{\mu\nu}  \ell^{-2}  \langle \Gamma \Phi \Gamma \rangle \ , \CR
 \tilde{A}_\mu &=& M\scal{ p^2 A_\mu - p_\mu p^\nu A_\nu  + \ell^{-2} |Z(\Gamma)|^2 A_\mu  - 4 \ell^{-2} \Gamma \times ( M^{-1}(\Gamma) \times A_\mu) - 2  \ell^{-1} \Gamma \times p^\nu B_{\mu\nu} } \CR
&&\hspace{45mm} - \ell^{-1} \Gamma \scal{ \ell^{-1} \langle \Gamma , A_\mu\rangle  + p^\nu h_{\mu\nu} - p_\mu h_{\nu}{}^\nu}   + \ell^{-1} M\scal{  p_\mu \Phi(\Gamma)} \ , \CR
\tilde{B}^{\mu\nu} &=& \ell^{-1} \varepsilon^{\mu\nu\sigma\rho\kappa} p_\kappa \Gamma \times B_{\sigma\rho} + 4\ell^{-1}  \Gamma \times M\scal{ p^{[\mu} A^{\nu]} + \ell^{-1} \Gamma \times B^{\mu\nu}} \ , \CR
\tilde{\Phi}_{MN} &=& \Scal{  p^2 + \ell^{-2} |Z(\Gamma)|^2 } M_{MP} M_{NQ} \Phi^{PQ} \\
&& \hspace{10mm} +6 \ell^{-1}  M_{P(M} \Scal{ -2\ell^{-1} \Gamma_{N)} \Gamma_Q \Phi^{PQ} -\ell^{-1}  \Gamma_{N)} \Gamma_Q M^{PQ}  h_\mu{}^\mu+ 2 \Gamma_{N)} p^\mu A_\mu^P }\Big|_{\mathfrak{e}_{6(6)}} \ , \ \nn
\eea
where we use the notation
\be
\begin{split} (M(A_\mu))_M &\equiv M_{MN} A_{\mu}^N \ , \\
\langle \Gamma, \Lambda \rangle &\equiv \Gamma_M \Lambda^M \ ,
\end{split}\hspace{10mm}\begin{split}
  (M^{-1}(B_{\mu\nu}))^M &\equiv M^{MN} B_{\mu\nu N}\ , \\
  |Z(\Gamma)|^2 &\equiv M^{MN} \Gamma_M \Gamma_N \ .
  \end{split}
  \ee
The linearised $\mathfrak{e}_{6(6)}$ current ${\cal M}_{MP} \partial_\mu {\cal M}^{PN}$ acting on $\Gamma$ is
\be 
\scal{M\scal{  p_\mu \Phi(\Gamma)}}{}_M  = M_{MN} p_\mu \Phi^{NP} \Gamma_P \ , 
\ee
and $G_{\mu\nu}(h)$ is the linearised Einstein tensor associated to $h_{\mu\nu}$, \ie 
\be  
2 G_{\mu\nu}(h) = p^2 h_{\mu\nu} - 2 p_{(\mu} p^\sigma h_{\nu)\sigma} + p_\mu p_\nu h_\sigma{}^\sigma + \eta_{\mu\nu} ( p^\sigma p^\rho h_{\sigma \rho} - p^2 h_\sigma{}^\sigma ) \ . 
\ee
Note that gauge invariance fixes all the corrections to the standard massless free field equations for all these fields, consistently with the property that generalised diffeomorphism invariance completely determines the bosonic action at the non-linear level  \cite{Hohm:2013vpa}. To show this one uses in particular 
 \be 
 4\,  \Gamma \times ( M^{-1}(\Gamma) \times ( \Gamma \times B)) = \Gamma \times B \,   |Z(\Gamma)|^2 \ , 
 \ee
 which is a consequence of the strong section constraint, as well as the property that $M\in E_{6(6)}$, such that \eg
\be 
M\scal{ \Gamma \times M(A)} = M^{-1}(\Gamma) \times A \ . 
\ee
Gauge invariance is translated into the property that the sources are not independent, but satisfy the constraints
\bea 
&& \ell \, p^\mu \tilde{A}_\mu = \tilde{\Phi}(M^{-1}(\Gamma)) - \frac{1}{3} \Gamma \tilde{h}_\mu{}^\mu \ , \quad p^\nu \tilde{h}_{\mu\nu} = - \ell^{-1} \langle \Gamma, M^{-1}(\tilde{A_\mu})\rangle \ , \quad  p_\nu \tilde{B}^{\mu\nu} = - 2 \ell^{-1} \Gamma \times \tilde{A}^\mu \ , \CR
&& 4 \, \Gamma \times ( M^{-1}(\Gamma) \times \tilde{B}_{\mu\nu}) = |Z(\Gamma)|^2 \tilde{B}_{\mu\nu}  \  .   
\eea
To exhibit the spectrum it is convenient to consider the unitary gauge 
\bea
&&\begin{split} h_\mu{}^\mu &= 0 \ , \\
\Phi(\Gamma) &= 0 \ , 
\end{split}\hspace{10mm}\begin{split}  p^\nu h_{\mu\nu} &= 0 \ , \\
\langle \Gamma , A_\mu \rangle &= 0 \ , 
\end{split}\hspace{10mm}\begin{split}  
p^\mu A_{\mu} &= 0\ ,  \\
\Gamma \times M(A_\mu) &= 0 \ , 
\end{split}\hspace{10mm}\begin{split}  
\end{split}\hspace{00mm}\begin{split}  
p^\nu B_{\mu\nu} &=0 \ , \\
&
\end{split} \CR
&&t_{MPR} t^{NQR} M^{PS} \Gamma_S B_{\mu\nu Q} = M^{NP} \Gamma_P B_{\mu\nu M} \ , 
\eea
where the first line corresponds to the usual space-time constraints whereas the two others define algebraic constraints in internal space. The first constraint on the scalar fields states that $M_{MP}\Phi^{PN}$ is in the semi-simple stabilizer of $\Gamma_M$ and $M^{MN} \Gamma_N$, and therefore parametrizes the symmetric space $SO(5,5)/(SO(5)\times SO(5))$. The two algebraic constraints on $A_\mu$ imply that it is only non-zero in the spinor representation of $Spin(5,5)$. The corresponding propagator reduces in this gauge to the standard massive vector field propagator projected on the spinor representation, \ie 
\be 
\Delta_{\mu\nu}^{MN} = \frac{\eta_{\mu\nu} +\ell^2  \frac{p_\mu p_\nu}{|Z(\Gamma)|^2}}{p^2 + \ell^{-2}  |Z(\Gamma)|^2 - i \epsilon } \Scal{ M^{MN} - \tfrac{1}{|Z(\Gamma)|^2} \scal{   t^{MPR} t^{NQS} M_{RS} \Gamma_P \Gamma_Q + M^{MP} M^{NQ} \Gamma_P \Gamma_Q  } } \ . 
\ee
The constraint on $B_{\mu\nu}$ implies that it belongs to the ${\bf 10}^\ord{-2}$. The tensor field equations can be inverted up to an $\mathcal{O}_{\mu\nu}$ type gauge transformation to 
\be 
B_{\mu\nu} = \frac{1}{p^2 + \ell^{-2}  |Z(\Gamma)|^2 - i \epsilon } \Scal{ - \frac{\ell }{   |Z(\Gamma)|^2} \varepsilon_{\mu\nu\sigma\rho\kappa} p^\kappa \, M^{-1}(\Gamma) \times \tilde{B}^{\sigma\rho} + M\scal{ \tilde{B}_{\mu\nu} - \tfrac{2\ell^2}{|Z(\Gamma)|^2} p_{[\mu} p^\sigma \tilde{B}_{\nu]\sigma} }}  \ .
\label{Propagators} 
\ee
The pole of the tensor field propagator
\bea
&&  \Delta_{\mu\nu M, \sigma\rho N}\\
 & =&  \frac{- \ell  \varepsilon_{\mu\nu\sigma\rho\kappa} p^\kappa \, t_{MNP} M^{PQ} \Gamma_Q  + M_{MP} M_{NQ} M_{RS} t^{PRT} t^{QSU} \Gamma_T \Gamma_U  \scal{ \eta_{\mu[\sigma} \eta_{\rho]\nu} - \tfrac{2\ell^2}{|Z(\Gamma)|^2} p_{[\mu} \eta_{\nu][\sigma} p_{\rho]}  }}{  |Z(\Gamma)|^2 ( p^2 + \ell^{-2}  |Z(\Gamma)|^2 - i \epsilon)  } \nn 
\eea
projects out the polarisations to self-dual and antiself-dual tensors with respect to the little group $SO(4)$, according to their representation in the $Sp(2)\times Sp(2)\subset Sp(4)$ stabilizer of $Z(\Gamma)$ in the ${\bf 27}$ of $Sp(4)$, such that the $10$ tensor field degrees of freedom decompose into $5$ spin $(1,0)$ and 5 spin $(0,1)$ polarisations. Note indeed that $t_{MNP}  M^{PQ} \Gamma_Q$ defines the split signature metric for $SO(5,5)$ vectors. The propagators project on the corresponding irreducible representations of $Spin(5,5)$ such that
\be 
- H_P{}^M  \Delta_{\mu\nu}^{PN} = -  \Delta_{\mu\nu}^{MN} \ , \qquad  H_M{}^P \Delta_{\mu\nu P, \sigma\rho N}  = - 2 \Delta_{\mu\nu M, \sigma\rho N} \ . 
\ee
The massive spectrum in five dimensions is therefore defined by the same representations  of $SU(2)\times SU(2) \times SO(5,5)$ as the massless spectrum in six dimensions, in agreement with super-Poincar\'e representations  \cite{Ferrara:1980ra}.

\subsection{Exceptional field theory amplitudes and locality}
\label{sec:EFTamps}

In this section, we will argue that on-shell amplitudes in exceptional field theory can be computed directly using unitarity, without referring to the explicit Lagrangian. For simplicity we will restrict ourselves to four-graviton scattering amplitudes, although we expect the method to be generalizable to any amplitude. Consider the $L$-loop 4-graviton amplitude in $11-d$ dimensions written in a double copy form \cite{Bern:2010ue}
\be 
\cM_{4,L}(1,2,3,4) = i^{1+L} \Scal{ \frac{\kappa}{2}}^{2+2L}  \sum_G \int \prod_{l=1}^L \frac{d^{11-d} p_l}{(2\pi)^{11-d}} \frac{1}{S_G} \frac{N^2_G(k_A,e_A,p_l)}{\prod_{I_G} \scal{ p_{I_G}(k_A,p_l)}^{2}} \ ,  
\ee
where the external momenta $k_A$ and helicities $e_A$ with $A$ ranging from $1$ to $4$ are restricted to massless states in $11-d$ dimensions. Here $S_G$ is a symmetry factor associated to the graph $G$ and $N^2_G(k_A,e_A,p_l)$ is the associated kinematic numerator, that is a polynomial in the helicities and the momenta. The latter is the square of the Bern--Carrasco--Johansson Yang--Mills numerator up to 4-loop \cite{Bern:2010ue}, but we shall not use this property in this paper. The amplitude integrand on $\mathds{R}^{11-d} \times T^d$ is obtained by discretising the component of the 11-dimensional loop momenta along the torus. The scalar products of the loop momenta $p_l$ with the external momenta and helicities are not modified, and only the terms involving the scalar product of loop momenta together is modified according to the substitution~\eqref{eq:loopmom}
\be 
( p_l,p_{l^\prime}) \rightarrow  ( p_l,p_{l^\prime}) + \ell^{-2} e^{-3\phi} M^{IJ} n_{l I} n_{l^\prime J} \ ,   
\ee
so that one obtains 
\be 
\cM_{4,L}^{\Td}(1,2,3,4) = i^{1+L} \Scal{ \frac{\kappa}{2}}^{2+2L}  \sum_G \sum_{n_l \in \mathds{Z}^{Ld}} \int \prod_{l=1}^L \frac{d^{11-d} p_l}{(2\pi)^{11-d}} \frac{1}{S_G} \frac{N_G^2(k_A,e_A,p_l\oplus n_l)}{\prod_{I_G} \scal{p_{I_G}( k_A,p_l\oplus n_l)}^{2}}   \ . 
\ee
These integrands are determined by the generalised unitary cuts and the tree-level 3-point amplitudes \cite{Bern:1994zx,Bern:1994cg}. In standard exceptional field theory, all the 3-point vertices between $\tfrac12$-BPS states are defined by restricted sums as in \eqref{3-point}, such that they are only non-zero if two adjacent charges satisfy the strong section constraint.\footnote{If one completes the theory with additional fields in order to ensure gauge invariance there are additional 3-point couplings involving states not satisfying the $\tfrac12$-BPS strong section constraint.}

Although all \textit{local} interactions are between $\tfrac12$-BPS fields satisfying the strong section constraint, there are 4-point tree-level amplitudes that violate it globally. Consider for example four component charge vectors $\Gamma = ( n_1,n_2,n_3,n^{12})$, where it is understood that all the other charge components vanish. One can have non-zero 4-point amplitudes with the charges 
 \def\xshift{5}
  \def\xmin{1}
 \def\ymin{-2}
\be
 \begin{tikzpicture}
  \draw (\xmin,\ymin) node{$(0,0,n_3,0)$};
\draw[-,draw=black,very thick](\xmin+1,\ymin) -- (\xmin+1.5,\ymin);
\draw[-,draw=black,very thick](\xmin+1.5,\ymin) -- (\xmin+2,\ymin+0.5);
 \draw[-,draw=black,very thick](\xmin+1.5,\ymin) -- (\xmin+2,\ymin-0.5);
  \draw (\xmin+3.2,\ymin+0.6) node{$(0,0,0,n^{12})$};
  \draw (\xmin+3.3,\ymin-0.6) node{$(0,0,n_3,-n^{12})$};
\draw[-,draw=black,very thick](\xmin-1,\ymin) -- (\xmin-1.5,\ymin);
\draw[-,draw=black,very thick](\xmin-1.5,\ymin) -- (\xmin-2,\ymin+0.5);
 \draw[-,draw=black,very thick](\xmin-1.5,\ymin) -- (\xmin-2,\ymin-0.5);
  \draw (\xmin-3.1,\ymin+0.6) node{$(n_1,0,0,0)$};
  \draw (\xmin-3.2,\ymin-0.6) node{$(-n_1,0,n_3,0)$};
\end{tikzpicture}
\label{4-pointMIIB}
\ee
which corresponds to the scattering  of two 11-dimensional Kaluza--Klein states into two type IIB supergravity Kaluza--Klein states. This is an explicit example of the statement below~\eqref{3-point} that fields not interacting directly can violate a given solution of the section constraint. As discussed before, unitarity requires also contributions to this process in the other channels involving an intermediate $\tfrac14$-BPS state.

In $D=4$ dimensions one can have even more extravagant scatterings at 5-point, with $\Gamma=(n_1,n^{23},n^{12345},n^{1,1234567})$,
\be 
 \begin{tikzpicture}
  \draw (\xmin-1.5,\ymin) node{$(0,n^{23},0,0)$};
\draw[-,draw=black,very thick](\xmin-0.5,\ymin) -- (\xmin+0.5,\ymin);
\draw[-,draw=black,very thick](\xmin-2.5,\ymin) -- (\xmin-3,\ymin);
\draw[-,draw=black,very thick](\xmin-3,\ymin) -- (\xmin-3.5,\ymin+0.5);
 \draw[-,draw=black,very thick](\xmin-3,\ymin) -- (\xmin-3.5,\ymin-0.5);
  \draw (\xmin-4.5,\ymin+0.6) node{$(n_1,0,0,0)$};
  \draw (\xmin-4.8,\ymin-0.6) node{$(-n_1,n^{23},0,0)$};
    \draw (\xmin+1.5,\ymin) node{$(0,0,\tilde n_{67},0)$};
\draw[-,draw=black,very thick](\xmin+2.5,\ymin) -- (\xmin+3,\ymin);
\draw[-,draw=black,very thick](\xmin+3,\ymin) -- (\xmin+3.5,\ymin+0.5);
 \draw[-,draw=black,very thick](\xmin+3,\ymin) -- (\xmin+3.5,\ymin-0.5);
  \draw (\xmin+4.5,\ymin+0.6) node{$(0,0,0,\tilde n^1)$};
  \draw (\xmin+4.8,\ymin-0.6) node{$(0,0,\tilde n_{67},-\tilde n^1)$};
  \draw[-,draw=black,very thick](\xmin,\ymin) -- (\xmin,\ymin-0.6);
  \draw (\xmin,\ymin-0.9) node{$(0,n^{23},-\tilde n_{67},0)$};
\end{tikzpicture}
\label{5-pointKK}
\ee
such that some outgoing charges do not satisfy \textit{any} constraint with respect to the incoming charges. One can think of the first diagram \eqref{4-pointMIIB} as describing a T-fold transition and the second \eqref{5-pointKK}, an S-fold transition, where outgoing states are S-dual to incoming states. Note that in these specific scattering processes with the shown choice of charges and only $\tfrac12$-BPS vertices, only one channel is permitted, whereas a scattering involving only 11-dimensional Kaluza--Klein states would involve all possible channels. Unitarity should imply nonetheless that there are contributions in the other channels involving $\tfrac14$-BPS and $\tfrac18$-BPS states. One must therefore consider separately loop diagrams whose unitarity cuts involve such processes from the ones that only involve external states with charges satisfying in a pairwise manner the strong section constraint. In particular, divergences for diagrams where not all charges satisfy the strong section constraint in pairs will not be treatable by counterterms that are constructed from a local counterterm (in the strong section constraint sense). This might have implications for the search for higher derivative corrections solely in standard exceptional field theory~\cite{Godazgar:2013bja,Coimbra:2014qaa} or double field theory~\cite{Hohm:2013jaa,Marques:2015vua}, see also~\cite{Lambert:2006he,Damour:2006ez,Bao:2007er}. 

\begin{figure}[htbp]
\begin{center}
 \begin{tikzpicture}
  \draw[-,draw=black,very thick] (\xmin-2,\ymin) circle (1cm);
    \draw[-,draw=black,very thick](\xmin-2,\ymin) -- (\xmin-2,\ymin+1);
        \draw[-,draw=black,very thick](\xmin-2,\ymin) -- (\xmin-2+0.866,\ymin-0.5);
                \draw[-,draw=black,very thick](\xmin-2,\ymin) -- (\xmin-2-0.866,\ymin-0.5);
 \draw[-,draw=black,dashed](\xmin-2-0.5,\ymin-0.6) -- (\xmin-2-0.5,\ymin+1.3);
 \draw[-,draw=black,dashed](\xmin-2+0.5,\ymin-0.6) -- (\xmin-2+0.5,\ymin+1.3);
  \draw[-,draw=black,very thick] (\xmin+2,\ymin) circle (1cm);
        \draw[-,draw=black,very thick](\xmin+2+0.33,\ymin-0.94) -- (\xmin+2+0.33,\ymin+0.94);
        \draw[-,draw=black,very thick](\xmin+2-0.33,\ymin-0.94) -- (\xmin+2-0.33,\ymin+0.94);
                \draw[-,draw=black,dashed](\xmin+2-1.5,\ymin) -- (\xmin+2+1.5,\ymin);
      \end{tikzpicture}
\end{center}
\caption{\small Example cuts of the 3-loop vacuum diagrams. The Mercedes diagram cut includes two 4-point tree diagrams, but the unconstrained external lines are connected to constrained ones such that the momenta satisfy the strong section constraint. On the contrary for the ladder diagram cut,  the unconstrained external lines are connected to unconstrained ones.}
\label{3loopGraph}
\end{figure}
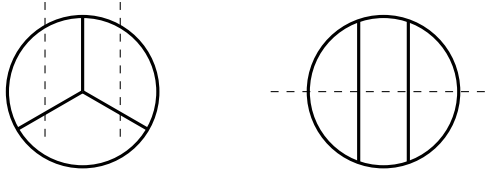

Because we restrict our analysis to scatterings of massless particles, one can forget about external states in analysing possible charge transitions, and consider only vacuum diagrams. A process like \eqref{4-pointMIIB} can only occur if there are at least three independent loop momenta, and so \textit{can only occur at 3-loop and beyond}. For the Mercedes 3-loop diagram all loop momenta are connected to common vertices, so there is no such a tree-level graph as \eqref{4-pointMIIB} involved in the unitarity cuts, see figure \ref{3loopGraph}. However, the 3-loop ladder diagram admits a unitarity cut involving such a tree-level graph. At 4-loop all graphs involved in the four-graviton amplitude \cite{Bern:2012uf} include a tree like  \eqref{4-pointMIIB}, and the 4-loop ladder diagram involves a tree amplitude of the same type as \eqref{5-pointKK}. 

We conclude that the exceptional field theory amplitude will involve a sum over all loop discrete charges $\Gamma_l$, such that \textit{all pairs of charges connected to a common vertex} satisfy the strong section constraint. If, however, all charges for a given diagram happen to satisfy the strong section constraint for all combinations the diagram describes a situation that could be realised in a fixed supergravity frame. This is crucially not required in general by exceptional field theory.

A general exceptional field theory amplitude at $L$ loops will read
\be 
\cM^{\rm E}_{4,L}(1,2,3,4) = i^{1+L} \Scal{ \frac{\kappa}{2}}^{2+2L}  \sum_G\hspace{-1.5mm} \sum_{\substack{\Gamma_l \in \mathds{Z}^{Ld(\alpha_d)} \\ \Gamma_l \times \Gamma_{l^\prime} = 0  \ \forall \langle l l^\prime\rangle  }} \int \prod_{l=1}^L \frac{d^{11-d} p_l}{(2\pi)^{11-d}} \frac{1}{S_G} \frac{N_G^E(k_A,e_A,p_l,\Gamma_l)}{\prod_{I_G} \scal{p_{I_G}( k_A,p_l\oplus \Gamma_l)}^{2}} \ , 
\ee
where the notation $\langle l l'\rangle$ in the summation over the loop charges $\Gamma_l$ indicates that the strong section constraint only has to be satisfied for adjacent (nearest neighbour) charges. The scalar propagators of loop momenta are promoted to the $E_{d(d)}$ invariant quadratic form 
\be 
( p_l,p_{l^\prime}) \rightarrow  ( p_l,p_{l^\prime}) + \ell^{-2} \langle Z(\Gamma_{l}) ,Z(\Gamma_{l^\prime})\rangle    \label{ScalarE} 
\ee
according to the discussion of the preceding section.

If the charges in the numerator satisfy the strong section constraint for any pairs (and not only for nearest neighbours) it will be identical to the numerator that arises in standard supergravity:
\be  
N_G^E(k_A,e_A,p_l,\Gamma_l)\big|_{\Gamma_l \times \Gamma_{l^\prime}=0}  =  N_G^2(k_A,e_A,p_l\oplus \Gamma_l) \ .\label{ExcKin}  
\ee
In exceptional field theory $N_G^E$ can differ from the supergravity expression by terms where non-neighbouring charges violate the strong section constraint.

We separate the dependence in $p_l$ and $\Gamma_l$ in the exceptional kinematic numerator $N_G^E$ to emphasise that it does not necessarily depend only on  $\Gamma_l$  through the scalar products  \eqref{ScalarE}. It appears therefore that the kinematic numerator is only determined up to monomials in the momenta that vanish when the charges are subjected to the strong section constraint, as for example 
\be 
8 Z_{ij}(\Gamma_1) Z^{jk}(\Gamma_4) Z_{kl}(\Gamma_1) Z^{li}(\Gamma_4) -\scal{ Z_{ij}(\Gamma_1)Z^{ij}(\Gamma_4) }^2  + \mbox{c.c.} \ , \qquad I_4(\Gamma_1+\Gamma_4)\ , 
\ee
in $\cN=8$ supergravity, where $Z_{ij}(\Gamma)$ is the antisymmetric rank two $SU(8)$ tensor central charge, and $I_4$ the $E_{7(7)}$ quartic invariant. One may expect nonetheless simplifications because the kinematic numerators are low order polynomials in the loop mementa, such that the number of possible corrections could be rather small. In particular, the kinematic numerators of ladder diagrams do not depend on the loop momenta  \cite{Bern:2012uf,Bern:2008pv}, and so one may expect that they will not depend on the charges in exceptional field theory. The three-loop amplitude is moreover expected to satisfy a non-renormalisation theorem such that the 3-loop kinematic numerators depend at most quadratically in the loop momenta, as it is indeed the case in supergravity \cite{Bern:2012uf}. Therefore we expect the kinematic numerators of the exceptional field theory 4-graviton 3-loop amplitude to be determined from the supergravity ones \cite{Bern:2012uf} by \eqref{ExcKin} for all charges. 

In this paper we will only consider the 4-graviton amplitude at $1$-loop and $2$-loop, such that this problem does not occur and the loop integrand is uniquely determined from the supergravity one by unitarity and $E_{d(d)}(\mathds{Z})$ invariance. In these cases the kinematic numerators do not depend on the loop momenta, and therefore do not depend on the charges. Moreover, these amplitudes do not involve $\tfrac14$-BPS states (of the type discussed around~\eqref{eq:NewFields}) and therefore are  gauge invariant within the standard exceptional field theory framework. 

\section{One-loop amplitude}
\label{sec:1loop}

\begin{figure}[t]
\centering
\begin{picture}(130,130)
\thicklines
\put(75,80){\line(-1,0){40}}
\put(35,40){\line(1,0){40}}
\put(35,80){\line(0,-1){40}}
\put(75,40){\line(0,1){40}}

\put(15,100){\vector(1,-1){20}}
\put(15,20){\vector(1,1){20}}
\put(95,100){\vector(-1,-1){20}}
\put(95,20){\vector(-1,1){20}}
\put(-15,100){$(k_1,0)$}
\put(95,100){$(k_2,0)$}
\put(-15,15){$(k_4,0)$}
\put(95,15){$(k_3,0)$}
\put(42.5,57){$(p,\Gamma)$}
\put(34.5,39){\CricArrowRight{\phantom{xix}}}
\end{picture}
\caption{\label{fig:1loop}\small The scalar box diagram that represents the one-loop amplitude in exceptional field theory. The external legs carry no charges in the extended space but the loop particle has a $D$-dimensional loop momentum $p$ as well as a charge $\Gamma\in \mathds{Z}^{d(\alpha_d)}$.}
\end{figure}
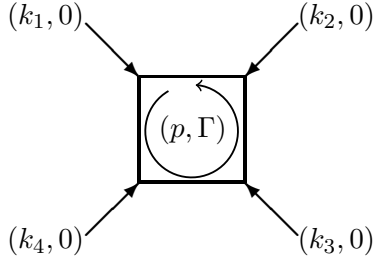

In this section we will consider the 1-loop amplitude of four gravitons in exceptional field theory without any charges on the external legs. For this purpose we proceed similarly as for the 11-dimensional supergravity amplitude \cite{Green:1997as}. We factorize the amplitude into the polarisation term quartic in the external momenta 
\be  
\cM^{\rm E}_{4}(1,2,3,4)  = \frac{i\kappa^2}{2} t_8 t_8 \prod_{A=1}^4R(k_A,e_A) A(k_1,k_2,k_3,k_4)  \ . 
\ee
The appearance of the universal $t_8t_8R^4$ term is a universal feature of supergravity and superstring theory~\cite{Deser:1977nt,Green:1981xx,Gross:1986iv}. At one loop, there is only the box diagram contribution of figure~\ref{fig:1loop}~\cite{Green:1997as}, and according to the discussion of the last section we get
\begin{align} 
\label{eq:1loopfull}
&\quad\quad A^{\scriptscriptstyle \mbox{\tiny 1-loop}}(k_1,k_2,k_3,k_4) &\\
&=2^6 \kappa^2 \int \frac{d^{11-d} p}{(2\pi)^{11-d}}\hspace{-1mm} \sum_{\substack{\Gamma \in \mathds{Z}^{d(\alpha_d)} \\ \Gamma \times \Gamma = 0   }}\hspace{-0mm} \frac{1}{{\fontsize{9pt}{9.5pt}\selectfont  \mbox{$ (p^2 +\frac{ |Z|^2}{\ell^2})((p-k_1)^2 + \frac{ |Z|^2}{\ell^2})((p-k_1-k_2)^2 +\frac{ |Z|^2}{\ell^2})((p+k_4)^2 + \frac{ |Z|^2}{\ell^2}) $}\fontsize{12pt}{14.5pt}\selectfont }}  +  \circlearrowleft &\nn
\end{align}
where $ \circlearrowleft$ stands for the sum over external legs permutations. The external momenta are restricted to $11-d$ dimensions, and the Newton coupling constant is $\kappa^2 =\frac{1}{2}(2\pi)^{8-d}\ell^{9-d}$ and $|Z|^2\equiv |Z(\Gamma)|^2$ was defined in~\eqref{Znorm} as the $E_{d(d)}$-invariant norm of $\Gamma$ as a function of the scalar moduli. We also recall the notation $d(\alpha_d)=\dim {\bf R}_{\alpha_d}$ for the rank of the charge lattice and the strong section constraint $\Gamma\times\Gamma=0$ from~\eqref{2ChargeRk1}.
The zero charge contribution defines by construction the supergravity amplitude in $11-d$ dimension, and constitutes therefore the non-analytic part of the amplitude. We shall thus define the Wilsonian component of the amplitude as the sum over non-zero charges, indicated by an asterisk on the lattice sum:
\begin{align}
\label{eq:1loopA}
&\quad A_{\rm \scriptscriptstyle W}^{\scriptscriptstyle \mbox{\tiny 1-loop}}(k_1,k_2,k_3,k_4)  &\\
&= 4\pi \ell^{9-d} \sum_{\substack{\Gamma \in \mathds{Z}^{d(\alpha_d)}_* \\ \Gamma \times \Gamma = 0   }} \int_0^\infty \frac{d\upsilon}{\upsilon^{\frac{d-1}{2}}} \int_0^1\hspace{-2mm} dx_1 \int_0^{x_1}\hspace{-2mm}  dx_2 \int_0^{x_2} \hspace{-2mm} dx_3 \, e^{\frac{\pi}{\upsilon} \scal{ (1-x_1)(x_2-x_3) s + x_3 (x_1-x_2) t - \frac{ |Z|^2}{\ell^2}}} +  \circlearrowleft , &\nn
\end{align}
where we have rewritten the amplitude in terms of Schwinger and Feynman parameters to bring out the (dimensionful) Mandelstam variables 
\begin{align}
\label{eq:Mandel}
s = -  (k_1+k_2)^2,\quad t = -(k_1+k_4)^2,\quad u= -s-t = - (k_1+k_3)^2.
\end{align}
The expression~\eqref{eq:1loopA} can be expanded in small values of the Mandelstam variables to obtain
\begin{align}
\label{1loopresult}
A_{\rm \scriptscriptstyle W}^{\scriptscriptstyle \mbox{\tiny 1-loop}}(k_1,k_2,k_3,k_4)  &= \pi  \ell^{6}  \Bigl(\xi(d-3) E_{\alpha_d,\frac{d-3}{2}} + \frac{\pi^2 \ell^4 (s^2+t^2+u^2)}{720}  \xi(d+1) E_{\alpha_d,\frac{d+1}{2}}  \Bigr .& \CR 
&\hspace{20mm} \Bigl . + \frac{\pi^3 \ell^6 (s^3+t^3+u^3)}{18144} \xi(d+3)E_{\alpha_d,\frac{d+3}{2}} + \dots \Bigr)  ,&
\end{align}
which represents the 1-loop contribution to the effective action in terms of $E_{d(d)}$ Eisenstein series multiplying  $\nabla^{2k} R^4$ type supersymmetry invariants. These Eisenstein series are defined as Epstein series in the fundamental representation $R_{\alpha_d}$ with $s=\frac{d-3}{2}+k$, in Langlands normalisation
\be 
E_{\alpha_d,s} =\frac{1}{2 \zeta(2s)} \sum_{\substack{\Gamma \in \mathds{Z}^{d(\alpha_d)} \\ \Gamma \times \Gamma = 0   }} |Z(\Gamma)|^{-2s} = \sum_{\gamma \in P_d(\mathds{Z})\backslash E_d(\mathds{Z})} e^{\langle \Lambda_d | H(\gamma \cV)\rangle }  \ . 
\ee
The series is only absolutely convergent for $k>\frac{3d}{10-d}$ (and $k>\frac{3}{2}$ for $d=3$), so we will define them in general as analytic functions in $d$ according to Langlands definition.  See appendix~\ref{app:ES} for more details on Eisenstein series. The function $\xi(s)$ appearing in the above expression is the completed Riemann zeta function
\begin{align}
\xi(s) = \pi^{-s/2} \Gamma(s/2) \zeta(s)
\end{align}
that is also discussed in the appendix.

To compare with the results of \cite{Green:2010wi,Green:2010kv} it is convenient to recall the definition 
\be 
A_{\rm \scriptscriptstyle W}(k_1,k_2,k_3,k_4)= \ell^6 \sum_{(p,q)\in \mathds{N}^2} \Scal{\frac{\ell}{2}}^{4p+6q} ( s^2 + t^2 + u^2 )^p  ( s^3 + t^3 + u^3 )^q \cE_\gra{p}{q} ,
\ee
such that we obtain for the first orders 
\be
\label{oneloopsplit}
\begin{split} \cE_\gra{0}{0}^{\scriptscriptstyle \mbox{\tiny 1-loop}} &= 4 \pi \xi(d-3) E_{\alpha_d,\frac{d-3}{2}}  \ , \\
 \cE_\gra{0}{1}^{\scriptscriptstyle \mbox{\tiny 1-loop}} &= \frac{8 \pi^4}{567} \xi(d+3) E_{\alpha_d,\frac{d+3}{2}} \ , \\
   \cE_\gra{1}{1}^{\scriptscriptstyle \mbox{\tiny 1-loop}} &= \frac{16 \pi^6}{66\, 825} \xi(d+7) E_{\alpha_d,\frac{d+7}{2}} \ , \\
       \cE_\gra{0}{2}^{\scriptscriptstyle \mbox{\tiny 1-loop}} &= \frac{6976 \pi^7}{638\, 512\, 875} \xi(d+9) E_{\alpha_d,\frac{d+9}{2}} \ , 
 \end{split}\hspace{10mm}\begin{split}
 \cE_\gra{1}{0}^{\scriptscriptstyle \mbox{\tiny 1-loop}} &= \frac{4 \pi^3}{45} \xi(d+1) E_{\alpha_d,\frac{d+1}{2}} \ , \\
  \cE_\gra{2}{0}^{\scriptscriptstyle \mbox{\tiny 1-loop}} &= \frac{16 \pi^5}{14\, 175} \xi(d+5) E_{\alpha_d,\frac{d+5}{2}} \ , \\ 
    \cE_\gra{3}{0}^{\scriptscriptstyle \mbox{\tiny 1-loop}} &= \frac{8 \pi^7}{945\, 945} \xi(d+9) E_{\alpha_d,\frac{d+9}{2}} \ , \\
       \cE_\gra{2}{1}^{\scriptscriptstyle \mbox{\tiny 1-loop}} &= \frac{16 \pi^8}{7\, 882\, 875} \xi(d+11) E_{\alpha_d,\frac{d+11}{2}} \ . 
  \end{split}
\ee
We will now discuss these functions in each dimension separately. 

\subsection{$D=8$ and $SL(2)\times SL(3)$}

Because the supergravity amplitude diverges logarithmically at 1-loop in eight dimensions~\cite{Fradkin}, one cannot directly disentangle the Wilsonian component from the non-analytic one as we did in~\eqref{eq:1loopA}, and we will therefore rather consider the complete amplitude~\eqref{eq:1loopfull}. For $D=8$, the representation $R_{\alpha_d}$ is associated to a linear combination of two simple roots corresponding to the decompactification limit to nine dimensions, \ie the $({\bf 2}, \overline{\bf 3})$ of $SL(2)\times SL(3)$. The section constraint enforces~\eqref{2ChargeRk1} the corresponding $2\times 3$ matrix $\Gamma$ to be of rank one, such that it factorises into the product of two vectors of $\mathds{Z}^2$ and $\mathds{Z}^3$, respectively. One can always rotate it with an $SL(2,\mathds{Z})\times SL(3,\mathds{Z})$ transformation to a standard form multiplying a relative integer, such that 
\be 
\Gamma =  \left( \begin{array}{ccc} \ n a_1 b_1 \ & \ n a_1 b_2\ &\ n a_1 b_3\ \\ n a_2 b_1  & n a_2 b_2&n a_2 b_3 \end{array} \right) = \left( \begin{array}{cc} \ a_1 \ & \ \times \ \\ a_2  & \times \end{array} \right)  \left( \begin{array}{ccc} \ n \ & \ 0\ &\ 0\ \\ 0  & 0&0  \end{array} \right) \left( \begin{array}{ccc} \ b_1 \ & \ b_2\ &\ b_3\ \\ \times   & \times &\times \\ \times   & \times &\times  \end{array} \right) \ , 
\ee
with $a_i$ and $b_i$ respectively relative primes, such that the non-zero charge contribution to the $R^4$ correction is
\be 
\sum_{\substack{\Gamma \in \mathds{Z}^{2\times3}_* \\ \Gamma \times \Gamma = 0   }}\int_0^\infty \frac{d\upsilon}{\upsilon^{1+s}} e^{-\frac{\pi}{\upsilon}|Z(\Gamma)|^2} = 2 \xi(2s) E_{[s]} E_{[0,s]} \ . 
\ee 
However, this series only converges absolutely for $s>\frac{3}{2}$. We shall therefore consider the series as the analytic extension of the function in $s$. Since this parameter is given by $s=\frac{d-3}{2}$ in general, one can think of this regularisation as a natural extension of dimensional regularisation. To regularise the 1-loop integral, we consider therefore the analytic continuation of $d$ in dimension $11-d=8-2\epsilon$ in the eight-dimensional expression while keeping the Cremmer-Julia group fixed. At zero momentum with a sliding scale $\mu$ to regularises the infrared divergence, one obtains 
\bea \label{1loop8D} 
A(0,0,0,0)^{\scriptscriptstyle \mbox{\tiny 1-loop}} &=& 2\pi \ell^6 \Scal{ \int_0^\infty \frac{d\upsilon}{\upsilon^{1+\epsilon}} e^{- \frac{\pi}{\upsilon} \mu^2} + 2 \xi(2\epsilon)  E_{[\epsilon]}   E_{[0,\epsilon]} } \CR
&=& 2\pi \ell^6 \Scal{ \frac{1}{\epsilon} - \gamma - \log (\pi \mu^2)  - \frac{1}{\epsilon} +4  \gamma -2 -2 \log (2) +2\xi(2) \hat{E}_{[1]} +2 \xi(3)  \hat{E}_{[\frac{3}{2},0] }} \CR
&=& \ell^6 \Scal{ 4 \zeta(2) \hat{E}_{[1]} +2 \zeta(3)  \hat{E}_{[\frac{3}{2},0]} + 2\pi ( 3\gamma -2 - \log(4\pi \mu^2) ) }  \ .  
\eea
It is remarkable that the ultra-violet divergence cancels out. Note that this cancelation is rather universal at this level, since the leading constant term in $E_{[\epsilon]}   E_{[0,\epsilon]}=1 + \mathcal{O}(\epsilon)$ is enough to ensure the finiteness of the amplitude. It follows that this cancelation would also hold if one was restricting the sum to either the 11-dimensional Kaluza--Klein momenta or the type IIB ones, with or without the string winding modes, consistently with the fact that it also cancels in perturbative string theory. Here we use the definition $\hat{E}$ of the regularised Eisenstein series used in \cite{Green:2010wi} where the pole term is subtracted. 

The cancelation of the logarithmic divergence is in fact to be expected. The $R^4$ type correction is $\tfrac12$-BPS, and as such is expected to receive contributions only from $\tfrac12$-BPS states and to be one-loop exact in perturbation theory~\cite{Bern:1998ug}. Exceptional field theory includes all the M-theory $\tfrac12$-BPS states and so this 1-loop amplitude is expected to be the exact M-theory result, which must be finite. This contribution defines indeed the exact threshold function in eight dimensions \cite{Kiritsis:1997em}.

Turning to the higher derivative corrections $\nabla^{2k} R^4$ we find in the standard normalisation that
\bea 
\hat{\cE}^{\scriptscriptstyle \mbox{\tiny 1-loop}}_\gra{0}{0} &=& 2( 2 \zeta(2) \hat{E}_{[1]} )+2 \zeta(3)  \hat{E}_{[\frac{3}{2},0]} + 2\pi ( 3\gamma -2 - \log(4\pi \mu^2) )   \ ,\CR
 \cE^{\scriptscriptstyle \mbox{\tiny 1-loop}}_\gra{1}{0} &=& - 4 ( 2 \zeta(4) E_{[2]} ) (2\zeta(-1) E_{[-\frac{1}{2},0]} )   \ ,\CR
 \cE^{\scriptscriptstyle \mbox{\tiny 1-loop}}_\gra{0}{1}  &=& \frac{40}{9} ( 2 \zeta(6) E_{[3]} ) (2\zeta(-3) E_{[-\frac{3}{2},0]} )    \ . 
 \eea
The functions all appear in the exact string theory effective action with the same coefficients \cite{Green:2010wi}. They are not the unique contributions however, but it is rather remarkable that those are precisely reproduced by a 1-loop computation whereas the higher derivative terms will also get corrections at higher loop order.

\subsection{$D=7$ and $SL(5)$}
\label{7D1-loop}
In $D=7$ dimensions the root $\alpha_4$ has to be understood as being the third root in the standard ordering for $SL(5)$ according to our $E_{d(d)}$ numbering convention of figure~\ref{fig:dynk}. For  $d=4+2\epsilon$ one then obtains a regular limit after using a functional relation~\eqref{eq:FR} on the Eisenstein series
\be 
\cE^{\scriptscriptstyle \mbox{\tiny 1-loop}}_\gra{0}{0} = \lim_{\epsilon\to 0} 4\pi \xi(1+2\epsilon) E_{[0,0,\frac{1}{2}+\epsilon,0]}  = 2 \zeta(3) E_{[\frac{3}{2},0,0,0]}\ .    
\ee
The superficially divergent behaviour at $\epsilon=0$ is an artefact of Langlands normalisation, whereas the lattice sum is itself finite, even if it is not absolutely convergent. However, the function at next order  in derivatives has a pole 
\be  \label{D4R47Ddiv}
\cE^{\scriptscriptstyle \mbox{\tiny 1-loop}}_\gra{1}{0}  = \frac{4\pi^3}{45} \xi(5+2\epsilon) E_{[0,0,\frac{5}{2}+\epsilon,0]}  = \frac{2\pi^2}{3 \epsilon} + \frac{\pi}{15} \zeta(5)\hat{E}_{[0,0,\frac{5}{2},0]} + \mathcal{O}(\epsilon)   
\ee
that could only be canceled by the supergravity amplitude divergence at 2-loop.  In fact there is no deep reason for the complete amplitude to be finite in this case, since we expect this function to get corrections associated to $\tfrac14$-BPS non-perturbative states that we have not taken into account in exceptional field theory. We will comment on this more in Section~\ref{sec:Specs}. Nonetheless, the regularised Eisenstein series above indeed appears in the exact threshold function with this specific coefficient \cite{Green:2010wi}. 

\subsection{$D=6$ and $SO(5,5)$}
\label{1loopD6}

Very similar results arise for $R^4$ in the $D=6$ case. For the $R^4$ correction one finds
\begin{align}
\cE^{\scriptscriptstyle \mbox{\tiny 1-loop}}_\gra{0}{0} 
= 4\pi \xi(2)  E_{\mbox{\DSOX0000{\mathnormal{1}}}}
= 4 \pi \xi(3) E_{\mbox{\DSOX{\mathnormal{\stfrac32}}0000}}
= 2 \zeta(3) E_{\mbox{\DSOX{\mathnormal{\stfrac32}}0000}}
\end{align}
after using a functional relation. This is the correct finite answer~\cite{Green:2010wi}.

The $\nabla^4 R^4$, however, exhibits a divergent behaviour at $d=5+2\epsilon$:
\begin{align}
\label{eq:D4R41loop}
\cE^{\scriptscriptstyle \mbox{\tiny 1-loop}}_\gra{1}{0} = \frac{2\zeta(3)}{\epsilon} E_{\mbox{\DSOX{\mathnormal{\stfrac32}}0000}} + \frac{8\zeta(6)}{45} \hat{E}_{\mbox{\DSOX0000{\mathnormal{3}}}} + O(\epsilon),
\end{align}
where importantly the divergence is in the minimal Eisenstein series. Despite the fact that the finite term appears in the exact string theory threshold function with this specific coefficient, this divergence indicates that the entire contribution to the $\nabla^4 R^4$ correction term that arises at one-loop should be removed by a counterterm in all dimensions as will be discussed in more detail in section~\ref{sec:D4R4oneloop}.

For the $\nabla^6 R^4$ higher derivative contributions one finds similarly a divergent contribution
\begin{align}
\cE^{\scriptscriptstyle \mbox{\tiny 1-loop}}_\gra{0}{1}  =  \frac{8\pi^8}{893025\, \epsilon} + \frac{16 \zeta(8)}{189}  \hat{E}_{\mbox{\DSOX0000{\mathnormal{4}}}}+ O(\epsilon)\ .  
\end{align}
As is known from~\cite{Green:2010wi,Pioline:2015yea} the coefficient functions $\cE_\gra{1}{0}$ and $\cE_\gra{0}{1}$ should have contributions from two (regularised) Eisenstein series, corresponding to the fact that there are two independent supersymmetric invariants~\cite{Bossard:2014aea,Bossard:2015uga}. Here, only one is recovered in the constant term albeit with the correct coefficient. 

\subsection{$3\leq D \leq 5$ and $E_{6(6)}$, $E_{7(7)}$ and $E_{8(8)}$}

As in higher dimensions,  the one-loop result from exceptional theory produces the correct $R^4$ correction term for $D\leq 5$:
\begin{subequations}
\begin{align}
E_{6(6)} & : &\cE^{\scriptscriptstyle \mbox{\tiny 1-loop}}_\gra{0}{0} &= 2\zeta(3) \EiEVI{\frac{3}{2}} ,&\\
E_{7(7)} & : &\cE^{\scriptscriptstyle \mbox{\tiny 1-loop}}_\gra{0}{0} &= 2\zeta(3) \EiEVII{\frac{3}{2}} ,&\\
E_{8(8)} & : &\cE^{\scriptscriptstyle \mbox{\tiny 1-loop}}_\gra{0}{0} &= 2\zeta(3) \EiEVIII{\frac{3}{2}} ,&
\end{align}
\end{subequations}
after using functional relations~\eqref{eq:FR}. We have formally included $E_{8(8)}$ although the application of our methods in this case is not fully justified as we discussed in section~\ref{sec:Frules}.

For $D\leq 5$ space-time dimensions, there is a unique $\nabla^4R^4$ supersymmetric invariant~\cite{Bossard:2014aea} and a single Eisenstein series contribution to the coefficient function $\cE_\gra{1}{0}$~\cite{Green:2010kv}. This complete function reproduced with the correct factor~\cite{Green:2010kv} from the one-loop calculation in exceptional field theory:
\begin{subequations}
\begin{align}
E_{6(6)} & : &\cE^{\scriptscriptstyle \mbox{\tiny 1-loop}}_\gra{1}{0} &= \zeta(5) \EiEVI{\frac{5}{2}} ,&\\
E_{7(7)} & : &\cE^{\scriptscriptstyle \mbox{\tiny 1-loop}}_\gra{1}{0} &= \zeta(5) \EiEVII{\frac{5}{2}} ,&\\
E_{8(8)} & : &\cE^{\scriptscriptstyle \mbox{\tiny 1-loop}}_\gra{1}{0} &= \zeta(5) \EiEVIII{\frac{5}{2}} .&
\end{align}
\end{subequations}
As we mentioned above and discuss in section~\ref{sec:D4R4oneloop}, these contributions should nonetheless be removed by renormalisation.

For $\nabla^6 R^4$, the one-loop calculation in exceptional field theory yields
\begin{subequations}
\begin{align}
E_{6(6)} & : &\cE^{\scriptscriptstyle \mbox{\tiny 1-loop}}_\gra{0}{1} &= \frac{5\zeta(9)}{54} \hat{E}_{\mbox{\DEVI00000{\mbox{$\frac{9}{2}$}}}} ,&\\
E_{7(7)} & : &\cE^{\scriptscriptstyle \mbox{\tiny 1-loop}}_\gra{0}{1} &= \frac{64\zeta(10)}{189} \hat{E}_{\mbox{\DEVII000000{\mbox{$5$}}}}  ,&\\
E_{8(8)} & : &\cE^{\scriptscriptstyle \mbox{\tiny 1-loop}}_\gra{0}{1} &= \frac{5\zeta(11)}{12\pi}   E_{\mbox{\DEVIII0000000{\mbox{$\frac{11}{2}$}}}}  .&
 \end{align}
\end{subequations}
The coefficients correctly reproduce the string theory effective action \cite{Green:2010kv}. The function $ \cE^{\scriptscriptstyle \mbox{\tiny 1-loop}}_\gra{0}{1}$ is nonetheless incompatible with string perturbations theory in three dimensions, but reproduces correctly the 3-loop contribution \cite{Bossard:2015oxa}. 

\subsection{Renormalisation of the $\nabla^4 R^4$ terms}
\label{sec:D4R4oneloop}

It is surprising that the 1-loop amplitude provides already so much information about the exact string theory effective action, whereas one would naively expect only $ \cE_\gra{0}{0}$ to be exact at this order. The contribution to the $\nabla^4 R^4$ correction term is divergent in $D=7$ and $D=6$, as we discussed in section~\ref{7D1-loop} and \ref{1loopD6}, see equation \eqref{D4R47Ddiv} and \eqref{eq:D4R41loop}, with a divergence in a non-trivial function of the moduli in the second case. The one-loop exceptional field theory effective action should be renormalised at this order to remove this divergence. However, it is a fundamental property of the theory that the effective action is consistent in all dimensions at the quantum level, and we should consider the theory for all $D$ as a whole.  Therefore, we will assume a renormalisation prescription in which the contribution to the $\nabla^4 R^4$ term from the one-loop exceptional field theory amplitude is consistently removed in \textit{all} dimensions by adding the corresponding $\nabla^4 R^4$ counterterm. 

Of course this renormalisation will affect the 2-loop amplitude through the 1-loop form factor of the supersymmetric $\nabla^4 R^4$ counterterm. These contributions can in principle be obtained by analysing the 3-loop sub-divergences. It appears that these counterterms only affect the $\nabla^{8+2k} R^4$ type 2-loop threshold functions \cite{Basu:2014hsa}, and one can avoid them for lower order couplings like $\nabla^6 R^4$. We will not carry out this analysis in this paper. The same renormalisation prescription implies that we must also renormalise the $\nabla^6 R^4$ threshold function, as well as other higher derivative couplings. But these additional renormalisations will only become essential for the 3-loop computation.

\subsection{General remarks on higher order terms}

All the functions $\cE^{\scriptscriptstyle \mbox{\tiny 1-loop}}_\gra{p}{q}$ produced in the computation~\eqref{1loopresult} admit a string theory limit consistent with string perturbation theory. They admit in general a 1-loop, a $k$-loop and $(2k-4)$-loop contribution to the $\nabla^{2k} R^4$ threshold function, that is compatible with the $\tfrac18$-BPS protected $F^{2k-4} \nabla^4 R^4$ threshold function for $k>4$  \cite{Bossard:2015uga}. Assuming that these functions do indeed contribute to the exact threshold functions would give some information about the ultra-violet divergences in supergravity. This computation predicts for instance a correction in
\be  
\cE^{\scriptscriptstyle \mbox{\tiny 1-loop}}_\gra{2}{1} = \frac{2048}{25025 \pi} \zeta(18)  \hat{E}_{\mbox{\DEVII0000009}} \ , 
\ee
to the $\nabla^{14} R^4$ threshold function in four dimensions. This function admits a pole and a corresponding logarithmic dependence in the string coupling constant at 10-loop which suggests the appearance of a logarithm divergence at 10-loop in $\cN=8$ supergravity. However, this conclusion has to be taken with care in view of the discussion of the preceding section.

\allowDbreak{The decompactification limit of Eisenstein series in the fundamental representation is simple because $\alpha_d$ is in the highest weight representation of $E_{d(d)}$ in the parabolic subgroup with Levi factor $\mathds{R}_+^* \times E_{d-1(d-1)}$, and the Poisson summation formula gives\footnote{This can also be evaluate using the constant term formula of M\oe glin--Waldspurger~\cite{MoeglinWaldspurger}.} 
\begin{align}
\label{1-loopDecomp}
E_{\alpha_{d},s} &=  \frac{1}{2\zeta(2s)} \sum_{\substack{\Gamma \in \mathds{Z}_*^{d(\alpha_d)} \\ \Gamma \times \Gamma = 0   }} |Z(\Gamma)|^{-2s} &\nn\\
&= r^{\frac{10-d}{9-d}2s}  +  \frac{1}{2\xi(2s)}  \sum_{{{\fontsize{6.8pt}{1.5pt}\selectfont  \mbox{$\begin{array}{c} \Gamma_{d\mbox{-}1} \ne 0 \vspace{1mm}\\ \Gamma_{d\mbox{-}1}\hspace{-1mm}  \times\hspace{-1mm} \Gamma_{d\mbox{-}1} =0\end{array}$} \fontsize{12pt}{14.5pt}\selectfont }}}  \sum_{N\in \mathds{Z}} \int \frac{dt}{t^{1+s}} e^{-\frac{\pi}{t} \scal{ r^{2\frac{10-d}{d-9}} ( N+\langle \Gamma_{d-1} ,a\rangle)^2 + r^{\frac{2}{d-9}}  |Z(\Gamma_{d-1})|^2}} + \dots &\nn\\
&= r^{\frac{(10-d)2s}{9-d}}  +  \frac{\xi(2s-1)}{\xi(2s)}  r^{\frac{2s}{9-d}+1 } E_{\alpha_{d-1},s-\frac{1}{2}}    + \dots 
\end{align}
such that the relevant Eisenstein series in this computation gives in the decompactification limit
\be 
E_{\alpha_d,\frac{d-3}{2}+k} = r^{\frac{(10-d)(d-3+2k)}{9-d}}  +  \frac{\xi(d-4+2k)}{\xi(d-3+2k)}  r^{\frac{2(3+k)}{9-d} }  E_{\alpha_{d-1},\frac{d-4}{2}+k}  + \dots .
\label{alphad,alphadlimit} 
\ee
However, note that the terms we neglect here are \textit{not} subleading for all values of $d$ and $k$, and this approximation is only meaningful for large enough $k$ (or $k=0$). }

Let us finally comment on the differential equation satisfied by the integrand function. With the normalisation $r = e^{-(9-d) \phi}$ one computes that 
\be 
\Delta F(\phi) = \frac{1}{2(9-d)(10-d)} \scal{ \partial_\phi^{\; 2} + ( d(19-d) - 30 ) \partial_\phi } F(\phi) \ , 
\ee
such that 
\be 
\Delta  E_{\alpha_d,\frac{d-3}{2}+k}  = \frac{(10k-(3+k)d)( d+2k-3)}{9-d}  E_{\alpha_d,\frac{d-3}{2}+k}\  . 
\ee
For the cases $k=0,2,3$ this reproduces the eigenvalues of the Laplace equations of~\cite{Green:2010wi}.

\section{Two-loop amplitude}
\label{sec:2loops}

In this section, we will evaluate the four-graviton amplitude at two loops in exceptional field theory in $D=11-d$ non-compact dimensions. As in the preceding section, we will start from the representation of the amplitude in terms of cubic scalar diagrams. According to \cite{Bern:1998ug} the 2-loop amplitudes is a sum of a planar and a non-planar diagram given by:

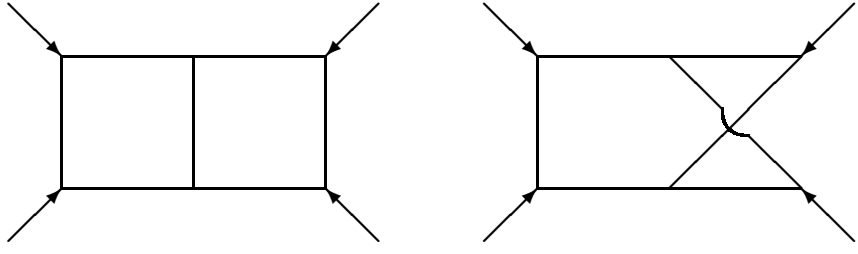
\begin{figure}[h!]
\centering
\begin{picture}(320,90)
\thicklines
\put(20,20){\line(1,0){100}}
\put(20,70){\line(1,0){100}}
\put(20,20){\line(0,1){50}}
\put(120,20){\line(0,1){50}}
\put(70,20){\line(0,1){50}}
\put(0,0){\vector(1,1){20}}
\put(0,90){\vector(1,-1){20}}
\put(140,0){\vector(-1,1){20}}
\put(140,90){\vector(-1,-1){20}}

\put(200,20){\line(1,0){100}}
\put(200,70){\line(1,0){100}}
\put(200,20){\line(0,1){50}}
\put(250,20){\line(1,1){50}}
\put(250,70){\line(1,-1){20}}
\put(280,40){\line(1,-1){20}}
\qbezier(270,50)(270,40)(280,40)
\put(180,0){\vector(1,1){20}}
\put(180,90){\vector(1,-1){20}}
\put(320,0){\vector(-1,1){20}}
\put(320,90){\vector(-1,-1){20}}

\end{picture}
\end{figure}

The calculation of the amplitude proceeds in a similar way to the one-loop case. There are now two internal momenta and associated charges $\Gamma_1$ and $\Gamma_2$. The manipulations here are similar to~\cite{Green:1999pu}. A very important point for us here is that the strong section condition has to be enforced for any pair of charges.

We start with the planar diagram that is expressed with Schwinger parameters for all seven propagators as
\begin{align}
A^{\scriptscriptstyle \mbox{\tiny 2-loop}}_{\textrm{p}}(k_1,k_2,k_3,k_4)& = 2^6 \kappa^4  \hspace{-5mm}\sum_{\substack{\Gamma_1, \Gamma_2\in \mathds{Z}^{d(\alpha_d)}\\ \Gamma_i\times \Gamma_j=0}} \int \frac{d^{11-d}p}{(2\pi)^{11-d}}  \frac{d^{11-d}q}{(2\pi)^{11-d}}
\left[\prod_{I=1}^7 \int_0^\infty d\sigma_I \right] \exp\left\{- \sum_{I=1}^7 \sigma_I k_I^2
\right\}
\nn\\
&\hspace{-20mm} \times\exp\left\{ -(\sigma_1+\sigma_2+\sigma_3) \ell^{-2} |Z(\Gamma_1)|^2 - (\sigma_4+\sigma_5+\sigma_6) \ell^{-2} |Z(\Gamma_2)|^2 - \sigma_7\ell^{-2} |Z(\Gamma_1+\Gamma_2)|^2 \right\}\nn\\
& + \circlearrowleft
\end{align}
with 
\begin{align}
k_I = (p, p-k_1, p-k_1-k_2, q, q-k_4, q-k_3-k_4, p+q),
\end{align}
and we have separated the non-compact integrals from the momentum sums. The symbol $\circlearrowleft$ represents the sum over the five additional non-trivial permutations of the momenta. We introduce the dimensionless combinations of Schwinger parameters
\begin{align}
L_1 = \ell^{-2} \lp \sigma_1 + \sigma_2 + \sigma_3\rp, \quad
L_2 = \ell^{-2} \lp \sigma_4 + \sigma_5 + \sigma_6\rp ,\quad
L_3 = \ell^{-2} \sigma_7
\end{align}
that we assemble into the $(2\times 2)$-matrix
\begin{align}
\label{SchwMatrix}
\Omega= \Omega^{ij} =\begin{pmatrix} L_1+L_3 & L_3 \\ L_3 & L_2+L_3\end{pmatrix}
,\quad
\det\Omega = L_1L_2 + L_2L_3 + L_3L_1,
\end{align}
and introduce Feynman parameters for the remaining four fractions of the $L_i$. After carrying out the Gaussian momentum integrals and using the definition of $\kappa$ the result is
\begin{align}
A_{\textrm{p}}^{\scriptscriptstyle \mbox{\tiny 2-loop}}(k_1,k_2,k_3,k_4)& = \frac{\pi^{5-d}}{4} \ell^{10}\hspace{-5mm} \sum_{\substack{\Gamma_1, \Gamma_2\in \mathds{Z}^{d(\alpha_d)}\\ \Gamma_i\times \Gamma_j=0}} 
\int \frac{d^3\Omega}{(\det\Omega)^{\frac{11-d}{2}}} L_1^2 L_2^2 e^{-\Omega^{ij} g(\Gamma_i, \Gamma_j)}\nn\\
&\hspace{-34mm} \times \int\limits_0^1 dv_2 dw_2 \int\limits_0^{v_2} dv_1 \int\limits_0^{w_2} dw_1 e^{\ell^2 t\frac{L_1L_2L_3}{\det\Omega} (v_2-v_1)(w_2-w_1) + \ell^2 s\left[ \frac{L_1L_2L_3}{\det\Omega} (v_1-w_1)(v_2-w_2) + L_1v_1(1-v_2)+L_2w_1(1-w_2)
\right]}\nn\\
&\quad+ \circlearrowleft
\end{align}
The Schwinger parameters $L_i$ in $\Omega$ are all integrated from $0$ to $\infty$. The integral over the Feynman parameters $v_i$ and $w_i$ can be done order by order in the dimensionless Mandelstam variables 
\begin{align}
s = - (k_1+k_2)^2 ,\quad
t = -(k_1+k_4)^2 ,\quad
u = - (k_1+k_3)^2 ,\quad
\end{align}
to obtain the low energy expansion of the amplitude.

The contribution from the non-planar diagram can be evaluated similarly to be
\begin{align}
A^{\scriptscriptstyle \mbox{\tiny 2-loop}}_{\textrm{np}}(k_1,k_2,k_3,k_4)& =\frac{\pi^{5-d}}{4} \ell^{10}\hspace{-5mm}  \sum_{\substack{\Gamma_1, \Gamma_2\in \mathds{Z}^{d(\alpha_d)}\\ \Gamma_i\times \Gamma_j=0}} 
\int \frac{d^3\Omega}{(\det\Omega)^{\frac{11-d}{2}}} L_1 L_2^2 L_3e^{-\Omega^{ij} g(\Gamma_i, \Gamma_j)}\nn\\
&\hspace{-35mm} \times \int\limits_0^1 du_1 dv_1 dw_2 \int\limits_0^{w_2} dw_1 e^{\ell^2 t\frac{L_1L_2L_3}{\det\Omega} (u_1-v_1)(w_2-w_1) +\ell^2  s\left[ \frac{L_1L_2L_3}{\det\Omega} (w_1(1-u_1)+v_1(u_1-w_2))+ \frac{L_2^2(L_1+L_3)}{\det\Omega}w_1(1-u_1)
\right]}\nn\\
&\quad+ \circlearrowleft
\end{align}

Combining the planar and the non-planar diagrams and summing over all permutations, one obtains the following result expanded at low orders in the Mandelstam variables
\begin{align}
\label{eq:2loops}
&\quad A^{\scriptscriptstyle \mbox{\tiny 2-loop}}(k_1,k_2,k_3,k_4)\CR
&=\frac{\pi^{5-d}\ell^6}{4} \hspace{-4mm}   \sum_{\substack{\Gamma_1,\Gamma_2 \in \mathds{Z}^{d(\alpha_d)} \\ \Gamma_i \times \Gamma_j = 0   }} \int \frac{d^3 \Omega}{(\det\Omega)^{\frac{7-d}{2}}} e^{- \Omega^{ij} g(\Gamma_i , \Gamma_j)}  \biggl(  \frac{\ell^4 (s^2 + t^2 + u^2)}{6} \Phi_\gra{1}{0}(\Omega) + \frac{\ell^6 (s^3 + t^3 + u^3)}{72} \Phi_\gra{0}{1}(\Omega) \biggr .\CR
&\quad \biggl.
+ \frac{\ell^8 (s^2 + t^2 + u^2)^2}{8640} \Phi_\gra{2}{0}(\Omega)  + \frac{\ell^{10} (s^2+t^2+u^2)(s^3+t^3+u^3)}{1088640} \Phi_\gra{1}{1}(\Omega)  + \dots \biggr) 
\end{align}
where the functions $\Phi_\gra{p}{q}(\Omega)$ of the Schwinger parameters~\eqref{SchwMatrix} can be computed iteratively from the low momenta expansion of the two-loop amplitude. At low orders, one finds
\begin{align} 
\label{Phis}
\Phi_\gra{1}{0}(\Omega) &= 1,&\CR
\Phi_\gra{0}{1}(\Omega)  &= L_1 + L_2 + L_3- 5 \frac{L_1 L_2 L_3}{\det \Omega} ,&\CR
\Phi_\gra{2}{0}(\Omega)&=4 ( L_1 + L_2 + L_3 )^2 - 22 ( L_1 + L_2 + L_3 ) \frac{L_1 L_2 L_3}{\det\Omega} - 3 \, \det\Omega+ 32 \Scal{\frac{L_1 L_2 L_3}{\det\Omega}}^2 ,&\\
\Phi_\gra{1}{1}(\Omega)&=45 ( L_1 + L_2 + L_3)^3  - 65 \, \det \Omega ( L_1  + L_2 + L_3 ) +  250 L_1 L_2 L_3 &\CR
& \quad- 285 \frac{ (L_1+L_2+L_3)^2 L_1 L_2 L_3}{\det\Omega}   + 347 \frac{ (L_1+L_2+L_3)( L_1 L_2 L_3)^2}{\det\Omega^2} - 145 \Scal{  \frac{L_1 L_2 L_3}{\det\Omega}}^3 .&\nn
\end{align}
We note that compared to the one-loop calculation~\eqref{1loopresult} the two-loop amplitude starts contributing at order $\ell^{10}$ rather than $\ell^6$ and therefore its lowest order correction is of the form $\nabla^4R^4$ rather than $R^4$. 

Let us briefly argue that this formula is consistent with the large radius limit in one dimension higher. At leading order in the large radius modulus, we will neglect the sum over the charges components of lower degree such that 
\begin{align}
&\quad \pi^{5-d} \hspace{-4mm}   \sum_{\substack{\Gamma_i\in \mathds{Z}^{2d(\alpha^d_d)}_* \\ \Gamma_i \times \Gamma_j = 0   }} \int \frac{d^3 \Omega}{(\det\Omega)^{\frac{7-d}{2}}}  \Phi_\gra{p}{q}(\Omega) e^{- \Omega^{ij} g(\Gamma_i , \Gamma_j)}  &\nn\\
&= \pi^{5-d}  \sum_{n_i\in \mathds{Z}^{2}_*} \int \frac{d^3 \Omega}{(\det\Omega)^{\frac{7-d}{2}}}  \Phi_\gra{p}{q}(\Omega) e^{- \Omega^{ij} e^{2(10-d)\phi} n_i n_j}  &\CR
&\quad  +\pi^{5-d} \hspace{-4mm}   \sum_{\substack{\Gamma_i\in \mathds{Z}^{2d(\alpha^{d-1}_d)}_* \\ \Gamma_i \times \Gamma_j = 0   }}  \sum_{n_i\in \mathds{Z}^{2}} \int \frac{d^3 \Omega}{(\det\Omega)^{\frac{7-d}{2}}}  \Phi_\gra{p}{q}(\Omega) e^{- \Omega^{ij} \scal{e^{2(10-d)\phi} (n_i + \langle a,\Gamma_i\rangle )  (n_j + \langle a,\Gamma_j\rangle )   + e^{2\phi} g(\Gamma_i , \Gamma_j)}}  + \dots &\nn\\
&=e^{-2(10-d)(d+2p+3q-6)\phi}\, \pi^{5-d}    \sum_{n_i\in \mathds{Z}^{2}_*} \int \frac{d^3 \Omega}{(\det\Omega)^{\frac{7-d}{2}}}  \Phi_\gra{p}{q}(\Omega) e^{- \Omega^{ij}  n_i n_j}  &\CR
& \quad +e^{-2(3+2p+3q)\phi}  \pi^{5-d+1}  \hspace{-4mm}   \sum_{\substack{\Gamma_i\in \mathds{Z}^{2d(\alpha^{d-1}_d)}_* \\ \Gamma_i \times \Gamma_j = 0   }} \int \frac{d^3 \Omega}{(\det\Omega)^{\frac{7-d+1}{2}}}  \Phi_\gra{p}{q}(\Omega) e^{- \Omega^{ij}  g(\Gamma_i , \Gamma_j)}  + \dots &
\end{align}
This implies that 
\be 
\cE_{\gra{p}{q}\, d}^{\scriptscriptstyle \mbox{\tiny 2-loop}}  = r^{\frac{2(10-d)(d+2p+3q-6)}{9-d}} c_{\gra{p}{q}\, d} + r^{\frac{2(3+2p+3q)}{9-d}}\cE_{\gra{p}{q}\, d-1}^{\scriptscriptstyle \mbox{\tiny 2-loop}} + \dots 
\ee
in agreement with the appropriate decompactification limit. However, similarly as in \eqref{alphad,alphadlimit}, the terms included in the dots are not subleading in general, and moreover the integral defining the constants $ c_{\gra{p}{q}\, d} $ and the threshold functions themselves must be regularised to avoid singularities associated to degenerate orbits.

In the following we will discuss the contributions to the threshold functions $\cE_\gra{p}{q}$ that are implied by~\eqref{eq:2loops} in various dimensions. The first point to be addressed is the sum over the charges $\Gamma_1$ and $\Gamma_2$ running in the loops and that have to satisfy the strong section constraint.

\subsection{Orbits of two charges}
\label{Nondege} 

We are interested in characterising the space of charges $\Gamma_1, \Gamma_2\in \mathds{Z}^{d(\alpha_d)}$ that satisfy the strong section constraint
\be 
\left.\Gamma_{(i} \times \Gamma_{j)} \right|_{{\bf R}_{\alpha_1}} = 0
\label{2ChargeRank1} 
\ee
for $i=1,2$. The `diagonal' components $i=j$ imply that the charges $\Gamma_i$ have to be of rank one each, meaning that they are $\tfrac12$-BPS charges. They can also vanish. It turns out that one can give a convenient parametrisation of the solution to~\eqref{2ChargeRank1} for arbitrary rank one charges. 

\subsubsection{Example: $E_{7(7)}$}

Before discussing the general case, we study the example of $E_{7(7)}$ where the charges $\Gamma_i$ are in $\mathds{Z}^{d(\alpha_d)}\cong \mathds{Z}^{56}$ and the representation for the constraint ${\bf R}_{\alpha_1}\cong {\bf 133}$ is the adjoint $\mf{e}_{7(7)}$. Note that the constraint in the ${\bf 133}$ implies by construction that the symplectic product of the two charges vanish, so the strong section constraint is indeed satisfied.   We consider the decomposition of representations of $\mf{e}_{7(7)}$ under its $\mf{e}_{6(6)}\oplus\mf{gl}_1$ subalgebra:
\begin{align}
\label{1ChargeGrad4}  
\mathfrak{e}_{7(7)} &\cong \overline{\bf 27}^{\ord{-2}} \oplus \scal{ \mathfrak{gl}_1 \oplus   \mathfrak{e}_{6(6)}}^\ord{0} \oplus  {\bf 27}^{\ord{2}} \ , \CR
{\bf 56} &\cong {\bf 1}^\ord{-3}  \oplus {\bf 27}^\ord{-1} \oplus \overline{\bf 27}^\ord{1} \oplus {\bf 1}^\ord{3} \ , 
\end{align}
where the superscript denotes the weight under $\mf{gl}_1$. As was proved in \cite{Krutelevich}, an integral element $\Gamma_1\in\mathds{Z}^{56}$ satisfying the constraint that its tensor product square vanishes in the adjoint representation can be rotated by an $E_{7(7)}(\mathds{Z})$ element such that it is a relative integer in the highest weight component (degree $3$) in the above decomposition of the ${\bf 56}$. The stabiliser in $E_{7(7)}(\mathds{Z})$ of such a charge is $E_{6(6)}(\mathds{Z})\ltimes \mathds{Z}^{27}$ \cite{Ferrara:1997uz}. 

The second charge $\Gamma_2 \in  \mathds{Z}^{56}$ has to satisfy the constraint
\be 
\Gamma_1 \times \Gamma_2 \big|_{{\bf 133}} = 0 \ .
\ee
If $\Gamma_1$ is chosen (without loss of generality) to have only the highest degree component, it follows directly by inspection of the degree $0$ and degree $2$ components in the adjoint that $\Gamma_2$ is only non-zero in the degree $1$ and $3$ components, \ie 
\be 
\Gamma_1 \in  {\bf 1}^\ord{3} \ , \qquad \Gamma_2 \in  \overline{\bf 27}^\ord{1} \oplus {\bf 1}^\ord{3} \ .  \ee
Note that this solution satisfies by construction $\langle \Gamma_1,\Gamma_2\rangle =0$, such that this additional section constraint \cite{Hohm:2013uia} is a consequence of $\Gamma_i \times \Gamma_j \big|_{{\bf 133}}= 0$. This justifies that we do not consider it separately in this paper. 

Using now that $\Gamma_2$ also has to have rank one, \ie
\be 
\Gamma_2 \times \Gamma_2 \big|_{{\bf 133}} = 0 \ , 
\ee
one obtains that the degree $1$ component $\Gamma_2^\ord{1} $ of $\Gamma_2$ in the $\overline{\bf 27}$ satisfies itself the constraint
\be 
\Gamma_2^\ord{1} \times \Gamma_2^\ord{1}\big|_{\bf 27}  = 0 \ , 
\ee
that is formally the same as the section constraint for $d=6$. As the stabiliser of $\Gamma_1$ contains $E_{6(6)}$, we are still free to use it to obtain a convenient representative of $\Gamma_2^\ord{1}$ in the same way as above for $\Gamma_1$.  That is, we consider the graded decomposition (associated to the  decompactification limit)
\bea 
\mathfrak{e}_{6(6)} &\cong& \overline{\bf 16}^{\ord{-3}} \oplus \scal{ \mathfrak{gl}_1 \oplus   \mathfrak{so}(5,5)}^\ord{0} \oplus  {\bf 16}^{\ord{3}} \ , \CR
\overline{\bf 27} &\cong& {\bf 10}^\ord{-2} \oplus \overline{\bf 16}^\ord{1} \oplus {\bf 1}^\ord{4}  \ .  \label{1ChargeGrad5}  
\eea
The results of~\cite{Krutelevich} imply now that one can rotate $\Gamma_2^\ord{1}\in \mathds{Z}^{27}$ by a discrete $E_{6(6)}(\mathds{Z})$ transformation such that it lies in the singlet highest weight component (degree $4$). In summary, using $E_{7(7)}(\mathds{Z})$ we can choose $\Gamma_1$ to have at most one non-zero component (in ${\bf 1}^\ord{3}$ of~\eqref{1ChargeGrad4}) and $\Gamma_2$ to have at most two non-zero components (in ${\bf 1}^\ord{3}$ of~\eqref{1ChargeGrad4} and ${\bf 1}^\ord{4}$ of~\eqref{1ChargeGrad5} in the $\overline{\bf 27}^\ord{1}$ of~\eqref{1ChargeGrad4}).

This statement can be made more uniform by considering yet another grading of $E_{7(7)}$. Denoting the $\mathfrak{gl}_1$ generator defining the graded decomposition \eqref{1ChargeGrad4} as $h_7$ and the one defining \eqref{1ChargeGrad5} as $h_6^6$, one finds that $h_6 = \frac{4}{3} h_7 + \frac{2}{3} h_6^6$ satisfies
\be 
h_6 \Gamma_i = 4\, \Gamma_i 
\ee
for $i=1,2$. The element $h_6$ defines the graded decomposition of $E_{7(7)}$ with respect to its next-to-last node $\alpha_6$
\begin{align} 
\label{alphadm14} 
\mathfrak{e}_{7(7)} &\cong {\bf 10}^{\ord{-4}} \oplus ({\bf 2}\otimes  {\bf 16})^{\ord{-2}} \oplus \scal{ \mathfrak{gl}_1 \oplus \mathfrak{sl}_2\oplus   \mathfrak{so}(5,5)}^\ord{0} \oplus  ( {\bf 2}\otimes \overline{\bf 16})^{\ord{2}} \oplus {\bf 10}^{\ord{4}}  \ , &\CR
{\bf 56} &\cong {\bf 2}^\ord{-4} \oplus \overline{\bf 16}^\ord{-2} \oplus ({\bf 2}\otimes {\bf 10})^\ord{0} \oplus {\bf 16}^\ord{2} \oplus {\bf 2}^\ord{4}  \  . &
\end{align}
We conclude in this way that any doublet of charges $\Gamma_i$ in $\mathds{Z}^{ 56}$ can be rotated using an appropriate $E_{7(7)}(\mathds{Z})$ element to the highest weight component (degree $4$) associated to this graded decomposition. In other words, we can consider the doublet of charges $\Gamma_i$ satisfying the section constraint as an integral $(2\times 2)$-matrix $M$ and this matrix transforms under left-multiplication by $SL(2,\mathds{Z})$. There are now different orbits depending on the rank of the matrix $M$ and we focus first on the generic case when the rank of $M$ is two. For such a generic doublet of linearly independent charges ($\varepsilon^{ij} \Gamma_i \Gamma_j\neq  0 $)  the stabilizer in $SL(2,\mathds{Z})$ is trivial. The full stabiliser of such a doublet of $\Gamma_i$ in $E_{7(7)}(\mathds{Z})$ is therefore defined as the discrete parabolic subgroup 
\be 
Spin(5,5;\mathds{Z})\ltimes \mathds{Z}^{2\times 16+10} \ . 
\ee
Choosing to leave the sum over the $\mathds{Z}^2$ doublets in $M$ unconstrained, we conclude that the sum over linearly independent charges of a given function $\Phi$ of the charges in $\mathds{Z}^{56}$ reduces to\footnote{The notation here is such that $M$ should be thought of as being embedded in two copies of the ${\bf 56}$ representation such that it lies only in the ${\bf 2}^\ord{4}$ component of the decomposition~\eqref{alphadm14}. The multiplication $\gamma M$ then represents the action of $\gamma \in E_{7(7)}$ on both copies of the ${\bf 56}$. In the last equality we have used an alternative notation for the action that is also used in appendix~\ref{app:ES}. Note that we have used a transpose on $\gamma$ in this equation since we prefer to write the duality coset sums with the stabilising parabolic appearing on the left below.}
\be 
\sum_{\substack{\Gamma_i \in \mathds{Z}^{2\times 56} \\ \Gamma_i \times \Gamma_j = 0 \\ \varepsilon^{ij} \Gamma_i \Gamma_j \ne 0  }} \Phi(\Gamma) = \sum_{\gamma \in E_{7(7)}/ P_{6}}  \sum_{\substack{M\in \mathds{Z}^{2\times 2} \\ {\rm det}(M)\ne 0}} \Phi( \gamma M) 
= \sum_{\gamma \in E_{7(7)}/ P_{6}}  \sum_{\substack{M\in \mathds{Z}^{2\times 2} \\ {\rm det}(M)\ne 0}} \gamma^T \left[\Phi( M) \right]
\   , 
\ee
where 
\be 
P_{6} \cong \scal{ SL(2,\mathds{Z})\times  Spin(5,5)(\mathds{Z})} \ltimes \mathds{Z}^{2\times 16+10}  \ . 
\ee 
Whenever the two charges are linearly dependent, \ie $M$ is of rank less than two, they can both be rotated to the highest weight component (degree $3$) associated to the graded decomposition \eqref{1ChargeGrad4}, such that their common stabilizer in $E_{7(7)}(\mathds{Z})$ is 
\be P_7\cong E_{6(6)}(\mathds{Z}) \ltimes \mathds{Z}^{27}  \ . \ee
In general, we therefore have  
\be 
\sum_{\substack{\Gamma_i \in \mathds{Z}^{2\times 56} \\ \Gamma_i \times \Gamma_j = 0 }} \Phi(\Gamma) = \sum_{\gamma \in P_6 \backslash E_{7(7)}}  \sum_{\substack{M\in \mathds{Z}^{2\times 2} \\ {\rm det}(M)\ne 0}}\gamma\left[ \Phi( M)\right]  +\sum_{\gamma \in P_7 \backslash E_{7(7)}}  \sum_{m \in \mathds{Z}^{2} } \gamma\left[\Phi( m)  \right]\   . 
\ee
The last sum could be further decomposed into the trivial case $m=(0,0)$, corresponding to the case when both charges vanish which is the strict four-dimensional supergravity calculation, and $m\neq (0,0)$. We will mainly be concerned with the contribution from the case of linearly independent charges $\det M\neq 0$, also called the non-degenerate orbit in~\cite{Green:1999pu}.

\subsubsection{Cases $E_{d(d)}$ for $3\le d \le 7$}

The same logic can be applied to $3\le d\le 7$.  In general, a rank 1 charge can always be rotated by $E_{d(d)}(\mathds{Z})$ to a maximal degree component with respect to the $\alpha_d$ decomposition of $e_{d(d)}$, \ie 
\begin{align}
\label{1ChargeGrad}  
 \mathfrak{e}_{d(d)} &\cong \overline{\bf R}_{\alpha_{d-1}}^{\ord{d-9}} \oplus \scal{ \mathfrak{gl}_1 \oplus   \mathfrak{e}_{d-1(d-1)}}^\ord{0} \oplus  {\bf R}_{\alpha_{d-1}}^{\ord{9-d}} \ , \CR
{\bf R}_{\alpha_d} &\cong { \updelta}_{d,7}^\ord{-3}  \oplus \overline{\bf R}_{\alpha_1}^\ord{d-8} \oplus \overline{\bf R}_{\alpha_{d-1}}^\ord{1} \oplus {\bf 1}^\ord{10-d} \ ,\CR
{\bf R}_{\alpha_1} &\cong\hspace{5mm}  \dots  \hspace{5mm} { \updelta}_{d\ge 6}^\ord{4(d-7)}   \oplus  \overline{\bf R}_{\alpha_2}^\ord{d-7} \oplus \overline{\bf R}_{\alpha_1}^\ord{2} \ ,  
\end{align}
where ${\bf R}_{\alpha_1}$ and ${\bf R}_{\alpha_{d-1}}$ on the right-hand-sides are the irreducible representations associated to the first and the last node of $E_{d-1(d-1)}$, with the labelling of roots associated to the convention we use for $E_{d}$.\footnote{For $d\le 3$ the expression becomes less uniform, but the decomposition always corresponds to the decompactification limit at large circle radius modulus.} The notation ${ \updelta}_{d,k}$ indicates a one-dimensional component for $d=k$ and empty space otherwise. This decomposition generalises~\eqref{1ChargeGrad4} for $d=7$ where the section constraint representation ${\bf R}_{\alpha_1}\cong {\bf 133}$ did not arise separately.

As above, we use $E_{d(d)}(\mathds{Z})$ to bring the first charge $\Gamma_1$ into the single component of top degree and analyse the consequences of the section constraint~\eqref{2ChargeRank1} for the second charge $\Gamma_2$. From the gradings displayed in~\eqref{1ChargeGrad}, one deduces that the second charge necessarily belongs to the positive degree components 
\be 
\Gamma_1 \in  {\bf 1}^\ord{10-d} \ , \qquad \Gamma_2 \in  \overline{\bf R}_{\alpha_{d-1}}^\ord{1} \oplus {\bf 1}^\ord{10-d} \ , \label{Sol2charges} 
\ee
with furthermore the degree $1$ component of $\Gamma_2$ being a rank 1 charge of $E_{d-1(d-1)}$. This implies that this degree $1$ component can itself be rotated by the stabiliser $E_{d-1(d-1)}(\mathds{Z})$ (of $\Gamma_1$) to a similar form. Altogether, one obtains that the two charges $\Gamma_i \in {\bf R}_{\alpha_d}$ can both be rotated to the degree $11-d$ (doublet) component  of the decomposition of $\mathfrak{e}_{d(d)}$ associated to next-to-last node $\alpha_{d-1}$, \ie 
\begin{align}
 \label{alphadm1} 
 \mathfrak{e}_{d(d)} &\cong \overline{\bf R}_{\alpha_{1}}^{\ord{2d-18}} \oplus ({\bf 2}\otimes  \overline{\bf R}_{\alpha_{d-2}})^{\ord{d-9}} \oplus \scal{ \mathfrak{gl}_1 \oplus \mathfrak{sl}_2\oplus   \mathfrak{e}_{d-2(d-2)}}^\ord{0} \oplus  ( {\bf 2}\otimes {\bf R}_{\alpha_{d-2}})^{\ord{9-d}} \oplus {\bf R}_{\alpha_{1}}^{\ord{18-2d}}  \ , \CR
{\bf R}_{\alpha_d} &\cong \hspace{25mm}  \dots  \hspace{20mm}   \oplus ({\bf 2}\otimes  \overline{\bf R}_{\alpha_1})^\ord{d-7} \oplus \overline{\bf R}_{\alpha_{d-2}}^\ord{2} \oplus {\bf 2}^\ord{11-d} \ , \CR
{\bf R}_{\alpha_1} &\cong \hspace{45mm}  \dots  \hspace{40mm}  \oplus \overline{\bf R}_{\alpha_1}^\ord{4}  \  . 
\end{align}
This doublet of charges is then stabilised by the subgroup 
\be E_{d-2(d-2)}( \mathds{Z}) \ltimes \mathds{Z}^{2d(\alpha_{d-2}) + d(\alpha_1)} \  .  \ee
Equivalently, the parabolic subgroup $P_{d-1}\subset E_{d(d)}$ (that includes then the $SL(2, \mathds{Z}) $ factor) generated by the non-negative degree pieces in the above decomposition of $\mathfrak{e}_{d(d)}$ with respect to $\alpha_{d-1}$, preserves a generic doublet of grad $(11-d)$ elements. 

For $d\le 7$ one can also check that the dimension of the space of two rank $1$ charges matches the dimension of the  orbit associated with the $P_{d-1}\subset E_{d(d)}$ parabolic. For this purpose, let us define the dimension of the highest weight irreducible representation of $E_{d(d)}$ associated to the simple root $\alpha_k$ as $d(\alpha_k^d)$, the second line of \eqref{1ChargeGrad} implies that 
\be 
\label{alphadDims}
d(\alpha_d^d) = 1 + d(\alpha_{d-1}^{d-1}) + d(\alpha_{1}^{d-1}) + \delta_{d,7} \ . 
\ee
The parabolic subgroup $P_k^d$ over the reals is generated by the non-negative degree component with respect to the Cartan generator associated to the simple root $\alpha_k^d$. The dimension of the parabolic coset over the reals $\mathcal{O}_{P_k^d}\cong E_{d(d)}/P_k$ equals the dimension of the negative degree component, and we have therefore according to \eqref{1ChargeGrad} and \eqref{alphadm1}
\be 
\label{cosDims}
\dim \mathcal{O}_{P_d^d}  = d(\alpha_{d-1}^{d-1}) \ , \qquad \dim \mathcal{O}_{P_{d-1}^d} =2 \, d(\alpha_{d-2}^{d-2})  + d(\alpha_1^{d-2}) \ . 
\ee
Now using \eqref{Sol2charges} one computes that the dimension of the space of doublets of charges in $\mathds{R}^{d(\alpha_d^d)}$ satisfying the strong section constraint is equal to the dimension of the space of a charges in $\mathds{R}^{d(\alpha^d_d)}$ satisfying the strong section constraint, plus 1, plus the dimension of the space of charges satisfying the strong section constraint in $\mathds{R}^{d(\alpha^{d-1}_{d-1})}$. For $E_{d(d)}$, the dimension of the space of (single) charges satisfying the strong section constraint is computed as follows. A rank $1$ charge in ${\bf R}_{\alpha_d}$ can always be rotated to the highest weight component associated to the grading under node $\alpha_d$ (cf.~\eqref{1ChargeGrad}) and is then stabilised by the subgroup of $P_d^d$ with the $GL(1)$ factor excluded. Therefore the corresponding space has the dimension of the parabolic coset $P_d^d\backslash E_{d(d)}$ plus $1$. 

It follows that the dimension of the space of doublets of charges satisfying the strong section constraint is 
\bea 
\dim \lp \mathcal{O}_{\Gamma_i|\Gamma_i\times \Gamma_j=0} \rp &=& \left(1 + \dim \lp \mathcal{O}_{P_d^d}\rp\rp +1 +\lp 1 + \dim \lp\mathcal{O}_{P_{d-1}^{d-1}}\rp\rp \  \CR
 &=& 3 + d(\alpha_{d-1}^{d-1}) + d(\alpha_{d-2}^{d-2}) \CR
   &=& 4 + d(\alpha_1^{d-2}) + 2d(\alpha_{d-2}^{d-2}) \CR
   &=& 4+ d(\mathcal{O}_{P_{d-1}^d})   \ ,
 \eea
 where we have used~\eqref{alphadDims},~\eqref{cosDims} and $d\leq 7$. The last equality shows that the dimension of the orbit of two charges satisfying the strong section constraint equals the dimension of the parabolic coset associated to $\alpha_{d-1}^{d}$ plus the dimension of the space of $2$ by $2$ matrices. 
 
From the above analysis we therefore conclude that for any sum over a doublet of rank $1$ charges the following rewriting is possible
\be 
\sum_{\substack{\Gamma_i \in \mathds{Z}^{2d(\alpha_d)} \\ \Gamma_i \times \Gamma_j = 0  }} \Phi(\Gamma) = \sum_{\gamma \in E_{d(d)}/ P_{d-1}}  \sum_{\substack{M\in \mathds{Z}^{2\times 2} \\ {\rm det}(M)\ne 0}}\gamma^T \left[ \Phi( M)\right]+\sum_{\gamma \in E_{d(d)}/ P_{d}}  \sum_{m\in \mathds{Z}^{2}} \gamma^T\left[\Phi( m)\right] \   , \label{2LaticeSum} 
\ee
where 
\bea 
P_{d} &\cong&  E_{d-1(d-1)}(\mathds{Z}) \ltimes \mathds{Z}^{d(\alpha_{d-1})} \CR
P_{d-1} &\cong& \scal{ SL(2,\mathds{Z})\times E_{d-2(d-2)}(\mathds{Z})} \ltimes \mathds{Z}^{2d(\alpha_{d-2}) + d(\alpha_1)}   \ . 
\eea
As before, the second sum in~\eqref{2LaticeSum} (over $m$) can be further decomposed into the trivial orbit $m=(0,0)$ (corresponding to the rank $0$ matrix $M=0$) and the non-trivial $m$-orbit (corresponding to rank $1$ matrices $M$).

In the following it will be also important to observe that the highest weight representation associated to the root $\alpha_{d-1}$ is the irreducible representation obtained from the antisymmetric tensor product of two copies of the $R(\alpha_d)$, with the symplectic trace removed for $d=7$, such that under the graded decomposition \eqref{alphadm1} associated with the node $\alpha_{d-1}$ one has
\be 
{\bf R}_{\alpha_{d-1}} \cong \hspace{5mm}  \dots  \hspace{5mm}   \oplus ( \overline{\bf R}_{\alpha_{d-3}} + {\bf 3}\otimes \overline{\bf R}_{\alpha_1} + \overline{\bf R}_{\alpha_1})^\ord{4} \oplus ({\bf 2}\otimes \overline{\bf R}_{\alpha_{d-2}})^\ord{13-d} \oplus {\bf 1}^\ord{22-2d} \ . \label{Alt2FRep}  
\ee
The singlet component is the highest weight space of the irreducible representation.

\subsubsection{Cases $E_{d(d)}$ for $d \geq 8$}

A similar construction also holds for $E_{8(8)}$, although in that case there are some extra components in the graded decompositions, and we do not have a theorem generalising the one in \cite{Krutelevich} to prove that all integral charges in the adjoint representation can be rotated using $E_{8(8)}(\mathds{Z})$ to the highest weight (degree $2$) component in the graded decomposition under  $E_{7(7)}$
\begin{align}
\label{E8Minimal}
\mathfrak{e}_{8(8)} &\cong  {\bf 1}^{\ord{-2}} \oplus {\bf 56}^{\ord{-1}} \oplus \scal{  \mathfrak{gl}_1 \oplus  \mathfrak{e}_{7(7)}}^\ord{0}  \oplus {\bf 56}^\ord{1} \oplus {\bf 1}^\ord{2} \ ,&\nn\\
{\bf 3875} &\cong {\bf 133}^\ord{-2} \oplus \scal{ {\bf 56} \oplus {\bf 912}}^\ord{-1} \oplus \scal{  {\bf 1} \oplus {\bf 133}\oplus {\bf 1539}}^\ord{0} \oplus \scal{ {\bf 56}\oplus {\bf 912}}^\ord{1} \oplus {\bf 133}^\ord{2} \ . &
\end{align}
The functional identities due to Langlands~\cite{LanglandsFE} applied to the constrained lattice sum in \eqref{1-loopDecomp} suggest, however, that one can still bring any rank $1$ charge to the top degree component. For the following analysis we shall assume that this is true. 
From the constraint in the ${\bf 3875}$, one directly obtains that the second charge can only lie in the positive degree components. $\Gamma_1 \times \Gamma_2|_{\bf 3875}$ vanishes for $\Gamma_2$ admitting a grad zero singlet component, but then $\Gamma_2 \times \Gamma_2|_{\bf 3875}$ would not vanish. We conclude that 
\be 
\Gamma_{(i} \times \Gamma_{j)} \big|_{{\bf 3875}} = 0 \ . \label{2ChargeRank1E8} 
\ee
is enough to imply the strong section constraint
\be 
\Gamma_{i} \times \Gamma_{j} \big|_{{\bf 3875}\oplus {\bf 248} \oplus {\bf 1}} = 0 \ . 
\label{StrongSecE8}  
\ee
So one obtains that a generic doublet of rank $1$ charges in the adjoint ${\bf 248}\cong \mathfrak{e}_{8(8)}$ can be rotated to the degree $3$ component of 
\bea
\mathfrak{e}_{8(8)} &\cong&  {\bf 2}^{\ord{-3}} \oplus {\bf 27}^{\ord{-2}} \oplus ( {\bf 2}\otimes \overline{\bf 27})^\ord{-1}\oplus  \scal{  \mathfrak{gl}_1 \oplus \mathfrak{sl}_2\oplus  \mathfrak{e}_{6(6)}}^\ord{0}  \oplus ( {\bf 2}\otimes {\bf 27})^\ord{1}\oplus \overline{\bf 27}^\ord{2} \oplus {\bf 2}^\ord{3} \ ,\CR
{\bf 3875} &\cong& \hspace{5mm} \dots \hspace{5mm}  \oplus \scal{  \overline{\bf 27} \oplus {\bf 3}\otimes \overline{\bf 27} \oplus  {\bf 351} }^\ord{2} \oplus \scal{ {\bf 2}\oplus {\bf 2}\otimes {\bf 78}}^\ord{3} \oplus {\bf 27}^\ord{4} \ . \CR
{\bf 30380} &\cong& \hspace{20mm} \dots \hspace{10mm} \oplus ( \overline{\bf 351} \oplus {\bf 3}\otimes {\bf 27}\oplus  {\bf 27})^\ord{4}  \oplus ( {\bf 2}\otimes \overline{\bf 27})^\ord{5} \oplus {\bf 1}^\ord{6} \ ,
\eea
where the last line reproduces indeed \eqref{Alt2FRep}. 

For $d\ge 9$, the (conjectured) symmetry groups $E_{d(d)}$ are of infinite-dimensional (indefinite) Kac--Moody type~\cite{Julia:1980gr}. The highest weight representations $R_{\alpha_d}$ and $R_{\alpha_1}$ are also infinite-dimensional, and it is not clear that one can make sense of the associated constrained lattice sum. Nevertheless, one can formally define the constrained lattice sum as the right-hand-side of \eqref{2LaticeSum} and use it as a definition for Kac--Moody groups. A definition of Eisenstein series in the sense of Langlands as orbit sums for affine $E_{9(9)}$ was given in~\cite{Garland1,Garland2} and an analysis of the minimal and next-to-minimal series for $E_{d(d)}$ with $9\le d \le 11$ was presented in~\cite{Fleig:2012xa}.

\subsubsection{Solutions to the strong section constraint}
\label{sec:sssols}

Before closing this section, let us mention that our analysis permits to prove that a general linear space solution to the strong section constraint is always associated to either M-theory or type IIB. If we consider tensors in the three-dimensional theory depending on one coordinate $Y^1$, we have seen above that we can always consider an embedding of $GL(1) \times E_{7(7)} \subset E_{8(8)}$ such that this coordinate lies in the highest ($GL(1)$ degree $2$) component in \eqref{E8Minimal}. The strong section constraint then implies that these tensors can depend additionally only on other coordinates $Y^M$ in the $GL(1)$ degree $1$ component ${\bf 56}$ of the same decomposition  \eqref{E8Minimal} these $Y^M$ coordinates have to satisfy the four-dimensional strong section constraint in order for the three-dimensional one to be fulfilled. Assuming then that the tensors depend on at least one additional coordinate $Y^2$ (chosen among the $Y^M$), one can always chose an embedding $GL(1) \times E_{6(6)}\subset E_{7(7)}$ such that this extra coordinate $Y^2$ lies in the $GL(1)$ degree $3$ component \eqref{1ChargeGrad4}. It follows then from the four-dimensional strong section constraint that these tensors can only depend then on the  $GL(1)$ degree $1$ component $\overline{\bf 27}^\ord{1}$ in \eqref{1ChargeGrad4} that in turn has to satisfy the five-dimensional strong section constraint. 

Iteratively, one concludes that each new coordinate that the tensor fields can depend on according to the strong section constraint always corresponds to an additional compactification circle up to duality. One obtains iteratively that tensors depending on six independent coordinates $Y^1,Y^2,Y^3,Y^4,Y^5,Y^6$ in the ${\bf 248}$ of $E_{8(8)}$ can always be defined in a frame such that the latter span the momenta along $T^6$ for the reduction of supergravity in nine dimensions, \ie the degree $7$ component of the graded decomposition with respect to the $\alpha_4$ node:
\bea
 \mathfrak{e}_{8(8)} &\cong&\  \cdots \ \oplus \scal{ \mathfrak{gl}_1\oplus  \mathfrak{gl}_1\oplus  \mathfrak{sl}_2\oplus  \mathfrak{sl}_6}^\ord{0} \oplus ( {\bf 6}_{\scriptscriptstyle 4} \oplus  {\bf 2}_{\scriptscriptstyle -3}\otimes {\bf 6})^\ord{1} \oplus ({\bf 2}_{\scriptscriptstyle 1}\otimes {\bf 15}   )^\ord{2} \oplus {\bf 20}_{\scriptscriptstyle -2}^\ord{3}\CR
&& \hspace{55mm}  \oplus \overline{\bf 15}_{\scriptscriptstyle 2}^\ord{4} \oplus ({\bf 2}_{\scriptscriptstyle -1}\otimes \overline{\bf 6})^\ord{5} \oplus ( {\bf 1}_{\scriptscriptstyle -4} \oplus {\bf 2}_{\scriptscriptstyle 3})^\ord{6} \oplus {\bf 6}^\ord{7} \ , \CR
{\bf 3875}&\cong&  \hspace{20mm} \cdots  \hspace{20mm} \oplus \overline{\bf 6}_{\scriptscriptstyle 2}^\ord{11} \oplus {\bf 2}_{\scriptscriptstyle -1}^\ord{12} \ .  
\eea
By the same argument, the dependence of the tensor fields on any additional coordinates requires them to belong to the degree $6$ component and to satisfy the strong section constraints in nine dimensions. In this case however, there are two independent solutions, either the additional coordinates belong to the ${\bf 1}_{\scriptscriptstyle -4}$ or to the ${\bf 2}_{\scriptscriptstyle 3}$. In the former case the tensor fields can be defined in type IIB supergravity, in the second the tensor fields may depend as well on the two additional coordinates in the  ${\bf 2}_{\scriptscriptstyle 3}$ and the tensor fields can be defined in eleven-dimensional supergravity.   The argument being inductive in the dimension, it follows that it applies for any $d\le 8$. Let us remark that the type IIB and eleven-dimensional solutions as maximal linear spaces are inequivalent under $E_{d(d)}$~\cite{Blair:2013gqa} although the strong section constraint is covariant. This is no contradiction with the fact that any single coordinate vector in the solution space can be thought of as a point in the associated generalised coordinate representation ${\bf R}_{\alpha_d}$ and that there is only a single $E_{d(d)}$ orbit of solutions to the strong section constraint in ${\bf R}_{\alpha_d}$ for a single charge for $d
\ge 3$. Our argument is similar to Kaluza--Klein oxidation and level decompositions of hidden symmetry groups~\cite{Cremmer:1999du,Keurentjes:2002xc,Kleinschmidt:2003mf}.

\subsection{The $\nabla^4 R^4$ non-degenerate orbit}
\label{sec:nd}

The analysis of the preceding section showed how to rewrite the sum over a pair of charges satisfying the strong section constraint~\eqref{2ChargeRank1} in terms of a sum over integral $(2\times 2)$-matrices $M$ and orbits of $P_{d-1}\backslash E_{d(d)}$. We will now apply this rewriting to the two-loop amplitude~\eqref{eq:2loops} expanded at low orders in the Mandelstam variables. The lowest non-trivial order in~\eqref{eq:2loops} is proportional to $s^2+t^2+u^2$, contributing to $\mathcal{E}_{\gra{1}{0}}$, and corresponds to the $\nabla^4 R^4$ correction. 
Recalling the definition 
\begin{align}
\sigma_n = \frac{1}{4^n} \left(s^n+t^n+u^n\right),
\end{align}
the threshold function $\cE_\gra{p}{q}$ is the coefficient of $\sigma_2^p \sigma_3^q$ in the expansion of the four-graviton amplitude in string theory.  It is convenient to consider the properties of the Schwinger integrand function, and we therefore define the exceptional field theory $2$-loop threshold functions $\cE^{\scriptscriptstyle \mbox{\tiny 2-loop}}_\gra{p}{q}$  as the integrals 
\be 
\cE_\gra{p}{q}^{\scriptscriptstyle \mbox{\tiny 2-loop}} = \int_{\mathds{R}_+^{\times 3}} d^3 \Omega\  \cF_\gra{p}{q} \ . 
\ee
The functions $\cF_\gra{p}{q}$ are given by sums over the charges $\Gamma_i$ and also involve the functions $\Phi_\gra{p}{q}$ from~\eqref{Phis}. The corresponding contribution to $\cE^{\scriptscriptstyle \mbox{\tiny 2-loop}}_\gra{1}{0}$ is the integral of
\begin{align}
\label{F10}
\cF_\gra{1}{0} = \frac{2\pi^{5-d}}{3}  \sum_{\substack{\Gamma_i \in \mathds{Z}^{2d(\alpha_d)} \\ \Gamma_i \times \Gamma_j = 0   }} \lp\det \Omega\rp^{\frac{d-7}{2}} e^{- \Omega^{ij} g(\Gamma_i , \Gamma_j)} \ .  
\end{align}

\allowDbreak{We now analyse the contribution to $\cE_\gra{1}{0}^{\scriptscriptstyle \mbox{\tiny 2-loop}}$ coming from the non-degenetate orbit corresponding to $\det M\neq 0$ in ~\eqref{2LaticeSum} in a fashion similar to~\cite{Green:1999pu}. Integrating~\eqref{F10} over Schwinger parameter space, this gives the $2$-loop contribution to the $\nabla^4 R^4$ type invariant in the Wilsonian effective action coming from the sum over linearly independent rank $1$ charges (indicated by `n.d.' for non-degenerate) 
\begin{align}
\label{2loopnd}
&\mathcal{E}_\gra{1}{0}^{\scriptscriptstyle \mbox{\tiny 2-loop, n.d.}} =\frac{2}{3} \pi^{5-d}   \sum_{\substack{\Gamma_i \in \mathds{Z}^{2d(\alpha_d)} \\ \Gamma_i \times \Gamma_j = 0 \\ \varepsilon^{ij} \Gamma_i \Gamma_j \ne 0  }} \int \frac{d^3 \Omega}{\lp\det \Omega\rp^{\frac{7-d}{2}}} e^{- \Omega^{ij} g(\Gamma_i , \Gamma_j)} &\nn \\
&= 8 \pi^{5-d}  \hspace{-5mm} \sum_{\gamma \in P_{d-1} \backslash E_{d(d)}} \int_0^\infty  \frac{dV}{V^{5-d}} \int_{\mathds{C}^+}\frac{d\tau_1 d\tau_2}{\tau_2^{\; 2}} \sum_{\substack{0\le j<m \\ m>0, \, n \ne 0}} \gamma \left[e^{-V e^{2(11-d)\phi }\scal{ \frac{|m \tau + ( j + n \taubis)|^2}{\tau_2 \taubis_2}- 2 mn}}\right] &\nn \\
&= 8 \pi^{\frac{11}{2}-d} \hspace{-5mm} \sum_{\gamma \in P_{d-1} \backslash E_{d(d)}} \hspace{-2mm} \gamma\left[ e^{-(11-d) \phi} \sqrt{ \taubis_2} \sum_{\substack{0\le j<m \\ m>0, \, n \ne 0}}  \frac{1}{m} \int_0^\infty   \frac{dV}{V^{\frac{11}{2}-d}} \int_0^\infty \frac{d\tau_2}{\tau_2{}^{\frac{3}{2}}}e^{-e^{2(11-d)\phi} \scal{ \frac{m^2}{\taubis_2} V \tau_2 + n^2 \taubis_2 \frac{V}{\tau_2}}} \right]&\nn \\
&= 4 \pi^{\frac{11}{2}-d}  \hspace{-5mm} \sum_{\gamma \in P_{d-1} \backslash E_{d(d)}}  \hspace{-2mm} \gamma\left[e^{-(11-d) \phi} \sqrt{ \taubis_2} \sum_{\substack{0\le j<m \\ m>0, \, n \ne 0}} \frac{1}{m} \int_0^\infty  \hspace{-3mm}dx \, x^{\frac{d-6}{2}} \int_0^\infty dy\,  y^{\frac{d-7}{2}} e^{-e^{2(11-d)\phi} \scal{ \frac{m^2}{\taubis_2} y  + n^2 \taubis_2 x}}\right]&\CR
&= 8\pi    \xi(d-5)\xi(d-4)  \sum_{\gamma \in P_{d-1}\backslash E_{d(d)}}\gamma\left[ e^{-2(11-d)(d-4) \phi} \right]&\CR
&= 8 \pi    \xi(d-5) \xi(d-4) E_{\alpha_{d-1},\frac{d-4}{2}} \ .&  
\end{align}
Unlike \cite{Green:1999pu} we do not use Poisson resummation but  consider the integral over the original Schwinger parameters. Let us explain the individual steps of this calculation.
In the first step we have used~\eqref{2LaticeSum} to rewrite the sum over charges using elements $\gamma$ of the coset $P_{d-1}\backslash E_{d(d)}$ and $\gamma[\cdot]$ denotes the action of $\gamma$ on the charges or dually on the symmetric space as in appendix~\ref{app:ES}. We have also redefined the Schwinger parameters contained in $\Omega$ as
\begin{align}
\label{COV}
\tau_1 = \frac{L_3}{L_2+L_3},\quad \tau_2 = \frac{\lp\det\Omega\rp^{1/2}}{L_2+L_3},\quad V= \lp\det\Omega\rp^{1/2}
\end{align}
that can be used to map the Schwinger parameters integral to the one of $V$ over $\mathds{R}_+^*$ and $\tau$ over the fundamental domain of the congruence subgroup $\Gamma_0(2)$ which is three times as large as that of the modular group $PSL(2,\mathds{Z})$. Permutation symmetry then allows to further reduce to the fundamental domain of $PSL(2,\mathds{Z})$ that also arises naturally in the corresponding string theory calculation~\cite{Pioline:2015nfa}. Going from $SL(2,\mathds{Z})$ to the $\Gamma_0(2)$ therefore yields a factor of $6$; the change of variables~\eqref{COV} additionally provides a factor of $2$ from the Jacobian. Afterwards, we have unfolded the $PSL(2,\mathds{Z})$ fundamental domain to the full upper half plane $\mathds{C}^+$ at the expense of taking a representative of the non-degenerate matrix $M$ appearing in the rewriting~\eqref{2LaticeSum} as follows. The sum over $j$, $m$ and $n$ is over representatives $\lp\begin{smallmatrix} m&j\\0&n\end{smallmatrix}\rp$ of the action of $SL(2,\mathds{Z})$ on integral $(2\times 2)$-matrices $M$ with $\det M\neq 0$. Furthermore, we have denoted by $\taubis$ the complex scalar field parametrising the Levi component $SL(2)/SO(2)$ in the parabolic subgroup, while $\phi$ is the dilaton associated to the root $\alpha_{d-1}$, normalised according to \eqref{alphadm1}. The change of variables $x=V/\tau_2$, $y=V\tau_2$ factorises the integral into two $\Gamma$-integrals and the integer sums can then be evaluated as Riemann zeta functions. }

In the last step, we have identified the coset sum with an Eisenstein series associated with the maximal parabolic for node $\alpha_{d-1}$. The normalisation of the Eisenstein series weight parameter $s=\frac{d-4}{2} $ is fixed by noting that for $6\le d\le 8$ the highest weight representation associated to the node $\alpha_{d-1}$ is the antisymmetric tensor product of ${\bf R}_{\alpha_d}$ with itself, minus the symplectic trace for $E_{7(7)}$. Therefore we conclude that the highest degree term in the decomposition of the ${\bf R}_{\alpha_{d-1}}$ in \eqref{alphadm1} is ${\bf 1}^\ord{22-2d}$, which gives
\be E_{\alpha_{d-1},s} =  \sum_{\gamma \in P_{d-1}\backslash E_{d(d)}}\gamma\left[ e^{-4s (11-d) \phi} \right] \ . \ee
 One must consider the case $d\le 4$ separately, although the decomposition still corresponds to the large $T^2$ volume decompactification limit. 

We will now discuss the $2$-loop result for linearly independent rank $1$ charges
\begin{align}
\label{twoloopind}
\mathcal{E}_\gra{1}{0}^{\scriptscriptstyle \mbox{\tiny 2-loop, n.d.}} = 
8\pi  \xi(d-4) \xi(d-5) E_{\alpha_{d-1},\tfrac{d-4}{2}}
\end{align}
in various dimensions. The manipulations in~\eqref{2loopnd} converge absolutely for $d>5$ where the result is rather uniform. The cases $d\leq 5$ are a bit less regular but they can be defined by analytic continuation, although there is a pole at $d=5$ as we will see.

\subsubsection{The non-degenerate orbit and $\nabla^4 R^4$ for $SL(2)\times SL(3)$}

In $D=8$ space-time dimensions ($d=3$) there are two distinct orbits, associated to the decompactification to type IIA and type IIB, coming from the fact that the constraint on the charge $\Gamma$ in $9$ dimensions has two distinguished orbits, either the charge is a Kaluza--Klein momentum, or an M2-brane. The M2-brane leads to the type IIB decompactification limit:
\begin{align}
\mathfrak{sl}_2 \oplus \mathfrak{sl}_3 &\cong ({\bf 1},{\bf 2}')^\ord{-12} \oplus \scal{ \mathfrak{gl}_1 \oplus \mathfrak{sl}_2  \oplus \mathfrak{sl}_2' }^\ord{0} \oplus ({\bf 1},{\bf 2}')^\ord{12} ,&\CR
\lp {\bf 2}, {\bf 3}\rp &\cong \lp{\bf 2},{\bf 2}'\rp^\ord{-4} \oplus \lp{\bf 2},{\bf 1}'\rp^\ord{8} \ . 
\end{align}
In this case the associated root is $2\alpha_2$ of $SL(3)$, and one obtains 
\be 
8\pi  \xi(2) \xi(-2)   E_{[0,-1]} =  \zeta(5) E_{[\stfrac{5}{2},0]} \ , 
\ee
which is the expected answer. 

The type IIA decompactification limit from $9$ dimensions corresponds instead to the removal of the nodes $\alpha_2$ and $\alpha_3$ of the $E_{3(3)}\cong SL(2)\times SL(3)$ Dynkin diagram~\ref{fig:dynk}. The associated graded decomposition is
\bea 
\mathfrak{sl}_2 \oplus \mathfrak{sl}_3 &\cong& {\bf 1}^\ord{-12} \oplus {\bf 2}^\ord{-6}\oplus  \scal{ \mathfrak{gl}_1 \oplus \mathfrak{sl}_2  \oplus \mathfrak{gl}_1 }^\ord{0} \oplus {\bf 2}^\ord{6} \oplus {\bf 1}^\ord{12} \  , \CR
\lp{\bf 2}, {\bf 3}\rp &\cong& {\bf 1}^\ord{-10} \oplus {\bf 2}^\ord{-4} \oplus {\bf 1}^\ord{2} \oplus {\bf 2}^\ord{8}\ ,
\eea
where the degree corresponding to $\phi$ is now given by the combination $2\alpha_2+\alpha_3$.  One obtains in this case 
\be 
8 \pi \xi(2) \xi(-2)   E_{[-1]} E_{[-\stfrac{1}{2},0]}= - 4   ( 2 \zeta(4)E_{[2]} )( 2\zeta(-1) E_{[-\stfrac{1}{2},0]})  \ . 
\ee
The sum of the two contributions  gives the complete $\nabla^4 R^4$ threshold function 
\be 
\mathcal{E}_\gra{1}{0}^{\scriptscriptstyle \mbox{\tiny 2-loop, n.d.}} = \zeta(5) E_{[\stfrac{5}{2},0]} - 4  ( 2 \zeta(4)E_{[2]} )( 2\zeta(-1) E_{[-\stfrac{1}{2},0]}) \ , 
\ee
whereas the one-loop amplitude already included the second factor that was, however, removed by the renormalisation at one loop within our prescription. This result is in agreement with~\cite{Basu:2007ru}.

\subsubsection{The non-degenerate orbit and $\nabla^4 R^4$ for $SL(5)$}

In $D=7$ space-time dimensions ($d=4$) one gets that the antifundamental representation decomposes in the decompactification limit as
\be 
\overline{\bf 5} \cong \hat{\bf 2}^\ord{-4} \oplus {\bf 2}^\ord{1} \oplus {\bf 1}^\ord{6} \ . 
\ee
In this case the function is evaluated at zero with a divergent coefficient, and one must consider the analytic continuation of the parameter. One works out that this graded decomposition is associated to $\alpha_2 + \alpha_4$ ($\alpha_2 + \alpha_3$ in the $E_4$ labeling), and the corresponding Eisenstein function is 
\be  8\pi   \xi(2\epsilon) \xi(2-2\epsilon)  E_{[0,\epsilon,0,\epsilon]}   = - \frac{2\pi^2}{3\epsilon}  + \zeta(5) \hat{E}_{[\stfrac{5}{2},0,0,0]} + \frac{6 \zeta(5) \zeta(4)}{\pi^3}\hat{E}_{[0,0,\stfrac{5}{2},0]} \ ,\label{5D2-loopThreshold}  \ee
which gives the correct answer. Note again that there would have been a double counting with the 1-loop computation without the renormalisation prescription to remove all the $\nabla^4 R^4$ contributions at one-loop. Note moreover that the 2-loop $\epsilon$ contribution cancels precisely the 2-loop logarithm divergence in seven dimensions. For this purpose we consider the infrared regularised component of the amplitude (with $d = 4 + 2 \epsilon$)
\be 
 4  \pi^{5-d}  \int_0^\infty \frac{dV}{V^{5-d}} \int_{\cF}\frac{d\tau d\bar \tau}{\tau_2^{\; 2}} e^{-V \mu^2}  = \frac{2\pi^2}{3\epsilon}  - \frac{4\pi^2}{3} \scal{ \gamma + \mbox{ln}(\pi \mu^2)} + \mathcal{O}(\epsilon) \ . \label{5D2-loopDiv} 
\ee

\subsubsection{The non-degenerate orbit and $\nabla^4 R^4$ for $SO(5,5)$}
\label{sec:6Dnd}

The case of $D=6$ space-time dimensions ($d=5$) is particularly interesting and subtle because in this case the degenerate orbits contributes non-trivially to the threshold function. This fact is related to the existence of a 1-loop form factor divergence of the $R^4$ type invariant into the $\nabla^4 R^4$ invariant, which is reflected in inhomogeneous terms in the differential equation satisfied by the threshold function \cite{Bossard:2014lra}. The associated form fact carries also an infrared divergence that is responsible for the non-trivial contribution of the degenerate orbit.

The prefactor in~\eqref{twoloopind} is  divergent for $d=5$; it has a double pole; and the Eisenstein series has a simple zero, such that the full expression will exhibit a single pole for $d\to 5$. We regularise by $d=5+2\epsilon$ and find
\begin{align}
\label{D4R46}
\mathcal{E}_\gra{1}{0}^{\scriptscriptstyle \mbox{\tiny 2-loop, n.d.}} =  8\pi  \xi(1+2\epsilon) \xi(2\epsilon) E_{\mbox{\DSOX000{\mbox{$\frac{1}{2}$+$\epsilon$}}0}}  
 = 8\pi   \xi (2 \epsilon ) \xi (3-2 \epsilon )  E_{\mbox{\DSOX{\mbox{$\frac{3}{2}$-$\epsilon$}}0{\mathnormal{\epsilon}}0{\mathnormal{\epsilon}}}}  
\end{align}
after using a functional relation for the Eisenstein series. This brings out the simple pole at $\epsilon\to 0$ in terms of the prefactor, multiplying a function that is regular for $\epsilon\to 0$. We determine the expansion of that function up to order $\epsilon$ as follows:\footnote{In the second step, we use the following general equality for the directional derivative of a function $f$ in the direction $\vec{W}=\sum_a \vec{w}_a$: $\partial_\epsilon f(\vec{v}+\epsilon \vec{W})|_{\epsilon=0} = \vec{W}\cdot \vec{\nabla} f(\vec{v})= \sum_a \vec{w}_a \cdot \vec{\nabla} f(\vec{v}) = \sum_a \partial_\epsilon f(\vec{v}+\epsilon \vec{w}_a)|_{\epsilon=0}$.}
\begin{align}
\label{vecex}
E_{\mbox{\DSOX{\mbox{$\frac{3}{2}$-$\epsilon$}}0{\mathnormal{\epsilon}}0{\mathnormal{\epsilon}}}}  &=E_{\mbox{\DSOX{\mbox{$\frac{3}{2}$}}0000}}  + \epsilon \Bigl( \partial_\epsilon  E_{\mbox{\DSOX{\mbox{$\frac{3}{2}$-$\epsilon$}}0{\mathnormal{\epsilon}}0{\mathnormal{\epsilon}}}} \Bigr)\Big|_{\epsilon=0} + \mathcal{O}(\epsilon^2)  &\CR
&=E_{\mbox{\DSOX{\mbox{$\frac{3}{2}$}}0000}}  + \epsilon \Bigl( \partial_\epsilon E_{\mbox{\DSOX{\mbox{$\frac{3}{2}$+$\epsilon$}}0000}} +\partial_\epsilon  E_{\mbox{\DSOX{\mbox{$\frac{3}{2}$-$\epsilon$}}000{\mathnormal{\epsilon}}}}  + \partial_\epsilon  E_{\mbox{\DSOX{\mbox{$\frac{3}{2}$-$\epsilon$}}0{\mathnormal{\epsilon}}00}} \Bigr)\Big|_{\epsilon=0} + \mathcal{O}(\epsilon^2)  &\CR
&= E_{\mbox{\DSOX{\mbox{$\frac{3}{2}$+$\epsilon$}}0000}}   +   E_{\mbox{\DSOX{\mbox{$\frac{3}{2}$-$\epsilon$}}000{\mathnormal{\epsilon}}}} +  E_{\mbox{\DSOX{\mbox{$\frac{3}{2}$-$\epsilon$}}0{\mathnormal{\epsilon}}00}} - 2 E_{\mbox{\DSOX{\mbox{$\frac{3}{2}$}}0000}} + \mathcal{O}(\epsilon^2)  &\CR
&= \frac{\xi(2-2\epsilon) \xi(5-2\epsilon)}{\xi(1-2\epsilon)\xi(3+2\epsilon)} E_{\mbox{\DSOX{\mbox{$\frac{5}{2}$-$\epsilon$}}0000}} +\frac{\xi(4-2\epsilon) \xi(6-2\epsilon)}{\xi(1-2\epsilon) \xi(3-2\epsilon)}E_{\mbox{\DSOX0000{\mbox{$3$-$\epsilon$}}}}  &\CR
& \hspace{20mm}+ \frac{\xi(4-2\epsilon) \xi(6-2\epsilon)}{\xi(1-2\epsilon) \xi(3-2\epsilon)} E_{\mbox{\DSOX00{\mbox{$3$-$\epsilon$}}00}}- 2 E_{\mbox{\DSOX{\mbox{$\frac{3}{2}$}}0000}} + \mathcal{O}(\epsilon^2).&   
\end{align}
In the last step, we have used the functional relation for Eisenstein series to bring three of the terms into simpler representatives in terms of vector and spinor Eisenstein series.

Putting the above rewriting together with~\eqref{D4R46}, one then finds 
\begin{align}
\mathcal{E}_\gra{1}{0}^{\scriptscriptstyle \mbox{\tiny 2-loop, n.d.}}  &  = 8\pi   \xi (2 \epsilon ) \xi (3-2 \epsilon )  E_{\mbox{\DSOX{\mbox{$\frac{3}{2}$-$\epsilon$}}0{\mathnormal{\epsilon}}0{\mathnormal{\epsilon}}}}  &\\
&= 8 \pi \biggl(   \frac{\xi(2-2\epsilon)  \xi (3-2 \epsilon )   \xi(5-2\epsilon)}{\xi(3+2\epsilon)} E_{\mbox{\DSOX{\mbox{$\frac{5}{2}$-$\epsilon$}}0000}} +\xi(4-2\epsilon) \xi(6-2\epsilon) E_{\mbox{\DSOX0000{\mbox{$3$-$\epsilon$}}}}  \biggr . &\CR
& \hspace{30mm}\biggl . + \xi(4-2\epsilon) \xi(6-2\epsilon) E_{\mbox{\DSOX00{\mbox{$3$-$\epsilon$}}00}}- 2  \xi (2 \epsilon ) \xi (3-2 \epsilon )  E_{\mbox{\DSOX{\mbox{$\frac{3}{2}$}}0000}} \biggr)&\CR
&=\Scal{ - \frac{2 \zeta(3)}{\epsilon} +c} E_{\mbox{\DSOX{\mbox{$\frac{3}{2}$}}0000}}  + \zeta(5) \hat{E}_{\mbox{\DSOX{\mbox{$\frac{5}{2}$}}0000}} + \frac{8\zeta(6)}{45} \hat{E}_{\mbox{\DSOX0000{\mbox{$3$}}}} + \frac{8\zeta(6)}{45} \hat{E}_{\mbox{\DSOX00{\mbox{$3$}}00}} + O(\epsilon)\ . &\nn
\label{NonDegene6D} 
\end{align}
This isolates the divergent piece in $\tfrac1\epsilon$ that we will discuss further in connection with the degenerate orbit $\textrm{rank}\,M<2$ in section~\ref{sec:IRdeg}. The finite terms contain the regularised vector and spinor Eisenstein series as well as a finite piece proportional to the series $E{\mbox{\DSOX{\mbox{$\frac{3}{2}$}}0000}}  $ that we do not specify.\footnote{One could straightforwardly compute the precise coefficient $c$ for a given prescription for the regularaised Eisenstein series, but its explicit value is rather complicated and will not be relevant in this paper.} The resulting threshold function as it stands in inconsistent with supersymmetry because the series $\hat{E}{\mbox{\DSOX00{\mbox{$3$}}00}} $ does not satisfy the appropriate differential equation \cite{Bossard:2014lra}. However, we will see that this contribution will be cancelled by the degenerate orbit contribution, together with the pole in $\frac{1}{\epsilon}$. 

\subsubsection{The non-degenerate orbit and $\nabla^4 R^4$ for $E_{6(6)}$}

In the case of $D=5$ space-time dimensions ($d=6$) the prefactor in~\eqref{twoloopind} still has a single pole, however, this is compensated by a single zero in the Eisenstein series for the relevant $s$-value. We demonstrate this by letting $d=6+2\epsilon$ and using a functional relation to find
\begin{align}
\mathcal{E}_\gra{1}{0}^{\scriptscriptstyle \mbox{\tiny 2-loop, n.d.}} = 8\pi  \xi(2+2\epsilon) \xi(1+2\epsilon) E_{\mbox{\DEVI0000{\mathnormal{1\mbox{+}\epsilon\, }}0}}  
= 8\pi  \frac{\xi(2\epsilon-1)\xi(2\epsilon-2)\xi(4\epsilon-4)}{\xi(4\epsilon-2)} E_{\mbox{\DEVI{\mathnormal{\frac{5}{2}\mbox{-}2\epsilon\, }}00{\mathnormal{\epsilon}}00}}\,  . 
\end{align}
Taking now the $\epsilon\to 0$ limit yields
\begin{align}
\mathcal{E}_\gra{1}{0}^{\scriptscriptstyle \mbox{\tiny 2-loop, n.d.}} =   \zeta(5) E_{\mbox{\DEVI{\mathnormal{\frac{5}{2}}}00000}}\ ,
\end{align}
the full $\nabla^4 R^4$ function with the correct coefficient since
\begin{align}
\lim_{\epsilon\to 0} \frac{\xi(2\epsilon-1)\xi(2\epsilon-2)\xi(4\epsilon-4)}{\xi(4\epsilon-2)}  = \frac{1}{8\pi} \zeta(5)\ .
\end{align}

\subsubsection{The non-degenerate orbit and $\nabla^4 R^4$ for $E_{7(7)}$}

For $D=4$ space-time dimensions ($d=7$) the prefactor in~\eqref{twoloopind} is finite. Putting $d=7+2\epsilon$ we find
\begin{align}
\mathcal{E}_\gra{1}{0}^{\scriptscriptstyle \mbox{\tiny 2-loop, n.d.}} = 8\pi  \xi(3+2\epsilon) \xi(2+2\epsilon) E_{\mbox{\DEVII00000{\mathnormal{\tfrac{3}{2}\mbox{+}\epsilon\, }}{\mathfrak{0}}}}  
=   8\pi  \frac{\xi(2\epsilon-1)\xi(2\epsilon-2)\xi(4\epsilon-4)}{\xi(4\epsilon-2)} E_{\mbox{\DEVII{\mathnormal{\frac{5}{2}\mbox{-}2\epsilon\, }}00{\mathnormal{\epsilon}}00{\mathfrak{0}}}}\  ,
\end{align}
the full $D^4 R^4$ threshold function for $\epsilon\to 0$ with the right coefficient.

\subsubsection{The non-degenerate orbit and $\nabla^4 R^4$ for $E_{8(8)}$}

For $D=3$ space-time dimensions ($d=8$) the prefactor in~\eqref{twoloopind} is finite. The prefactor in~\eqref{twoloopind} now is finite. Putting $d=8+2\epsilon$ we find
\begin{align}
\mathcal{E}_\gra{1}{0}^{\scriptscriptstyle \mbox{\tiny 2-loop, n.d.}} = 8\pi  \xi(4+2\epsilon) \xi(3+2\epsilon) E_{\mbox{\DEVIII000000{\mathnormal{2\mbox{+}\epsilon\, }}0}}  
=   8\pi  \frac{\xi(2\epsilon-1)\xi(2\epsilon-2)\xi(4\epsilon-4)}{\xi(4\epsilon-2)} E_{\mbox{\DEVIII{\mathnormal{\frac{5}{2}\mbox{-}2\epsilon\, }}00{\mathnormal{\epsilon}}0000}}\  ,
\end{align}
the full $D^4 R^4$ threshold function for $\epsilon\to 0$ with the right coefficient.

\subsubsection{The non-degenerate orbit and $\nabla^4 R^4$ for $E_{d(d)}$ and $d>8$}

We can also formally consider the case $d>8$ where the hidden symmetry group becomes of Kac--Moody type. Using functional relations one can map the Eisenstein series on node $\alpha_{d-1}$ to a series on node $\alpha_1$ for $9\leq d \leq 11$. The resulting two-loop threshold function is 
\begin{align}
\mathcal{E}_\gra{1}{0}^{\scriptscriptstyle \mbox{\tiny 2-loop, n.d.}} =  \zeta(5)   E_{\mbox{\DEVIII{\mathnormal{\frac{5}{2}}}000000{\mathnormal{\ldots}}}} 
\end{align}
in all cases, corresponding to the correctly normalised Eisenstein series discussed in~\cite{Fleig:2012xa} for the $\nabla^4 R^4$ correction. This `next-to-minimal' series has the special property that it only possesses a finite number of constant terms. 

\subsection{Differential equations for $\nabla^4 R^4$}
\label{sec:DE}

The full threshold function $\cE_\gra{1}{0}$ for the $\nabla^4 R^4$ has to satisfy differential constraints in order to be consistent with supersymmetry~\cite{Bossard:2014lra,Bossard:2014aea}. In particular, the threshold function should be an (almost) eigenfunction of the Laplace operator on $E_{d(d)}/K_d$ with eigenvalue $\tfrac{5(4-d)(d+1)}{9-d}$~\cite{Green:2010wi}. 

In order to investigate this, one computes in a first step that the Laplace operator $\Delta$ on $E_{d(d)}/K_d$ acts on the integrand function $\cF_\gra{1}{0}$ defined in~\eqref{F10} as
\bea 
&& \Scal{ \Delta - \frac{5(4-d)(d+1)}{9-d}} \cF_\gra{1}{0} \CR
&=&  \frac{\partial\, }{\partial \Omega^{ij}} \Scal{2 \Omega^{ik} \Omega^{jl}  \frac{\partial\, }{\partial \Omega^{kl}} + \frac{2}{9-d} \Omega^{ij} \Omega^{kl}  \frac{\partial\, }{\partial \Omega^{kl}} + \frac{82-13d+d^2}{9-d} \Omega^{ij}} \cF_\gra{1}{0}  \ . 
\eea
One obtains therefore that the integrated threshold function $\cE^{\scriptscriptstyle \mbox{\tiny 2-loop}}_\gra{1}{0}$ satisfies a Laplace equation, up to potential boundary contributions. The ultraviolet boundary terms at $L_I=0$ all vanish, and the infrared boundary terms at $L_I\rightarrow\infty$ vanish for the non-degenerate orbit $\det M\neq 0$. For the degenerate orbit the infrared boundary terms only vanish for $d\le 4$. For $d=5$ one obtains a boundary term that simplifies to the minimal representation function
\be
\Scal{ \Delta  + \frac{15}{2} } \int d^3 \Omega\  \cF_\gra{1}{0} = 14 \zeta(3)E_{\mbox{\DSOX{\mbox{$\frac{3}{2}$}}0000}} \ . 
\ee
In space-time dimensions $D<6$, corresponding to $d>5$, the boundary term diverges in $L_i^{\frac{d-5}{2}}$ times the minimal representation function $E_{\alpha_d,\frac{d-3}{2}}$, plus additional subleading terms.

In the preceding section, we have carried out the integral and obtained the contribution from the non-degenerate orbit at two loops in terms of an Eisenstein series, cf.~\eqref{2loopnd}. Let us also analyse the differential equation satisfied by this partial answer. This Eisenstein series was induced from a parabolic decomposition of $E_{d(d)}$ with Levi factor $GL(2)\times E_{d-2(d-2)}$. For any function of $\phi$ and $\taubis$ in the $GL(2)$ subgroup of $E_{d(d)}$, the Laplace operator reduces to 
\be 
\Delta F(\phi,\taubis) = \frac{1}{4(9-d)(11-d)} \scal{ \partial_\phi^{\; 2} + 2( d(20-d) - 39 ) \partial_\phi } F(\phi,\taubis) - ( \taubis - \bar \taubis)^2 \partial_{\taubis} \partial_{\bar \taubis} F(\phi,\taubis)   \ . 
\ee
The Eisenstein series $E_{\alpha_{d-1},s}$ is obtained from a function on $GL(2)$ by summing over $E_{d(d)}(\mathds{Z})$ images and invariance of the Laplace operator then implies that 
\be 
\Delta E_{\alpha_{d-1},s}  = \frac{ 2s( 2(11-d)s + 39-20d + d^2)}{9-d} E_{\alpha_{d-1},s} \ .
\ee
In particular, the function $E_{\alpha_{d-1},\frac{d-4}{2}}$ that arises in the two-loop calculation~\eqref{2loopnd} satisfies indeed
\be 
\Delta E_{\alpha_{d-1},\frac{d-4}{2}}  = \frac{5(4-d)(d+1)}{9-d}  E_{\alpha_{d-1},\frac{d-4}{2}}  \ , 
\ee
as required by supersymmetry. This function satisfies the correct Laplace eigen value equation for all $d$, but actually fails to satisfy the additional tensorial equation derived in \cite{Bossard:2014lra} in six dimensions. We will see that the complete six-dimensional threshold function including the degenerate orbit contribution does indeed satisfy the tensorial differential equation required by supersymmetry.

\subsection{Infrared divergences and the degenerate $\nabla^4 R^4$ orbit at 2-loop}
\label{sec:IRdeg}

In section~\ref{sec:nd} we have only considered the orbit of non-degenerate charges satisfying the strong section constraint. After the rewriting~\eqref{2LaticeSum} this corresponded to $(2\times 2)$-matrices $M$ of full rank, \ie $\det M\neq 0$. We saw that the non-degenerate orbit produces the full and correctly normalised $\nabla^4 R^4$ threshold for $d\ne 5$. In this section, we discuss the relevance of the degenerate orbits with $\det M =0$. In terms of the charges $\Gamma_1$ and $\Gamma_2$ circulating in the loops this means that they are collinear. Our main focus will the case of $D=6$ space-time dimensions ($d=5$) where we found a divergent contribution from the non-degenerate orbit in~\eqref{NonDegene6D}.

The first task is to separate the amplitude into its analytic and non-analytic parts. This analysis was carried out in detail in string theory in \cite{Pioline:2015nfa}. For this one must consider an infrared regularisation and a sliding scale $\mu$ which will appear in the Wilsonian action that defines the analytic part of the amplitude.  We will therefore consider the lowest order term $\Phi_\gra{1}{0}$ of the two-loop amplitude~\eqref{eq:2loops} with charges $\Gamma_i$ satisfying the strong section constraint and the collinearity condition $\varepsilon^{ij} \Gamma_i\Gamma_j=0$ of the degenerate orbits (denoted by `d.' for degenerate):
\begin{align}
\mathcal{E}_\gra{1}{0}^{\scriptscriptstyle \mbox{\tiny 2-loop, d.}} = \frac23 \pi^{5-d}    \sum_{\substack{\Gamma_i \in \mathds{Z}^{2d(\alpha_d)}_* \\ \Gamma_i \times \Gamma_j = 0 \\ \varepsilon^{ij} \Gamma_i \Gamma_j = 0  }} \int \frac{d^3 \Omega}{{\rm det}\Omega^{\frac{7-d}{2}}} e^{- \Omega^{ij} g(\Gamma_i , \Gamma_j)} 
\end{align}
After manipulations as in~\eqref{2loopnd}, the sliding scale is introduced in the infrared domain of Schwinger parameter space according to 
\begin{align}
\mathcal{E}_\gra{1}{0}^{\scriptscriptstyle \mbox{\tiny 2-loop, d.}} &= 4 \pi^{5-d}  \hspace{-2mm} \sum_{\gamma \in P_{d-1} \backslash E_{d(d)}} \gamma\left[\int\limits_0^\infty  \frac{dV}{V^{5-d}} \int\limits_{-1/2}^{1/2} d\tau_1 \int\limits_0^\infty \frac{ d\tau_2}{\tau_2^{\; 2}}  \sum_{(m,n)\in \mathds{Z}^2_*} e^{-V  \scal{ e^{2(11-d)\phi} \frac{|m+ n \taubis|^2}{\tau_2 \taubis_2} + \tau_2 \mu^2}} \right]&\CR
&= 2 \pi^{5-d}  \hspace{-2mm} \sum_{\gamma \in P_{d-1} \backslash E_{d(d)}} \gamma\left[\sum_{(m,n)\in \mathds{Z}^2_*} \int_0^\infty  \hspace{-3mm}dx \, x^{\frac{d-5}{2}} \int_0^\infty dy\,  y^{\frac{d-7}{2}} e^{-x e^{2(11-d)\phi} \frac{|m+ n \taubis|^2}{ \taubis_2} - y \mu^2 }\right]&\CR
&= 2 \pi^{5-d}  \mu^{5-d} \Gamma(\tfrac{d-5}{2}) \Gamma(\tfrac{d-3}{2}) \hspace{-2mm} \sum_{\gamma \in P_{d-1} \backslash E_{d(d)}} \gamma\left[\sum_{(m,n)\in \mathds{Z}^2_*} \Scal{ e^{2(11-d)\phi} \frac{|m+ n \taubis|^2}{\taubis_2} }^{-\frac{d-3}{2}}\right]&\CR
&= 4 \pi^{\frac{7-d}{2}}   \mu^{5-d} \Gamma(\tfrac{d-5}{2}) \xi(d-3) E_{\alpha_d,\frac{d-3}{2}}  &\label{Degenerate} 
\end{align}
To consider the contribution in $D=6$ space-time dimensions, we will regularise the expression using $d=5+2\epsilon$, such that this reduces to 
\be 
\mathcal{E}_\gra{1}{0}^{\scriptscriptstyle \mbox{\tiny 2-loop, d.}} = 4  \Scal{ \frac{1}{\epsilon} - 2 \scal{ \gamma + \mbox{ln}(\pi \mu)}} \zeta(2+2\epsilon) E_{\mbox{\DSOX0000{\mbox{$1$+$\epsilon$}}}}\  .  
\ee
Note that the $ \mbox{ln}(\mu)$ divergence is associated to the infrared divergence one obtains by expanding the degenerate orbit contribution in the external momenta whereas it is not an analytic function, while the $\frac{1}{\epsilon}$ pole corresponds to the ultra-violet divergence of the integral over the massless momentum. To understand the second, note that the integral over the massive momentum attached to the non-zero charge effectively produces a local $R^4$ type coupling equal to the one we derived at one loop, so that the subsequent integral over the remaining massless momentum corresponds effectively to the 1-loop $R^4$ type form factor in supergravity. This ultra-violet divergence then corresponds to this form factor divergence in supergravity.

We note moreover that we can rewrite the non-degenerate result \eqref{NonDegene6D} as
\be 
\mathcal{E}_\gra{1}{0}^{\scriptscriptstyle \mbox{\tiny 2-loop, n.d.}} = 8 \pi \xi(1-2\epsilon) \xi(2+2\epsilon) E_{\mbox{\DSOX0000{\mbox{$1$+$\epsilon$}}}} + \zeta(5) \hat{E}_{\mbox{\DSOX{\mbox{$\frac{5}{2}$}}0000}} + \frac{8\zeta(6)}{45} \hat{E}_{\mbox{\DSOX0000{\mbox{$3$}}}} 
\ee
by reabsorbing the conjugate spinor representation function in the first derivative of the minimal series with respect to $\epsilon$. Summing up the two contributions, one obtains the expected finite result 
\be  
\zeta(5) \hat{E}_{\mbox{\DSOX{\mbox{$\frac{5}{2}$}}0000}} + \frac{8\zeta(6)}{45} \hat{E}_{\mbox{\DSOX0000{\mbox{$3$}}}}  - 4  \, \mbox{ln}(2\pi \mu) \zeta(3) E_{\mbox{\DSOX{\mbox{$\frac{3}{2}$}}0000}}  \ . 
\ee
As emphasised before, we have not been careful in computing the constant part in section~\ref{sec:6Dnd} since this precise coefficient would only be relevant to define the complete amplitude, and we do not work out the non-analytic component of the amplitude in this paper. The complete answer for the 2-loop threshold function now reproduces the exact answer in string theory \cite{Green:2010wi}.\\

\allowDbreak{To further justify the infrared regulator we have used, let us consider the specific components of the amplitude that diverges in the infrared, \ie when either one of the charge $\Gamma_i$ or their sum vanish. In this case we will give a soft mass $\mu^2$ to the massless propagator to regularise the integral. One obtains 
\begin{align}
& \quad \frac{2 \pi^{5-d}}{3}  \sum_{i=1}^3\sum_{\substack{\Gamma \in \mathds{Z}^{d(\alpha_d)} \\ \Gamma \times \Gamma = 0   }}\int\limits_{0}^\infty dL \int\limits_0^1 dx \int\limits_0^\infty dL_i L (L (L_i  +  x(1-x)L))^{\frac{d-7}{2}} e^{-L |Z(\Gamma)|^2-L_i \mu^2 } &\CR
&= 2\pi^{5-d} \sum_{\substack{\Gamma \in \mathds{Z}^{d(\alpha_d)} \\ \Gamma \times \Gamma = 0   }}\int_{0}^\infty \frac{dL}{L^{\frac{5-d}{2}}}     \int_0^1 dx \,   \mu^{5-d} \Gamma\scal{ \tfrac{d-5}{2},Lx(1-x) \mu^2} e^{L \scal{  x(1-x)\mu^2 -|Z(\Gamma)|^2} }  \CR
&= 2\pi^{5-d} \sum_{\substack{\Gamma \in \mathds{Z}^{d(\alpha_d)} \\ \Gamma \times \Gamma = 0   }}\int_{0}^\infty dL     \int_0^1 dx \,  \Scal{   \Gamma\scal{ \tfrac{d-5}{2}} \frac{\mu^{5-d}}{L^{\frac{5-d}{2}}}+ \frac{2}{5-d} \scal{ L^2 x(1-x)}^{\frac{d-5}{2}} }  e^{L \scal{  x(1-x)\mu^2 -|Z(\Gamma)|^2} }  \CR
 &= 2\pi^{5-d} \sum_{\substack{\Gamma \in \mathds{Z}^{d(\alpha_d)} \\ \Gamma  \times \Gamma = 0   }}\Scal{ \mu^{5-d} \Gamma(\tfrac{d-5}{2}) \Gamma(\tfrac{d-3}{2}) |Z(\Gamma)|^{-\frac{d-3}{2}} - 2 \frac{ \Gamma(d-5) \Gamma(\tfrac{d-3}{2})^2}{\Gamma(d-3)} |Z(\Gamma)|^{-(d-4)}   } \CR
 &= 4 \pi^{\frac{7-d}{2}} \mu^{5-d} \Gamma(\tfrac{d-5}{2}) \xi(d-3) E_{\alpha_d,\frac{d-3}{2}} - 8 \pi \frac{ \Gamma(d-5) \Gamma(\tfrac{d-3}{2})^2}{\Gamma(d-3)\Gamma(d-4)} \xi(2d-8) E_{\alpha_d,d-4}.
\end{align}
We see therefore that we get the same dependence on the infrared regulator $\mu$, and therefore our prescription in \eqref{Degenerate} is equivalent to including a soft mass to the propagator whenever it becomes massless. Note that the second term does not depend on the infrared regulator, and is an artefact of the specific truncation to infrared divergent contributions carried out in this computation. }

Because we have disregarded power law divergences, the degenerate orbit contribution vanishes in dimension $D>6$. In dimension $D$ lower than six, the degenerate orbit contributes a term that diverges as a power of the infrared regulator when the latter vanishes, and it does not contribute to the Wilsonian effective action. In $D<6$ one understands therefore that the degenerate orbit only contribute to the non-analytic part of the amplitude. 

It is important to note that the degenerate orbit contribution vanishes in $d=4$, because this exhibit that the cancelation of the ultraviolet divergence in the sum of \eqref{5D2-loopThreshold} and \eqref{5D2-loopDiv} indeed extends to the complete 2-loop amplitude.

\subsection{The $\nabla^6 R^4$ threshold function}

We will now consider the higher derivative corrections, understanding that the infrared divergent contributions have already been reabsorbed in the non-analytic component of the amplitude according to the discussion of the preceding section. In particular, we will find that the function 
\begin{align}
\cE_\gra{0}{1}^{\scriptscriptstyle \mbox{\tiny 2-loop}} =  \frac{2\pi^{5-d}}{9}\sum_{\substack{\Gamma_i \in \mathds{Z}^{2d(\alpha_d)}_* \\ \Gamma_i \times \Gamma_j = 0   }}  \int_{\mathds{R}_+^{\times 3}} d^3 \Omega \ \left(\det\Omega\right)^{\frac{d-7}{2}} \ \left( L_1 + L_2 + L_3- 5 \frac{L_1 L_2 L_3}{\det \Omega}\right) e^{- \Omega^{ij} g(\Gamma_i , \Gamma_j)}
\end{align}
satisfies an inhomogeneous differential equation as required by supersymmetry \cite{Green:2010kv,Bossard:2015uga}, and provides an explicit integral formula for this function, defining an alternative  formula to \cite{Pioline:2015yea} valid in all dimensions $D\ge  4$. The sum over the pair of charges is restricted to $\mathds{Z}^{2d(\alpha_d)}_*$ which excludes the trivial orbit $\Gamma_1=\Gamma_2=0$ but includes the degenerate orbit discussed in section~\ref{Nondege}.

\subsubsection{The Poisson equation}

As written out above, the two-loop contribution to the $\nabla^6 R^4$ threshold function is defined by the integral of 
\be 
\cF_\gra{0}{1}= \frac{2\pi^{5-d}}{9}  \sum_{\substack{\Gamma_i \in \mathds{Z}^{2d(\alpha_d)}_* \\ \Gamma_i \times \Gamma_j = 0   }} \det\Omega^{\frac{d-7}{2}} \Phi_\gra{0}{1}(\Omega)e^{- \Omega^{ij} g(\Gamma_i , \Gamma_j)} \ ,  
\ee
over three copies of the positive half real line, where $\Phi_\gra{0}{1}(\Omega)$ is the function defined in \eqref{Phis}, and satisfies
\be  
\frac{\partial\, }{\partial \Omega^{ij}} \Scal{2 \Omega^{ik} \Omega^{jl}  \frac{\partial\, }{\partial \Omega^{kl}} - \Omega^{ij}}  \Phi_\gra01(\Omega) = 12  \, \Phi_\gra01(\Omega) \ . 
\ee
This equation exhibits that  the function $\Phi_\gra{0}{1}(\Omega)$, pulled back to $SL(2)/SO(2)$, satisfies a Poisson equation, which implies that $\cE_\gra{0}{1}^{\scriptscriptstyle \mbox{\tiny 2-loop}} $ itself satisfies a Poisson equation as obtained in 9 dimensions in \cite{Green:1999pu}. One computes similarly as in section~\ref{sec:DE} that 
\bea \label{LapF01}
&& \Scal{ \Delta - \frac{6(5-d)(d+3)}{9-d}} \cF_\gra{0}{1} \\
&=&  \frac{\partial\, }{\partial \Omega^{ij}} \Scal{2 \Omega^{ik} \Omega^{jl}  \frac{\partial\, }{\partial \Omega^{kl}} + \frac{2}{9-d} \Omega^{ij} \Omega^{kl}  \frac{\partial\, }{\partial \Omega^{kl}} + \frac{(d-2)(d-3)}{9-d} \Omega^{ij}} \cF_\gra{0}{1}  \CR&& \  + \sum_{I=1}^3 \frac{\partial\,  }{\partial L_I} \biggl(  \frac{8\pi^{5-d}}{9}  \hspace{-2mm} \sum_{\substack{\Gamma_i \in \mathds{Z}^{2d(\alpha_d)}_* \\ \Gamma_i \times \Gamma_j = 0   }} {\rm det}\Omega^{\frac{d-7}{2}} \Scal{L_I^{\; 2} + 3 L_{I+1} L_{I+2} - L_I ( L_{I+1}+L_{I+2})}e^{- \Omega^{ij} g(\Gamma_i , \Gamma_j)} \biggr) \nn \ . 
\eea
The right-hand-side gives a boundary term in the Laplace equation for  $\cE_\gra{0}{1}^{\scriptscriptstyle \mbox{\tiny 2-loop}} $ that we are going to analyse now. For degenerate orbits one gets a potentially divergent boundary term at $L_I\rightarrow \infty$  in space-time dimensions $D$ lower than eight, but we assume here the use of an appropriate infrared regulator such that these contributions vanish. The lower boundary term at $L_I \rightarrow 0$ has contributions from the first and second line in~\eqref{LapF01} and they are equal for the three $L_I$. They result in an expression that is an integral over only two remaining Schwinger parameters and that factorizes such that the total boundary term is 
\bea \label{BoundaryD6} 
&&-4\pi^{5-d} \sum_{\substack{\Gamma_i \in \mathds{Z}_*^{2d(\alpha_d)} \\ \Gamma_i \times \Gamma_j = 0   }}  \int_0^\infty \frac{dL_1}{L_1^{\frac{5-d}{2}}}  e^{-L_1 g(\Gamma_1,\Gamma_1)}\int_0^\infty \frac{dL_2}{L_2^{\frac{5-d}{2}}}  e^{-L_2 g(\Gamma_2,\Gamma_2)}   \CR
&=& - 4\pi^{5-d} \sum_{\substack{\Gamma_1 \in \mathds{Z}_*^{d(\alpha_d)} \\ \Gamma_1 \times \Gamma_1 = 0   }}  \int_0^\infty \frac{dL_1}{L_1^{\frac{5-d}{2}}}  e^{-L_1 g(\Gamma_1,\Gamma_1)} \sum_{\substack{\Gamma_2 \in \mathds{Z}^{d(\alpha_d)}_* \\ \Gamma_i \times \Gamma_2 = 0   }}  \int_0^\infty \frac{dL_2}{L_2^{\frac{5-d}{2}}}  e^{-L_2 g(\Gamma_2,\Gamma_2)}   \CR
&& \quad  -  8\pi^{5-d} \sum_{\substack{\Gamma \in \mathds{Z}_*^{d(\alpha_d)} \\ \Gamma \times \Gamma = 0   }}  \int_0^\infty \frac{dL_1}{L_1^{\frac{5-d}{2}}}  e^{-L_1 g(\Gamma,\Gamma)}  \int_0^\infty \frac{dL_2}{L_2^{\frac{5-d}{2}}}  e^{-L_2 \mu^2}\  ,
\eea
where we have separated the degenerated orbit piece $\Gamma_i=0$ (for one of $\Gamma_1$ or $\Gamma_2$) in the last line and have regularised the corresponding expression by introducing a sliding scale $\mu$ in the second integral, such that 
\bea 
&& -  8\pi^{5-d} \sum_{\substack{\Gamma \in \mathds{Z}_*^{d(\alpha_d)} \\ \Gamma \times \Gamma = 0   }}  \int_0^\infty \frac{dL_1}{L_1^{\frac{5-d}{2}}}  e^{-L_1 g(\Gamma,\Gamma)}  \int_0^\infty \frac{dL_2}{L_2^{\frac{5-d}{2}}}  e^{-L_2 \mu^2} \CR
&=&   -16  \pi^{\frac{7-d}{2}} \Gamma(\tfrac{d-3}{2}) \mu^{3-d}  \xi(d-3) E_{\alpha_d,\frac{d-3}{2}} \ .  
\eea
The sliding scale defines the splitting of the amplitude into its analytic and non-analytic components. This term can always be reabsorbed in a redefinition of the function $ \cE_\gra{0}{1}^{\scriptscriptstyle \mbox{\tiny 2-loop}} $ by adding a term proportional to $ E_{\alpha_d,\frac{d-3}{2}}$ that will be considered to be part of the non-analytic component of the amplitude, except for $d=6$. In $D=5$ dimensions the Laplace  eigenvalue of $ \cE_\gra{0}{1}^{\scriptscriptstyle \mbox{\tiny 2-loop}} $ and the minimal representation series are the same, but one can still reabsorb the infrared divergent source term in the expression analytically continued to $d=6+2\epsilon$. In eight dimensions this term contributes to a logarithmic infrared divergence, and one cannot disentangle unambiguously the analytic and the non-analytic components of the amplitude. We will discuss this case separately in section~\ref{sec:2loops8}. We define therefore the infrared regularised 2-loop threshold function as
\begin{multline}
 \hat{{\cE}}_\gra{0}{1}^{\mbox{\tiny 2-loop}} = \frac{2\pi^{5-d}}{9}  \sum_{\substack{\Gamma_i \in \mathds{Z}^{2d(\alpha_d)}_* \\ \Gamma_i \times \Gamma_j = 0   }} \int_{\mathds{R}_+^{\times 3}} \frac{d^3 \Omega}{{\rm det}\Omega^{\frac{7-d}{2}}} \Phi_\gra{0}{1}(\Omega)e^{- \Omega^{ij} g(\Gamma_i+\mu^2 , \Gamma_j+\mu^2)} \\ - \frac{9-d}{3(6-d)(5+d)}\, 4\pi^{\frac{5-d}{2}} \Gamma(\tfrac{d-3}{2}) \mu^{3-d}  \scal{ 4\pi  \xi(d-3) E_{\alpha_d,\frac{d-3}{2}}} \ ,  
\end{multline}
that should be understood as a meromorphic function of $d\rightarrow d+2\epsilon$ evaluated at $\epsilon =0$. 

\allowDbreak{We now analyse the first term in \eqref{BoundaryD6} for $d>3$ using the decomposition \eqref{1ChargeGrad} of ${\bf R}_{\alpha_d}$ and the constraints \eqref{Sol2charges}
\begin{align}
&\quad 4\pi^{5-d} \sum_{\substack{\Gamma_1 \in \mathds{Z}_*^{d(\alpha_d)} \\ \Gamma_1 \times \Gamma_1 = 0   }}  \int_0^\infty \frac{dL_1}{L_1^{\frac{5-d}{2}}}  e^{-L_1 g(\Gamma_1,\Gamma_1)} \sum_{\substack{\Gamma_2 \in \mathds{Z}^{d(\alpha_d)}_* \\ \Gamma_i \times \Gamma_2 = 0   }}  \int_0^\infty \frac{dL_2}{L_2^{\frac{5-d}{2}}}  e^{-L_2 g(\Gamma_2,\Gamma_2)}   &\nn\\
&= 4 \pi^{2}  \hspace{-2mm} \sum_{\gamma \in P_d \backslash E_{d(d)}} \gamma\Biggl[ 2  \xi(d-3) r^{\frac{(10-d)(d-3)}{9-d}}   \biggl(2 \xi(d-3) r^{\frac{(10-d)(d-3)}{9-d} } \biggr. \Biggr. &\nn\\
&  \hspace{40mm} \Biggl.\biggl .   + \sum_{\substack{\Gamma \in \mathds{Z}_*^{d(\alpha_{d-1}^{d-1})} \\ \Gamma \times \Gamma = 0}} \sum_{n\in \mathds{Z}} \int_0^\infty \frac{dt}{t^{\frac{d-1}{2}}}  e^{-\frac{\pi}{t} \scal{ r^{2\frac{10-d}{d-9}} ( n + \langle \Gamma , a\rangle )^2  + r^{\frac{2}{d-9}} |Z(\Gamma)|^2 }}    \biggr)         \Biggr] &\nn\\
&= 4 \pi^{2}  \hspace{-2mm} \sum_{\gamma \in P_d \backslash E_{d(d)}} \gamma\Biggl[ 2  \xi(d-3) r^{\frac{(10-d)(d-3)}{9-d}}  \biggl(2 \xi(d-3) r^{\frac{(10-d)(d-3)}{9-d}} + 2\xi(d-4) r^{\frac{6}{9-d}} E_{\alpha_{d-1}^{d-1},\frac{d-4}{2}} \biggr. \Biggr. &\nn\\
&  \hspace{40mm} \Biggl.\biggl .   + \, 4 \hspace{-5mm}\sum_{\substack{\Gamma \in \mathds{Z}_*^{d(\alpha_{d-1}^{d-1})} \\ \Gamma \times \Gamma = 0}} \Bigl( \sum_{n|{\rm gcd}(\Gamma)} n^{d-4} \Bigr) \frac{r^{\frac{6}{9-d} + \frac{d-4}{2}}}{|Z(\Gamma)|^{\frac{d-4}{2}}} K_{\frac{d-4}{2}}( 2\pi r|Z(\Gamma)|) e^{2\pi i \langle \Gamma, a\rangle }   \biggr)         \Biggr] &\nn\\
&= 4 \pi^{2}  \hspace{-2mm} \sum_{\gamma \in P_d \backslash E_{d(d)}} \gamma\Biggl[ 2  \xi(d-3) r^{\frac{(10-d)(d-3)}{9-d}}  \, 2\xi(d-3) E_{\alpha_d,\frac{d-3}{2}} \Biggr] &\nn\\
\label{BT1}
&= \Scal{ 4 \pi \xi(d-3) E_{\alpha_d,\frac{d-3}{2}}}^2 \  . &
\end{align}
Let us explain the various steps in this calculation. First, we have written out the sum over the pair of charges using~\eqref{Sol2charges} up to $E_{d(d)}(\mathds{Z})$ transformations modulo the stabiliser of the representative charge $\Gamma_1\in {\bf 1}^\ord{10-d}$ and the two terms in the parenthesis correspond to vanishing and non-vanishing component of $\Gamma_2\in \overline{\bf R}_{\alpha_{d-1}}^\ord{1}$. The next step is the Poisson resummation of $n$ together with a Bessel integration. The resulting terms in the parenthesis are then recognised as the minimal automorphic series. The property that it is enough to sum over the positive degree components of the lattice in the parabolic decomposition associated to $\alpha_d$ to get an automorphic function is a distinguishing feature of the minimal unitary representation Eisenstein series. One can indeed check that it is the case for $d=5$ using \cite{Kazhdan:2001nx}. More generally one proves this identity using Langlands constant term formula, and the explicit form of the Whittaker vector given in  \cite{Fleig:2013psa}. We exhibit in appendix \ref{MinimalOrtho}  through an explicit computation that this property holds in particular for orthogonal groups. Note that this proof requires the Fourier decomposition be abelian to match all Fourier modes to simple root representatives determined by Whittaker vectors. Therefore the before to last line in \eqref{BT1} is not correct for $d=8$,  since in this case the minimal series also admit non-abelian Fourier coefficients with a non-zero Kaluza--Klein monopole charge, which are not included in the third line of \eqref{BT1}.  Supersymmetry nevertheless implies the differential equation to be satisfied, so for the exceptional field theory amplitude to be consistent with supersymmetry in this case, one would require the last line to remain correct for $d=8$. 
}

As
\begin{align}
\label{BT2}
\Scal{ 4 \pi \xi(d-3) E_{\alpha_d,\frac{d-3}{2}}}^2 = 4\pi^{5-d} \sum_{\substack{\Gamma_1 \in \mathds{Z}_*^{d(\alpha_d)} \\ \Gamma_1 \times \Gamma_1 = 0   }}  \int_0^\infty \frac{dL_1}{L_1^{\frac{5-d}{2}}}  e^{-L_1 g(\Gamma_1,\Gamma_1)} 
\times \sum_{\substack{\Gamma_2 \in \mathds{Z}^{d(\alpha_d)}_* \\ \Gamma_2 \times \Gamma_2 = 0   }}  \int_0^\infty \frac{dL_2}{L_2^{\frac{5-d}{2}}}  e^{-L_2 g(\Gamma_2,\Gamma_2)}  ,
\end{align}
the equality~\eqref{BT1} means that for the specific powers of $L_1$ and $L_2$ appearing in $\nabla^6R^4$, the contribution from $\Gamma_2$ such $\Gamma_1\times \Gamma_2=0$ is not satisfied in~\eqref{BT1} vanishes. This is only expected to be true in the minimal representation.

\subsubsection{$\hat{{\cal E}}^{\mbox{\tiny 2-loop}}_{\gra{0}{1}}$ in eight dimensions}
\label{sec:2loops8} 

Because the above series is not absolutely convergent, this identity is only true for the analytic continuation of the series as a function of $d$, evaluated at the specific value. Moreover, one must therefore be careful in eight dimensions because the series has a pole at $d=3$. 
 
\allowDbreak{In eight dimensions  
\be 
r^{\frac{(10-d)}{9-d}}  = r^\frac{7}{6} = e^{-\phi_2+2 \phi_3} 
\ee
with $a + i e^{-2\phi_2}$ parametrizing $SL(2)/SO(2)$ and $e^{2\phi_3}$ the scalar associated to the first root  in $SL(3)$, and one computes for $d-3 = 2\epsilon$ 
\begin{align}
&\quad  4\pi^{2-2 \epsilon} \sum_{\substack{\Gamma_1 \in \mathds{Z}_*^{2\times 3} \\ \Gamma_1 \times \Gamma_1 = 0   }}  \int_0^\infty \frac{dL_1}{L_1^{1-\epsilon}}  e^{-L_1 g(\Gamma_1,\Gamma_1)} \sum_{\substack{\Gamma_2 \in \mathds{Z}^{d(\alpha_d)}_* \\ \Gamma_i \times \Gamma_2 = 0   }}  \int_0^\infty \frac{dL_2}{L_2^{1-\epsilon}}  e^{-L_2 g(\Gamma_2,\Gamma_2)}   &\nn\\
&= 4 \pi^{2}  \hspace{-2mm} \sum_{\gamma \in P_3 \backslash E_{3(3)}} \gamma\Biggl[ 2  \xi(2\epsilon ) e^{-2\epsilon \phi_2 + 4 \epsilon \phi_3}   2 \xi(2\epsilon)  \biggl(e^{-2\epsilon \phi_2 } E_{[0,\epsilon]} + e^{ 4 \epsilon \phi_3} E_{[\epsilon]}    -e^{-2\epsilon \phi_2 + 4 \epsilon \phi_3}      \biggr)         \Biggr] &\nn\\
&= 4 \pi^{2} \scal{ 2  \xi(2\epsilon )}^2   \biggl(  E_{[2\epsilon]}  (E_{[0,\epsilon]} )^2 + (E_{[\epsilon]})^2 E_{[0,2\epsilon]} - E_{[2\epsilon]} E_{[0,2\epsilon]} \biggr)   &\nn\\
&= \Scal{ 4 \pi \xi(2\epsilon) E_{[\epsilon]} E_{[0,\epsilon]} }^2  + \mathcal{O}(\epsilon) \  . &
\end{align}
The sum of the two terms in \eqref{BoundaryD6} finally gives 
\bea 
&& -\Scal{ 4\pi \xi(2\epsilon) E_{[\epsilon]} E_{[0,\epsilon]} }^2   -16  \pi^{2-\epsilon} \Gamma(\epsilon) \mu^{-2\epsilon}  \xi(2\epsilon) E_{[\epsilon]} E_{[0,\epsilon]}  \CR
&=& 4\pi^2 \Scal{ \frac{\Gamma(\epsilon)}{(\pi \mu^2)^\epsilon}  }^2  - \scal{ 4\zeta(2) \hat{E}_{[1]} + 2\zeta(3) \hat{E}_{[\frac{3}{2},0]} +2\pi(3 \gamma-2- \log( 4\pi \mu^2))}^2  \ ,
\eea
where the first term must be reabsorbed in the threshold function to regularise it, 
\be \hat{{\cal E}}^{\mbox{\tiny 2-loop}}_{\gra{0}{1}} = 
\lim_{\epsilon\rightarrow 0}\Biggl(  \frac{2\pi^{2}}{9\pi^{2\epsilon}} \hspace{-2mm} \sum_{\substack{\Gamma_i \in \mathds{Z}^{2\times 3}_* \\ \Gamma_i \times \Gamma_j = 0   }}\int \frac{d^3 \Omega}{ {\rm det}\Omega^{1-\epsilon}} \Phi_\gra{0}{1}(\Omega)e^{- \Omega^{ij} g(\Gamma_i + \mu^2 , \Gamma_j+\mu^2)} 
  - \frac{\pi^2}{3} \frac{3-4\epsilon}{3-5\epsilon} \Scal{ \frac{\Gamma(\epsilon)}{(\pi \mu^2)^\epsilon}  }^2 \Biggr) \ ,  
\ee
whereas the second defines the regularised source term, which is precisely the 1-loop amplitude square \eqref{1loop8D} such that 
\be 
(\Delta-12) \hat{{\cal E}}^{\mbox{\tiny 2-loop}}_{\gra{0}{1}}  = - \scal{ \ell^{-6} A(0,0,0,0)^{\scriptscriptstyle \mbox{\tiny 1-loop}} }^2\ . 
\ee}

\subsubsection{Sliding scale independent formula}

Coming back to the general case, it would be more satisfying to have a definition of the function $ \cE_\gra{0}{1}^{\scriptscriptstyle \mbox{\tiny 2-loop}} $ independent of the sliding scale $\mu$. We will therefore consider the infrared safe sum over non-zero charges of non-vanishing sum, understanding that the remaining contribution must be included in the non-analytic component of the amplitude as pointed out in section \ref{sec:IRdeg}. One obtains then the boundary term
\bea
&& - 4\pi^{5-d} \sum_{\substack{\Gamma_1 \in \mathds{Z}_*^{d(\alpha_d)} \\ \Gamma_1 \times \Gamma_1 = 0   }}  \int_0^\infty \frac{dL_1}{L_1^{\frac{5-d}{2}}}  e^{-L_1 g(\Gamma_1,\Gamma_1)} \sum_{\substack{\Gamma_2 \in \mathds{Z}^{d(\alpha_d)}_* \\ \Gamma_i \times \Gamma_2 = 0   }}  \int_0^\infty \frac{dL_2}{L_2^{\frac{5-d}{2}}}  e^{-L_2 g(\Gamma_2,\Gamma_2)}   \CR
&& \hspace{20mm}  + 4 \pi^{5-d} \sum_{\substack{\Gamma  \in \mathds{Z}_*^{d(\alpha_d)} \\ \Gamma \times \Gamma = 0   }}  \int_0^\infty \frac{dL_1}{L_1^{\frac{5-d}{2}}}  \int_0^\infty \frac{dL_2}{L_2^{\frac{5-d}{2}}}  e^{-(L_1+L_2) g(\Gamma,\Gamma)} \CR
&=& -\Scal{ 4\pi \xi(d-3) E_{\alpha_d,\frac{d-3}{2}}}^2  + \frac{2^{7-d} \pi^{\frac{5}{2}} \Gamma(\tfrac{d-3}{2}) }{ \Gamma(\tfrac{d-2}{2})} \xi(2d-6)   E_{\alpha_d,d-3} \ ,  
\eea
where the second term removes the sum over opposite charges.  The latter does not appear in the infrared regularised amplitude, and must therefore be reabsorbed in a redefinition of the threshold function, such that there is no such boundary term. This is always possible, provided it is not an eigenfunction of the Laplace operator with the same eigenvalue, \ie unless 
\be 
\frac{(3-d)(30-7d+d^2)}{9-d} - \frac{6(5-d)(d+3)}{9-d} = d(d-7) = 0  \ . 
\ee
This signals that the sum actually diverges in four dimensions, and the logarithmic term gives rise to such a right-hand-side. However, the correct coefficient is not reproduced by this formula, because the series itself diverges in that case, such that one cannot safely assume that the sum and the integral commute. Nevertheless, the formul\ae{}  derived by assuming that the sum and the integral commute should be valid, up to anomalous corrections associated to logarithm terms in the threshold function, which only correct the differential equations by anomalous linear sources proportional to Eisenstein series. Taking into account the possible anomalous corrections,  we conclude that an appropriate infrared regularisation should permit to define the threshold function such that \cite{Green:2010kv,Pioline:2015yea,Bossard:2015oxa}
\bea  
\Delta  \hat{\cE}_\gra{0}{1}^{\mbox{\tiny 2-loop}} &=&  \frac{6(5-d)(d+3)}{9-d} \hat{\cE}_\gra{0}{1}^{\mbox{\tiny 2-loop}} - \Scal{ 4\pi \xi(d-3) E_{\alpha_d,\frac{d-3}{2}}}^2  \CR
&&  + \gamma_1 \delta_{d,7} \frac{4 \pi^3}{45} \xi(d+1) E_{\alpha_d,\frac{d+1}{2}} + \gamma_2 \delta_{d,6} 4\pi  \xi(d-3)    E_{\alpha_d,\frac{d-3}{2}} + \gamma_3 \delta_{d,5}  \ . \label{PoissonD6R4} 
\eea
We will not derive the precise coefficients here (they can be found in \cite{Pioline:2015yea,Bossard:2015oxa}), which would require a careful analysis of these integrals and the analytic continuation of the divergent series to define the consistent split between the analytic and the non-analytic component of the amplitude. From the point of view of string theory, the source terms arise from boundary degeneration contributions when integrating over the moduli space of Riemann surfaces~\cite{Basu:2015dqa,Pioline:2015nfa}.

\subsubsection{Tensorial equation}

Now we would like to consider the tensorial equation satisfied by the threshold function in order to disentangle the inhomogeneous solution to \eqref{PoissonD6R4} from the homogeneous solution defined by the Eisenstein function we obtained at $1$-loop. The two functions satisfy two distinct tensorial equations \cite{Bossard:2015uga}, and in dimensions $D=4$, $D=5$ and $D=6$, the tensorial equation satisfied by the inhomogeneous solution is
\be 
\Scal{ {\bf D}_{\alpha_d}^{\; 3} - \frac{3}{2} \frac{2+ d(5-d)}{9-d} {\bf D}_{\alpha_d}} {\cE}_\gra{0}{1}^{\mbox{\tiny 2-loop}} = - \frac{1}{4}   {\bf D}_{\alpha_d}\Scal{ 4\pi \xi(d-3) E_{\alpha_d,\frac{d-3}{2}}}^2  \ ,  \label{D3E} 
\ee
where we have neglected the anomalous terms for simplicity. In order to  check this equation on the integrand function in the form of a sum over $P_{d-1}(\mathds{Z})\backslash E_{d(d)}(\mathds{Z})$ as in section~\ref{sec:DE}, it is convenient to decompose the covariant derivative in a parabolic gauge associated with the parabolic subgroup $P_{d-1}$. On a function that only depends on the subspace  $\mathds{R}_+^*\times SL(2) /SO(2)\times E_{d-2(d-2)}/K_{d-2}$ of the Levi factor, one obtains straightforwardly that the restriction of the covariant derivative   ${\bf D}_{\alpha_d}$ to the highest weight component ${\bf 2}^\ord{11-d}$ in \eqref{alphadm1}, reduces to 
\be
{\bf D}_{\alpha_d}\Big|_{{\bf 2}^\ord{11-d}} = \left( \begin{array}{cc}\  \frac{1}{4(9-d)} \partial_\phi  +  \frac{1}{2 u_2} \partial_{u_2}  \ & \  \frac{1}{2 u_2} \partial_{u_1} \ \\ \  \frac{1}{2 u_2} \partial_{u_1} \  &  \frac{1}{4(9-d)} \partial_\phi  -    \frac{1}{2 u_2} \partial_{u_2}  \ \end{array}\right)\ . 
\ee
To compute the third order differential operator $ {\bf D}_{\alpha_1}^{\; 3}$ restricted to the same component, we use the known constant terms associated to the parabolic subgroup $P_{d-2}$ of the Eisenstein series that are solutions to the homogeneous equation obtained from \eqref{D3E} by setting the source to zero. They are
\bea 
{\bf D}_{56}^{\;3}   E_{\mbox{\DEVII{\mathnormal{s}}00000{\mathfrak{0}}}} &=& \Scal{ \frac{s(2s-17)}{2}+6}   {\bf D}_{56} E_{\mbox{\DEVII{\mathnormal{s}}00000{\mathfrak{0}}}}  \ ,  \CR 
 {\bf D}_{27}^{\; 3}  E_{\text{\DEVI0{\mathnormal{s}}000{\mathfrak{0}}}} &=&  \frac{1}{2} (s-5) (2 s-1)     {\bf D}_{27} E_{\text{\DEVI0{\mathnormal{s}}000{\mathfrak{0}}}} \ ,  \CR
  {\bf D}_{16}^{\; 3} E_{\mbox{\DSOX{0}{\mathnormal{s}}000}}&=&  \frac{2s(2s-7)+3}{4} {\bf D}_{16} E_{\mbox{\DSOX{0}{\mathnormal{s}}000}} \ . 
\eea
This permits to determine (where we write the 2 by 2 matrix in terms of Pauli matrices $\sigma_i$)
\bea  
{\bf D}^{\; 3}_{56}\Big|_{{\bf 2}^\ord{4}} &=&  \mathds{1}_{2} \Scal{  \frac{1}{8^3} \partial^{\; 3}_\phi +  \frac{5}{32} \partial^{\; 2}_\phi+  \frac{3}{2} \partial_\phi   + \scal{   \tfrac{3}{32} \partial_\phi +2} \Delta_{A_1}  - \frac{3}{4} \Delta_{D_5}   }   \CR
&& \qquad + \Scal{\sigma_3   \frac{1}{2 u_2} \partial_{u_2}  +  \sigma_1 \frac{1}{2 u_2} \partial_{u_1} } \Scal{\frac{3}{8^2} \partial^{\; 2}_\phi +  \frac{43}{16} \partial_\phi  +   26 + \frac{1}{4}  \Delta_{A_1} } , \CR
 {\bf D}^{\; 3}_{27}\Big|_{{\bf 2}^\ord{5}} &=&  \mathds{1}_{2} \Scal{  \frac{1}{12^3} \partial^{\; 3}_\phi +  \frac{7}{160} \partial^{\; 2}_\phi+  \frac{3}{8} \partial_\phi  + \frac{1}{4} \scal{   \tfrac{1}{4} \partial_\phi +5} \Delta_{A_1}  - \frac{1}{2} \Delta_{A_4}  }   \CR
&& \qquad + \Scal{\sigma_3   \frac{1}{2 u_2} \partial_{u_2}  +  \sigma_1 \frac{1}{2 u_2} \partial_{u_1} } \Scal{\frac{1}{48} \partial^{\; 2}_\phi +  \frac{9}{8} \partial_\phi  +   10 + \frac{1}{4}  \Delta_{A_1} } , \CR
  {\bf D}^{\; 3}_{16}\Big|_{{\bf 2}^\ord{6}} &=&  \mathds{1}_{2} \Scal{  \frac{1}{16^3} \partial^{\; 3}_\phi +  \frac{1}{64} \partial^{\; 2}_\phi+  \frac{3}{32} \partial_\phi  +       \frac{3}{4}\scal{   \tfrac{1}{16} \partial_\phi +1} \Delta_{A_1}  - \frac{3}{8} \scal{ \Delta_{A_1^\prime} + \Delta_{A_2} }}   \CR
&& \qquad + \Scal{\sigma_3   \frac{1}{2 u_2} \partial_{u_2}  +  \sigma_1 \frac{1}{2 u_2} \partial_{u_1} } \Scal{\frac{3}{16^2} \partial^{\; 2}_\phi +  \frac{17}{32} \partial_\phi  +   \frac{15}{4} + \frac{1}{4}  \Delta_{A_1} } .
\eea
To determine how much the component ${{\bf 2}^\ord{11-d}}$ of the differential equation \eqref{D3E} constrains the function, it is important to consider the general homogeneous solutions to this differential equation
\be \label{HomoD6} {\bf D}_{\alpha_d}^{\; 3}\Big|_{{\bf 2}^\ord{11-d}} {\cE}_d  =  \frac{3}{2} \frac{2+ d(5-d)}{9-d} {\bf D}_{\alpha_d}\Big|_{{\bf 2}^\ord{11-d}}  {\cE}_d  \ . 
\ee 
Taking the expansion of the adjoint Eisenstein series at $s=6,\, \frac{9}{2},\, \frac{7}{2}$ for $d=7,\, 6,\, 5$, one finds a term in $e^{-2(11-d)(d-3)\phi} E_{[4]}(u,\bar u)$, which always satisfies \eqref{HomoD6} by construction. One can argue that equation \eqref{HomoD6} together with the Laplace equation \eqref{PoissonD6R4} is enough to determine uniquely the independent modular invariant solutions as being either  $e^{-2(11-d)(d-3)\phi} E_{[4]}(u,\bar u)$ or the series $1,\, e^{-15\phi} E_{[\stfrac{3}{2}]}(u,\bar u)$ and $e^{-24\phi} E_{[1]}(u,\bar u)$ for $d=5,6$ and $7$, respectively, that are solutions corresponding to the  anomalous source term in $1,\, E{\mbox{\DEVI00000{\mbox{$\frac{3}{2}$}}}}$ and  $E{\mbox{\DEVII000000{4}}}$. We conclude that this component of the differential equation is strong enough to disentangle the solutions to \eqref{D3E} from the one of the homogeneous differential equation satisfied by $\hat{E}_{\alpha_d,\frac{d+3}{2}}$ for $5\le d\le7$. 

The equation linear in the $SL(2)/SO(2)$ covariant derivative is more complicated to implement. Therefore we would like to only compute the component of the  third order equation proportional to the identity. We note indeed that the latter, together with the Laplace equation, is enough to determine the correct solution, up to a solution of  type $e^{-a\phi} E_{[ \stfrac{1}{2} \pm i r ]}(u,\bar u)$ for some specific positive number  $r$ (\ie $r= \frac{3}{2} \sqrt{151}$, $r=\frac{\sqrt{5015}}{10}$ and $r=\frac{\sqrt{119}}{6}$ for $d=7,\, 6$ and $5$, respectively). The homogeneous solution Eisenstein series we want to disregard involve instead as a solution $e^{-(11-d)(d+3)\phi} E_{[\stfrac{d+3}{2}]}(u,\bar u)$, which is ruled out by the singlet component of \eqref{HomoD6}. It is therefore enough to check the singlet third order equation to identify the complete third order differential equation satisfied by the function $ {\cE}_\gra{0}{1}^{\mbox{\tiny 2-loop}}$.\footnote{Even though the Eisenstein series  $\hat{E}_{\alpha_d,\frac{d+3}{2}}$  is indeed ruled out by this differential equation, we note that the general solution we obtain does not permit to distinguish the general homogeneous solution to the inhomogeneous equation in $d=7$, and that all solutions depending only on $\phi$ and $SL(2)/SO(2)$ then belong to terms that already appear in the $\nabla^4 R^4$ threshold function, \ie $E{\mbox{\DEVII{\mbox{$\frac{5}{2}$}}00000{\mathfrak{0}}}} $ in this case. This does not alter the conclusion that the series  $\hat{E}{\mbox{\DEVII000000{{5}}}} $ is ruled out.}

For $E_{7(7)}$, one computes that\footnote{\label{fn:56}The notation here is such that the components of the 56-dimensional vector $\{ ( {\bf D}_{56}^{\; 3} +9  {\bf D}_{56})_{i}|_{i=1}^{56}\}$ are ordered in such a way that the singlet component in the ${\bf 2}^\ord{4}$ of~\eqref{alphadm1} is the first entry $ ( {\bf D}_{56}^{\; 3} +9  {\bf D}_{56})_{1}$}
\bea 
&&  \Scal{ {\bf D}_{56}^{\; 3} +9  {\bf D}_{56}} {\cF}_\gra{0}{1} \CR
\hspace{-5mm}&=&  \frac{2}{9\pi^2} \hspace{-1mm} \sum_{\gamma \in P_6\backslash E_{7(7)}}\hspace{-3mm} \Phi_\gra{0}{1}(\Omega)\hspace{-2mm}  \sum_{M\in \mathds{Z}^{2\times 2}_*} \hspace{-0mm}\biggl\{   \tfrac{1}{512} \partial^{\; 3}_\phi +  \tfrac{5}{32} \partial^{\; 2}_\phi+  \tfrac{21}{8} \partial_\phi   + \scal{   \tfrac{3}{32} \partial_\phi +2} \Delta_u ,\dots \biggr\}  e^{-\Omega^{ij}  e^{8\phi} \langle M_i , M_j \rangle_{u} }\CR
&=& \biggl\{   \frac{\partial\, }{\partial \Omega^{pq}} \Scal{  \Omega^{pq}   \frac{\partial\, }{\partial \Omega^{ij}} \Scal{\frac{3}{2} \Omega^{ik} \Omega^{jl}  \frac{\partial\, }{\partial \Omega^{kl}} + \frac{1}{4} \Omega^{ij} \Omega^{kl}  \frac{\partial\, }{\partial \Omega^{kl}} - \frac{3}{2} \Omega^{ij}}-  \Omega^{pq}  + 2 \varepsilon^{pi} \varepsilon^{qj} \det(\Omega)   \frac{\partial\, }{\partial \Omega^{ij}}} \cF_\gra{1}{0} \biggr . \CR
&& +3 \sum_{I=1}^3 \frac{\partial\, }{\partial L_I}  \Scal{ \frac{ L_I \scal{ -2 L_I^{\; 2} ( L_{I+1}^{\; 2} + L_{I+2}^{\; 2}) +L_1 L_2 L_3 ( L_1\hspace{-0.5mm}+\hspace{-0.5mm}L_2 \hspace{-0.5mm}+\hspace{-0.5mm}L_3) +3  L_{I+1}^{\; 2}  L_{I+2}^{\; 2}}}{\det\Omega}  \frac{\partial\, }{\partial L_I}  \frac{\cF_\gra{1}{0}}{\Phi_\gra{1}{0}} } \CR 
&&\biggl .  +\frac{3}{2} \sum_{I\ne J\ne K}  \frac{\partial\, }{\partial L_{I}}  \Scal{ \frac{(L_I+L_J) \scal{\scriptstyle  - 2  L_I L_J( L_I L_J + 2 L_K^{\; 2}) + L_1 L_2 L_3 (L_I +L_J)  + 3 L_K^{\; 2}(  L_I^{\; 2}  +L_J^{\; 2})}}{\det \Omega}  \frac{\partial\, }{\partial L_J}   \frac{\cF_\gra{1}{0}}{\Phi_\gra{1}{0}} } \CR 
&& - 2 \sum_{I=1}^3 \frac{\partial\, }{\partial L_I}   \Scal{ \frac{\scriptstyle - 3 L_{I+1}^{\; 2} L_{I+2}^{\; 2} + 3 ( L_{I+1}^{\; 2} -L_{I+1} L_{I+2}+L_{I+2}^{\; 2}) L_I^{\; 2} + ( L_{I+1} + L_{I+2}) L_I^{\; 3}  }{\det\Omega}   \frac{\cF_\gra{1}{0}}{\Phi_\gra{1}{0}} }\, , \, \dots  \biggl\}  
\eea
where 
\be  \langle M_i , M_j \rangle_{u}  = \frac{ (M_{i1} + u_1 M_{i2})(M_{j1} + u_1 M_{j2})}{u_2} + u_2 M_{i2} M_{j2} \ . 
\ee
The corresponding boundary term at $L_I=0$ gives the right-hand-side 
\bea
&&  -  \frac{1}{\pi^2}   \sum_{\substack{\gamma \in P_6 \backslash E_{7(7)}\\M\in \mathds{Z}^{2\times 2}_*}}  \int dL_1 dL_2  \biggl\{  \frac{1}{8}\partial_\phi  + 2\frac{ \partial \, }{\partial L_{1} } L_{1} +2\frac{ \partial \, }{\partial L_{2} }  L_{2}  , \dots \biggr\}  L_{1} L_{2}   e^{-  e^{8\phi} ( L_1 \langle M_1 , M_1 \rangle_{u} +L_2 \langle M_2 , M_2 \rangle_{u} )}\CR
&=& - {\bf D}_{56} \   \frac{1}{\pi^2} \sum_{\substack{\Gamma_i \in \mathds{Z}_*^{2\times 56} \\ \Gamma_i \times \Gamma_j = 0  \\\Gamma_1 \ne -\Gamma_2 \ne 0}}  \int_0^\infty dL_1 L_1   e^{-L_1 g(\Gamma_1,\Gamma_1)} \int_0^\infty dL_2L_2 e^{-L_2 g(\Gamma_2,\Gamma_2)}  \CR
&=& - \frac{1}{4} {\bf D}_{56}   \Scal{ 4\pi \xi(4) E_{\mbox{\DEVII000000{2}}}}^2  +  {\bf D}_{56}   \Scal{\frac{ \pi^2}{3} \xi(8) E_{\mbox{\DEVII000000{4}}}}  \ .  
 \eea
In these equalities we have only spelt out the first component of the ${\bf 56}$ according to footnote~\ref{fn:56} and left the remaining ones unspecified as $\dots$. This is justified as the covariant derivative is restricted to this first component. Indeed, only an $E_{7(7)}$ left invariant differential operator can safely be passed through the sum over the discrete parabolic coset. The total derivative terms in this equation are therefore uniquely determined to rearrange the integrand into a linear derivative in $\phi$, such as to reproduce the form of the covariant derivative in this component.\\

 For $E_{6(6)}$, one computes similarly that 
\bea 
&&  \Scal{ {\bf D}_{27}^{\; 3} +2  {\bf D}_{27}} {\cF}_\gra{0}{1} \CR
\hspace{-5mm}&=&  \frac{2}{9\pi} \hspace{-1mm} \sum_{\gamma \in P_5 \backslash E_{6(6)}}\hspace{-3mm} \Phi_\gra{0}{1}(\Omega)\hspace{-2mm}  \sum_{M\in \mathds{Z}^{2\times 2}_*} \biggl\{  \tfrac{1}{1728} \partial^{\; 3}_\phi +  \tfrac{7}{160} \partial^{\; 2}_\phi+  \tfrac{13}{24} \partial_\phi   +\tfrac{1}{4} \scal{   \tfrac{1}{4} \partial_\phi +5} \Delta_u ,\dots \biggr\}  e^{-\Omega^{ij}  e^{10\phi} \langle M_i , M_j \rangle_{u} }\CR
&=& \biggl\{ \frac{5}{4} \frac{\partial\, }{\partial \Omega^{pq}} \Scal{  \Omega^{pq}   \frac{\partial\, }{\partial \Omega^{ij}} \Scal{ \Omega^{ik} \Omega^{jl}  \frac{\partial\, }{\partial \Omega^{kl}} - \frac{1}{27} \Omega^{ij} \Omega^{kl}  \frac{\partial\, }{\partial \Omega^{kl}} - \frac{7}{9} \Omega^{ij}}+\frac{5}{6} \Omega^{pq}  + \varepsilon^{pi} \varepsilon^{qj} \det(\Omega)   \frac{\partial\, }{\partial \Omega^{ij}}} \cF_\gra{1}{0} \CR
&& \hspace{-0mm}+\frac{5}{2} \sum_{I=1}^3 \frac{\partial\, }{\partial L_I}  \Scal{ \frac{ L_I \scal{ \scriptstyle -2 L_I^{\; 2} ( L_{I+1}^{\; 2} + L_{I+2}^{\; 2}) +L_1 L_2 L_3 ( L_1+L_2  + L_3) +3  L_{I+1}^{\; 2}  L_{I+2}^{\; 2}}}{\det \Omega^\frac{3}{2}}  \frac{\partial\, }{\partial L_I}  \frac{\det \Omega^\frac{1}{2} \cF_\gra{1}{0}}{\Phi_\gra{1}{0}} } \CR 
&& +\frac{5}{4} \sum_{I\ne J\ne K}  \frac{\partial\, }{\partial L_{I}}  \Scal{ \frac{(L_I+L_J) \scal{\scriptstyle  - 2  L_I L_J( L_I L_J + 2 L_K^{\; 2}) + L_1 L_2 L_3 (L_I +L_J)  + 3 L_K^{\; 2}(  L_I^{\; 2}  +L_J^{\; 2})}}{\det \Omega^\frac{3}{2}}  \frac{\partial\, }{\partial L_J}   \frac{\det \Omega^\frac{1}{2} \cF_\gra{1}{0}}{\Phi_\gra{1}{0}} } \CR 
&& - \frac{5}{4} \sum_{I=1}^3 \frac{\partial\, }{\partial L_I}   \Scal{ \frac{\scriptstyle - 3 L_{I+1}^{\; 2} L_{I+2}^{\; 2} + 3 ( L_{I+1}^{\; 2} -L_{I+1} L_{I+2}+L_{I+2}^{\; 2}) L_I^{\; 2} + ( L_{I+1} + L_{I+2}) L_I^{\; 3}  }{\det\Omega }   \frac{\cF_\gra{1}{0}}{\Phi_\gra{1}{0}} } \, , \, \dots \biggr\} \ .  \eea
The corresponding boundary term at $L_I=0$ gives the right-hand-side 
\bea&&\hspace{-5mm}  -  \frac{1}{\pi}  \hspace{-2mm} \sum_{\substack{\gamma \in P_5 \backslash E_{6(6)}\\M\in \mathds{Z}^{2\times 2}_*}}   \int dL_1 dL_2  \biggl\{  \frac{1}{12}\partial_\phi  + \frac{5}{3}\frac{ \partial \, }{\partial L_{1} } L_{1} +\frac{5}{3} \frac{ \partial \, }{\partial L_{2} }  L_{2}  ,\dots \biggr\}  ( L_{1} L_{2} )^\frac{1}{2} e^{-  e^{10\phi} ( L_1 \langle M_1 , M_1 \rangle_{u} +L_2 \langle M_2 , M_2 \rangle_{u} )}\CR
&=& - {\bf D}_{27} \   \frac{1}{\pi} \sum_{\substack{\Gamma_i \in \mathds{Z}_*^{2\times 27} \\ \Gamma_i \times \Gamma_j = 0  \\\Gamma_1 \ne -\Gamma_2 \ne 0}}  \int_0^\infty dL_1 \sqrt{L_1}   e^{-L_1 g(\Gamma_1,\Gamma_1)} \int_0^\infty dL_2\sqrt{L_2} e^{-L_2 g(\Gamma_2,\Gamma_2)}  \CR
&=& - \frac{1}{4} {\bf D}_{27}   \Scal{ 4\pi \xi(3) E_{\mbox{\DEVI00000{\mbox{$\frac{3}{2}$}}}}}^2  +  {\bf D}_{27}   \Scal{\frac{ \pi^3}{4} \xi(6) E_{\mbox{\DEVI00000{3}}}}  \ .  
 \eea
 Note that the second term in the right-hand-side is an artefact of the restriction on the sum used for the infrared regularisation, and should be reabsorbed in the threshold function.\\
 
\allowDbreak{For $SO(5,5)$, one computes that 
\begin{align}
&\quad  \Scal{ {\bf D}_{16}^{\; 3} - \frac{3}{4} {\bf D}_{16}} {\cF}_\gra{0}{1} &\nn\\
\hspace{-5mm}&=  \frac{2}{9} \hspace{-3mm} \sum_{\gamma \in P_3 \backslash SO(5,5)}\hspace{-3mm} \Phi_\gra{0}{1}(\Omega)\hspace{-2mm}  \sum_{M\in \mathds{Z}^{2\times 2}_*} \biggl\{   \tfrac{1}{4096} \partial^{\; 3}_\phi +  \tfrac{1}{64} \partial^{\; 2}_\phi+  \tfrac{3}{64} \partial_\phi   +\tfrac{3}{4} \scal{   \tfrac{1}{32} \partial_\phi +1} \Delta_u ,\dots \biggr\} e^{-\Omega^{ij}  e^{12\phi} \langle M_i , M_j \rangle_{u} }&\nn\\
&=\biggl\{  \frac{3}{8} \frac{\partial\, }{\partial \Omega^{pq}} \Scal{  \Omega^{pq}   \frac{\partial\, }{\partial \Omega^{ij}} \Scal{3 \Omega^{ik} \Omega^{jl}  \frac{\partial\, }{\partial \Omega^{kl}} - \frac{3}{8} \Omega^{ij} \Omega^{kl}  \frac{\partial\, }{\partial \Omega^{kl}} - \frac{11}{8} \Omega^{ij}}+\Omega^{pq}  +2 \varepsilon^{pi} \varepsilon^{qj} \det(\Omega)   \frac{\partial\, }{\partial \Omega^{ij}}} \cF_\gra{1}{0} &\nn\\
&\quad +\frac{3}{4} \sum_{I=1}^3 \frac{\partial\, }{\partial L_I}  \Scal{ \frac{ L_I \scal{ \scriptstyle -2 L_I^{\; 2} ( L_{I+1}^{\; 2} + L_{I+2}^{\; 2}) +L_1 L_2 L_3 ( L_1+L_2 +L_3) +3  L_{I+1}^{\; 2}  L_{I+2}^{\; 2}}}{\det\Omega^2}  \frac{\partial\, }{\partial L_I}  \frac{\det\Omega\cF_\gra{1}{0}}{\Phi_\gra{1}{0}} } &\nn\\
&\quad +\frac{3}{8} \sum_{I\ne J\ne K}  \frac{\partial\, }{\partial L_{I}}  \Scal{ \frac{(L_I+L_J) \scal{\scriptstyle  - 2  L_I L_J( L_I L_J + 2 L_K^{\; 2}) + L_1 L_2 L_3 (L_I +L_J)  + 3 L_K^{\; 2}(  L_I^{\; 2}  +L_J^{\; 2})}}{\det \Omega^2}  \frac{\partial\, }{\partial L_J}   \frac{\det\Omega \cF_\gra{1}{0}}{\Phi_\gra{1}{0}} } &\nn\\
&\quad - \frac{3}{4} \sum_{I=1}^3 \frac{\partial\, }{\partial L_I}   \Scal{ \frac{\scriptstyle - 3 L_{I+1}^{\; 2} L_{I+2}^{\; 2} + 3 ( L_{I+1}^{\; 2} -L_{I+1} L_{I+2}+L_{I+2}^{\; 2}) L_I^{\; 2} + ( L_{I+1} + L_{I+2}) L_I^{\; 3}  }{\det\Omega}   \frac{\cF_\gra{1}{0}}{\Phi_\gra{1}{0}} }\, , \, \dots \biggr\}  \ .  &
\end{align}
The corresponding boundary term at $L_I=0$ gives the right-hand-side 
\bea
&&-\hspace{-4mm} \sum_{\substack{\gamma \in P_3\backslash SO(5,5)\\M\in \mathds{Z}^{2\times 2}_*}}  \int dL_1 dL_2 \biggl\{ \frac{1}{16}\partial_\phi  + \frac{3}{2}\frac{ \partial \, }{\partial L_{1} } L_{1} +\frac{3}{2} \frac{ \partial \, }{\partial L_{2} }  L_{2}  , \dots \biggr\}    e^{-  e^{12\phi} ( L_1 \langle M_1 , M_1 \rangle_{u} +L_2 \langle M_2 , M_2 \rangle_{u} )}\CR
&=& - {\bf D}_{16} \   \frac{1}{\pi} \sum_{\substack{\Gamma_i \in \mathds{Z}_*^{2\times 16} \\ \Gamma_i \times \Gamma_j = 0  \\\Gamma_1 \ne -\Gamma_2 \ne 0}}  \int_0^\infty dL_1   e^{-L_1 g(\Gamma_1,\Gamma_1)} \int_0^\infty dL_2 e^{-L_2 g(\Gamma_2,\Gamma_2)}  \CR
&=& - \frac{1}{4} {\bf D}_{16}   \Scal{ 4\pi \xi(2) E_{\mbox{\DSOX0000{\mbox{$1$}}}} }^2  +  {\bf D}_{16}   \Scal{2 \zeta(4) E_{\mbox{\DSOX0000{\mbox{$2$}}}}  }  \ .  
 \eea}

Up to the terms due to the infrared regularisation, we see therefore that we have indeed verified formula  \eqref{D3E}. This exhibits that the homogeneous solution Eisenstein series appearing at 1-loop does not appear in the 2-loop threshold function. Comparing with the literature \cite{Green:2010kv,Pioline:2015yea}, we conclude that the sum of the 1-loop and the 2-loop threshold function (appropriately regularised) must reproduce the exact string theory threshold function 
\be  {\cE}_\gra{0}{1}  = \hat{{\cE}}_\gra{0}{1}^{\mbox{\tiny 1-loop}}  + \hat{{\cE}}_\gra{0}{1}^{\mbox{\tiny 2-loop}}  \ . \ee
The first function indeed appears in the exact string theory threshold function with the correct coefficient \cite{Bossard:2015oxa}, and the second satisfies the correct Poisson equation that determines the normalisation. However, the latter equation admits an automorphic homogeneous solution that is in contradiction with string perturbation theory \cite{Bossard:2015oxa}. Assuming that the 2-loop threshold function is consistent with string perturbation theory would rule it out, but we have not proved at this level that it is the case. It is nonetheless likely, given that this threshold function admits by construction a decompactification limit to nine dimensions that reproduces the threshold function computed in \cite{Green:2005ba}. 

In three dimensions ${\cE}_\gra{0}{1}^{\mbox{\tiny 1-loop}} $ is inconsistent with string perturbation theory, and there is a unique supersymmetry invariant such that one cannot disentangle the series $\frac{5\zeta(11)}{12\pi}   E{\mbox{\DEVIII0000000{\mbox{$\frac{11}{2}$}}}}$  from the solution to \eqref{PoissonD6R4} by a tensorial differential equation. We do not prove that the 2-loop threshold function satisfies \eqref{PoissonD6R4} with the correct right-hand-side, however, it is likely to do so. We therefore expect this integral formula to reproduce the exact string theory threshold function up to a term in  $\frac{5\zeta(11)}{12\pi}   E{\mbox{\DEVIII0000000{\mbox{$\frac{11}{2}$}}}}$ that could be determined by computing the perturbative contributions of ${{\cE}}_\gra{0}{1}^{\mbox{\tiny 2-loop}} $ in string theory.  

 \subsection{Higher order threshold functions}
At higher order we expect the threshold functions to get corrections to arbitrary high orders in perturbation theory. It is therefore not clear that one can extract informations about the exact string theory threshold functions from these amplitudes. Note moreover that the renormalisation prescription we adopted in section \ref{sec:D4R4oneloop} would require to consider corrections associated to the inclusion of the relevant $\nabla^4 R^4$ type supersymmetric counterterm at 1-loop. 

We shall  nonetheless study the corresponding 2-loop corrections in this section.  For this purpose we define an $E_{d(d)}$ automorphic function depending on two parameters by
\bea\label{Eis12} 
E_{\alpha_d,s_1,s_2} &= &  \frac{1}{4\xi(2s_1)\xi(2s_2)}  \sum_{\substack{\Gamma_1 \in \mathds{Z}_*^{d(\alpha_d)} \\ \Gamma_1 \times \Gamma_1 = 0   }}  \int_0^\infty \frac{dt_1}{t_1^{1+s_1}}   e^{-\frac{\pi}{t_1}  g(\Gamma_1,\Gamma_1)} \sum_{\substack{\Gamma_2 \in \mathds{Z}^{d(\alpha_d)}_* \\ \Gamma_i \times \Gamma_2 = 0   }}  \int_0^\infty \frac{dt_2}{t_2^{1+s_2}}  e^{-\frac{\pi}{t_2} g(\Gamma_2,\Gamma_2)} \\
&=&\hspace{-3mm} \sum_{\gamma \in P_d \backslash E_{d(d)}} \gamma\Biggl[  r^{\frac{2(10-d)}{9-d}s_1}  \biggl( r^{\frac{2(10-d)}{9-d} s_2} + \frac{\xi(2s_2-1)}{\xi(2s_2)} r^{\frac{2}{9-d}s_2+1} E_{\alpha_{d-1}^{d-1},s_2-\frac{1}{2}} \biggr. \Biggr. \CR
&&  \hspace{20mm} \Biggl.\biggl .   + \, \frac{2}{\xi(2s_2)} \hspace{-5mm}\sum_{\substack{\Gamma \in \mathds{Z}_*^{d(\alpha_{d-1}^{d-1})} \\ \Gamma \times \Gamma = 0}} \Bigl( \sum_{n|{\rm gcd}(\Gamma)} n^{2s_2-1} \Bigr) \frac{r^{\frac{11-d}{9-d}s_2 +\frac{1}{2}}}{|Z(\Gamma)|^{s_2-\frac{1}{2}}} K_{s_2-\frac{1}{2}}( 2\pi r|Z(\Gamma)|) e^{2\pi i \langle \Gamma, a\rangle }   \biggr)         \Biggr] \nn  
  \eea 
  which is symmetric in $s_1$ and $s_2$ by construction. According to the discussion in the preceding section, this function reduces to a product of two Eisenstein series if and only if $s_1=\frac{d-3}{2}$ or  $s_2=\frac{d-3}{2}$ (for $d\le 7$), in which case
  \be E_{\alpha_d,\frac{d-3}{2}+k,\frac{d-3}{2}}  = E_{\alpha_d,\frac{d-3}{2}+k} E_{\alpha_d,\frac{d-3}{2}} \ , \label{DoubleEreduce} \ee
  for almost every $k$. In general the series are defined as meromorphic functions of $d$ and  one must carry out a more detail analysis if there is a pole in the specific dimension. In eight dimensions one can compute it exactly, and one obtains 
  \be  E_{\alpha_3,s_1,s_2} = E_{[s_1+s_2]} E_{[0,s_1]} E_{[0,s_2]} +   E_{[s_1]} E_{[s_2]} E_{[0,s_1+s_2]} - E_{[s_1+s_2]} E_{[0,s_1+s_2]} \ . \ee
  In this case \eqref{DoubleEreduce} is reproduced by the infrared regularised series $ \hat{E}_{\alpha_3,k+\epsilon,\epsilon}$ including the degenerate contribution when $\Gamma_2$ vanishes (and is set equal to $\mu^2$) in \eqref{Eis12}
  \bea\label{8DReducesEi12}  && (4\pi \xi(2\epsilon) ) ( 4\pi \xi(2k+2\epsilon))   \hat{E}_{\alpha_3,k+\epsilon,\epsilon} \CR
  &=&  \scal{ 4\pi  \xi(2\epsilon) E_{[\epsilon]} E_{[0,\epsilon]}+ 2 \pi^{1-\epsilon} \Gamma(\epsilon) \mu^{-2\epsilon} 
 } \scal{  4\pi  \xi(2k+2\epsilon) E_{[k+\epsilon]} E_{[0,k+\epsilon]} }  \CR
 &=&\scal{ 4\zeta(2) \hat{E}_{[1]} + 2\zeta(3) \hat{E}_{[\frac{3}{2},0]} +2\pi(3 \gamma-2- \log( 4\pi \mu^2))}\scal{  4\pi  \xi(2k) E_{[k]} E_{[0,k]} } \ .    \eea
 Note that the pole in $\epsilon$ cancels out, consistently with the property that the 2-loop logarithmic divergence is in $\nabla^6 R^4$, and does not contribute to higher derivative corrections. 
 
  \subsubsection{$\nabla^{8}R^4$}

The function $\cE_\gra{2}{0}^{\mbox{\tiny 2-loop}}$ is defined from the integral of
\be 
\cF_\gra{2}{0}= \frac{\pi^{5-d}}{135}  \sum_{\substack{\Gamma_i \in \mathds{Z}^{2d(\alpha_d)} \\ \Gamma_i \times \Gamma_j = 0   }} {\rm det}\Omega^{\frac{d-7}{2}} \Phi_\gra{2}{0}(\Omega)e^{- \Omega^{ij} g(\Gamma_i , \Gamma_j)} \ . 
\ee
In this case the function $\Phi_\gra{2}{0}(\Omega)$ does not satisfy a projective Poisson equation on $GL(2)/SO(2)$, and one must split it into four eigenfunctions as in \cite{Green:2008bf}. We define accordingly 
\be 
\Phi_\gra{2}{0}(\Omega) = \sum_{n=0}^{3} \Phi^{\ord{2n+1}}_\gra{2}{0}(\Omega) \ , 
\ee
such that the functions
\bea
\Phi^{\ord{1}}_\gra{2}{0}(\Omega) &=& - \frac{65}{21} \det \Omega \ , \\
\Phi^{\ord{3}}_\gra{2}{0}(\Omega) &=& \frac{50}{21} \scal{ ( L_1 + L_2 + L_3)^2 - \det \Omega } \ , \CR
\Phi^{\ord{5}}_\gra{2}{0}(\Omega) &=& \frac{10}{77} \Scal{ 5 ( L_1 + L_2 + L_3)^2 - 35 \frac{ L_1 L_2 L_3}{\det \Omega} ( L_1 + L_2 + L_3) + 2 \det \Omega }  \ , \CR
\Phi^{\ord{7}}_\gra{2}{0}(\Omega) &=& \frac{32}{231} \Scal{ 7 ( L_1 + L_2 + L_3)^2 - 126 \frac{ L_1 L_2 L_3}{\det \Omega} ( L_1 + L_2 + L_3) +231 \Scal{\frac{ L_1 L_2 L_3}{\det \Omega}}^2  +16  \det \Omega } \ , \nn  
\eea
satisfy 
\be 
 \frac{\partial\, }{\partial \Omega^{ij}} \Scal{2 \Omega^{ik} \Omega^{jl}  \frac{\partial\, }{\partial \Omega^{kl}} - 2 \Omega^{ij} } \Phi^{\ord{s}}_\gra{2}{0}(\Omega) = s(s-1)  \Phi^{\ord{s}}_\gra{2}{0}(\Omega)  \ . 
 \ee
The 2-loop threshold function splits accordingly such that 
\be 
\cF^\ord{s}_\gra20 = \frac{\pi^{5-d}}{135}  \sum_{\substack{\Gamma_i \in \mathds{Z}^{2d(\alpha_d)} \\ \Gamma_i \times \Gamma_j = 0   }} {\rm det}\Omega^{\frac{d-7}{2}} \Phi^\ord{s}_\gra{2}{0}(\Omega)e^{- \Omega^{ij} g(\Gamma_i , \Gamma_j)} \ . 
\ee
One computes accordingly that 
\be 
\Scal{ \Delta - \frac{(17-7d)(d-2)}{9-d}-s(s-1) } \cF^\ord{s}_\gra{2}{0} = \frac{ \partial \cB^{\ord{s} ij}_\gra20 }{\partial \Omega^{ij}}  \ , 
\ee 
where $\cB^{\ord{s} ij}_\gra20$ is a vector field that we will not display explicitly. We shall only discuss the ultraviolet boundary term that contributes to the Poisson equation. For $s=1$ this boundary term vanishes, and $\cE_\gra20^\ord{1}$ is an Eisenstein series. Repeating the computation of section \ref{Nondege}, one obtains similarly that 
\be 
\cE_\gra20^\ord{1} = - \frac{52\pi^3}{189}\xi(d-3) \xi(d-2) E_{\alpha_{d-1},\frac{d-2}{2}} \ . 
\ee
For  $\cE_\gra20^\ord{3}$, one obtains a boundary term 
\bea
&&  - \frac{40}{189}  \pi^{5-d} \sum_{\substack{\Gamma_1 \in \mathds{Z}_*^{d(\alpha_d)} \\ \Gamma_1 \times \Gamma_1 = 0   }}  \int_0^\infty \frac{dL_1}{L_1^{\frac{5-d}{2}}}  e^{-L_1 g(\Gamma_1,\Gamma_1)} \sum_{\substack{\Gamma_2 \in \mathds{Z}^{d(\alpha_d)}_* \\ \Gamma_i \times \Gamma_2 = 0   }}  \int_0^\infty \frac{dL_2}{L_2^{\frac{3-d}{2}}}  e^{-L_2 g(\Gamma_2,\Gamma_2)} \CR
&=& - \frac{160 \pi^3 }{189} \xi(d-3) \xi(d-1) \, E_{\alpha_{d},\frac{d-3}{2},\frac{d-1}{2}}  \CR
&=& - \frac{160 \pi^3 }{189} \xi(d-3) \xi(d-1) \, E_{\alpha_{d},\frac{d-3}{2}}  E_{\alpha_{d},\frac{d-1}{2}} \ ,   \eea
where we assume that the function is appropriately regularised in the infrared. In eight dimensions the appropriately regularised function appearing on the right-hand-side is finite according to \eqref{8DReducesEi12}. Note that the diagonal contribution associated to the charges $\Gamma_1+\Gamma_2=0$ in $ E_{\alpha_{d},d-2}$ can always be removed, but in dimension six, in which case the function is in fact $E_{\alpha_{d},\frac{d+1}{2}}$. This is the sign of a  logarithmic divergence of the 1-loop $\nabla^4 R^4$ form factor  in $\nabla^8 R^4$, together with a logarithmic divergence of the 2-loop $R^4$ form factor. We conclude that 
\bea 
\Delta  \hat{\cE}^\ord{3}_\gra{2}{0} &=& \Scal{ \frac{(17-7d)(d-2)}{9-d}+6}  \hat{\cE}^\ord{3}_\gra{2}{0}   - \frac{160 \pi^3 }{189} \xi(d-3) \xi(d-1) \, E_{\alpha_{d},\frac{d-3}{2}}  E_{\alpha_{d},\frac{d-1}{2}} \CR
&& \qquad  + \delta_{d,5} \Scal{ \gamma_1  \frac{8}{45} \zeta(6) \hat{E}_{\alpha_5,3} + \gamma_1^\prime \zeta(5) \hat{E}_{\alpha_1,\frac{5}{2}} + \gamma_2 2 \zeta(3)  E_{\alpha_1,\frac{3}{2}} } \  . 
\eea
One computes in the same way that 
\bea 
\Delta  \hat{\cE}^\ord{5}_\gra{2}{0} &=& \Scal{ \frac{(17-7d)(d-2)}{9-d}+20}  \hat{\cE}^\ord{3}_\gra{2}{0}   - \frac{80 \pi^3 }{77}\xi(d-3) \xi(d-1) \, E_{\alpha_{d},\frac{d-3}{2}}  E_{\alpha_{d},\frac{d-1}{2}} \CR
&& \qquad  + \delta_{d,7} \Scal{ \gamma_1 \frac{64\zeta(10)}{189} E_{\alpha_7,5}  + \gamma_1^\prime \hat{\cE}_\gra01+ \gamma_2  \zeta(5)  E_{\alpha_1,\frac{5}{2}} } \  . 
\eea
Here again we included a possible anomalous term allowed by the Laplace eigenvalue, and which is also suggested by the property that the infrared diagonal contribution in $E_{\alpha_d,d-2}$ is proportional to  $E_{\alpha_d,\frac{d+3}{2}}$ in four dimensions. Such a contribution is associated to the logarithm divergence of the 1-loop $\nabla^6 R^4$ form factor, respectively 2-loop $\nabla^4 R^4$ form factor, in $\nabla^8 R^4$.  

The last function satisfies 
\be 
\Delta  \hat{\cE}^\ord{7}_\gra{2}{0} = \Scal{ \frac{(17-7d)(d-2)}{9-d}+42}  \hat{\cE}^\ord{7}_\gra{2}{0}   - \frac{1024 \pi^3 }{297}  \xi(d-3) \xi(d-1) \, E_{\alpha_{d},\frac{d-3}{2}}  E_{\alpha_{d},\frac{d-1}{2}} \CR
\  . 
\ee
In this case there is no potential anomalous contribution in dimension lower than eight. 

\subsubsection{$\nabla^{10}R^4$}

\allowDbreak{The function $\cE_\gra{1}{1}^{\scriptscriptstyle \mbox{\tiny 2-loop}}$ is defined from the integral of
\begin{align}
\cF_\gra{1}{1} = \frac{2\pi^{5-d}}{8505}  \sum_{\substack{\Gamma_i \in \mathds{Z}^{2d(\alpha_d)} \\ \Gamma_i \times \Gamma_j = 0   }} {\rm det}\Omega^{\frac{d-7}{2}} \Phi_\gra{1}{1}(\Omega)e^{- \Omega^{ij} g(\Gamma_i , \Gamma_j)} \ . 
\end{align}
In this case on must split the function $\Phi_\gra{1}{1}(\Omega)$ into five  eigenfunctions of the projective $SL(2)/SO(2)$ Laplace operator \cite{Green:2008bf}
\be 
\Phi_\gra{1}{1}(\Omega) = \sum_{n=1}^{5} \Phi^{\ord{2n}}_\gra{1}{1}(\Omega) \ , 
\ee
such that the functions $ \Phi^{\ord{s}}_\gra11(\Omega)$ defined as
\begin{align} 
\Phi^{\ord{2}}_\gra11(\Omega) &= - \frac{245}{33} (L_1+L_2+L_3) \det \Omega \ , &\nn\\
\Phi^{\ord{4}}_\gra11(\Omega) &= \frac{14}{429} \Scal{ 679 ( L_1 + L_2 + L_3)^3+6714 \, L_1 L_2 L_3 -2565 (L_1+L_2+L_3) \det \Omega } \ , &\nn\\
\Phi^{\ord{6}}_\gra11(\Omega) &= \frac{98}{39} \Bigl(  7 ( L_1 + L_2 + L_3)^3- 63   \frac{ L_1 L_2 L_3}{\det \Omega} ( L_1 + L_2 + L_3)^2   + 21L_1 L_2 L_3 \Bigr . &\CR
&\quad \Bigl . \hspace{80mm}+ 3(L_1+L_2+L_3) \det \Omega \Bigr)  \ , &\nn\\
\Phi^{\ord{8}}_\gra11(\Omega) &=  \frac{3724}{7293} \Bigl(  9 ( L_1 + L_2 + L_3)^3  - 198   \frac{ L_1 L_2 L_3}{\det \Omega} ( L_1 + L_2 + L_3)^2 \Bigr . &\nn\\
&\quad \Bigl . \hspace{15mm}+ 429 \Scal{\frac{ L_1 L_2 L_3}{\det \Omega} }^2  ( L_1 + L_2 + L_3) -12 \, L_1 L_2 L_3+  28(L_1+L_2+L_3) \det \Omega \Bigr)   \ , &\nn\\
\Phi^{\ord{10}}_\gra11(\Omega) &=      \frac{145}{2431} \Bigl(  11 ( L_1 + L_2 + L_3)^3 - 429   \frac{ L_1 L_2 L_3}{\det \Omega} ( L_1 + L_2 + L_3)^2\Bigr .& \\
&\quad \Bigl . \hspace{20mm}+ 2145 \Scal{\frac{ L_1 L_2 L_3}{\det \Omega} }^2  ( L_1 + L_2 + L_3)  -2431 \Scal{\frac{ L_1 L_2 L_3}{\det \Omega} }^3\Bigr . &\nn\\
&\quad \Bigl . \hspace{55mm} -264 \, L_1 L_2 L_3+  72(L_1+L_2+L_3) \det \Omega \Bigr)   \ ,\nn&
\end{align}
satisfy 
\be 
\frac{\partial\, }{\partial \Omega^{ij}} \Scal{2 \Omega^{ik} \Omega^{jl}  \frac{\partial\, }{\partial \Omega^{kl}} -  3 \Omega^{ij}}  \Phi^{\ord{s}}_\gra{1}{1}(\Omega) = s(s-1) \Phi^{\ord{s}}_\gra{1}{1}(\Omega)  \ . 
\ee
The 2-loop threshold function splits accordingly such that 
\be 
\cF^\ord{s}_\gra11 = \ \frac{2\pi^{5-d}}{8505}   \sum_{\substack{\Gamma_i \in \mathds{Z}^{2d(\alpha_d)} \\ \Gamma_i \times \Gamma_j = 0   }} {\rm det}\Omega^{\frac{d-7}{2}} \Phi^\ord{s}_\gra11(\Omega)e^{- \Omega^{ij} g(\Gamma_i , \Gamma_j)} \ . 
\ee}

\allowDbreak{One computes accordingly that 
\be 
\Scal{ \Delta - \frac{4(7-2d)(d-1)}{9-d}-s(s-1) } \cF^\ord{s}_\gra{1}{1} = \frac{ \partial \cB^{\ord{s} ij}_\gra11 }{\partial \Omega^{ij}}  \ , 
\ee 
where $\cB^{\ord{s} ij}_\gra20$ is a vector field that we will not display explicitly. We shall only discuss the ultraviolet boundary term that contributes to the Poisson equation. Similarly as in the preceding section one computes that 
\begin{align}
\Delta  \hat{\cE}^\ord{2}_\gra11 &= \Scal{\frac{4(7-2d)(d-1)}{9-d}+2}  \hat{\cE}^\ord{2}_\gra11   + \frac{7}{66}  \Scal{ \frac{4\pi^2}{9} \xi(d-1)  }^2 E_{\alpha_{d},\frac{d-1}{2},\frac{d-1}{2}} + \delta_{d,3} \gamma_1 \frac{4\pi\zeta(4)}{45} E_{\alpha_3,2}  &\nn\\
\Delta  \hat{\cE}^\ord{4}_\gra11 &= \Scal{\frac{4(7-2d)(d-1)}{9-d}+12}  \hat{\cE}^\ord{4}_\gra11   + \frac{347 }{143}  \Scal{ \frac{4\pi^2}{9} \xi(d-1) }^2 E_{\alpha_{d},\frac{d-1}{2},\frac{d-1}{2}}  &\CR
&\hspace{-8mm}  - \frac{1356}{1287}\Scal{4 \pi  \xi(d-3) E_{\alpha_{d},\frac{d-3}{2}} }\Scal{ \frac{4\pi^3}{45} \xi(d+1) E_{\alpha_{d},\frac{d+1}{2}} }  + \delta_{d,5} \Scal{ \gamma_1 \frac{16\zeta(8)}{189}\hat{E}_{\alpha_5 , 4} + \gamma^\prime_1 \hat{\cE}_\gra01  + \gamma_2} \ , &\nn\\
\Delta  \hat{\cE}^\ord{6}_\gra11 &= \Scal{\frac{4(7-2d)(d-1)}{9-d}+30}  \hat{\cE}^\ord{6}_\gra11   - \frac{70 }{13}  \Scal{ \frac{4\pi^2}{9} \xi(d-1) }^2 E_{\alpha_{d},\frac{d-1}{2},\frac{d-1}{2}}  &\CR
&\quad \quad - \frac{392}{117}\Scal{4 \pi  \xi(d-3) E_{\alpha_{d},\frac{d-3}{2}} }\Scal{ \frac{4\pi^3}{45} \xi(d+1) E_{\alpha_{d},\frac{d+1}{2}} }  + \delta_{d,7}  \gamma_1 \frac{16\pi^5}{14175} \xi(d+5) {E}_{\alpha_d , \frac{d+5}{2}}  \ , &\nn\\
\Delta  \hat{\cE}^\ord{8}_\gra11 &=1 \Scal{\frac{4(7-2d)(d-1)}{9-d}+56}  \hat{\cE}^\ord{8}_\gra11   - \frac{26068}{7293}  \Scal{ \frac{4\pi^2}{9} \xi(d-1) }^2 E_{\alpha_{d},\frac{d-1}{2},\frac{d-1}{2}}  &\CR
&\quad \hspace{30mm}- \frac{13300}{7293}\Scal{4 \pi  \xi(d-3) E_{\alpha_{d},\frac{d-3}{2}} }\Scal{ \frac{4\pi^3}{45} \xi(d+1) E_{\alpha_{d},\frac{d+1}{2}} }    \ , &\nn\\
\Delta  \hat{\cE}^\ord{10}_\gra11 &= \Scal{\frac{4(7-2d)(d-1)}{9-d}+90}  \hat{\cE}^\ord{10}_\gra11   - \frac{2610}{2431}  \Scal{ \frac{4\pi^2}{9} \xi(d-1) }^2 E_{\alpha_{d},\frac{d-1}{2},\frac{d-1}{2}}  &\CR
&\quad \hspace{30mm}- \frac{290}{663}\Scal{4 \pi  \xi(d-3) E_{\alpha_{d},\frac{d-3}{2}} }\Scal{ \frac{4\pi^3}{45} \xi(d+1) E_{\alpha_{d},\frac{d+1}{2}} }    \ .  
\end{align}
In dimensions seven and six the series $E_{\alpha_{d},\frac{d+1}{2}} $ is divergent, so one cannot necessarily use identity \eqref{DoubleEreduce}, such that one should rather write the corresponding function as $E_{\alpha_d,\frac{d-3}{2},\frac{d+1}{2}}$.} 

\subsection{Wavefront set of $  \cE_{\gra{p}{q}}^{\scriptscriptstyle \mbox{\tiny 2-loop}}$}

In this section we would like to describe the wavefront set of the functions appearing at 2-loop. In order to do this, let us first determine the homogeneous part of the differential equations satisfied by the threshold functions. 

The function $\Phi_\gra{p}{q}(\Omega)$ is homogeneous of degree $2p+3q-2$ in $\Omega$, and defines a function on $SL(2)/SO(2)$ through the change of variable 
\be 
\Phi_\gra{p}{q}(\Omega) = V^{2p+3q-2} \Phi_\gra{p}{q}(\tau,\bar\tau) \ , 
\ee
where we use the same symbol for the rescaled function for short. $ \Phi_\gra{p}{q}(\tau,\bar\tau)$ can be decomposed into eigenfunctions of the Laplace operator on the upper complex plane as 
\be 
\Delta  \Phi^\ord{s}_\gra{p}{q}(\tau,\bar\tau) = s(s-1)  \Phi^\ord{s}_\gra{p}{q}(\tau,\bar\tau) \ , 
\ee
such that the non-degenerate orbit 2-loop contribution to the threshold function $\cE_\gra{p}{q}^{\scriptscriptstyle \mbox{\tiny 2-loop}}$ decomposes into functions of the form
\be 
\sum_{\gamma \in P_{d-1} \backslash E_{d(d)}} \int_0^\infty  \frac{dV}{V^{7-d-2p-3q}} \int_{\mathds{C}^+}\frac{d\tau_1 d\tau_2}{\tau_2^{\; 2}}  \Phi^\ord{s}_\gra{p}{q}(\tau,\bar\tau) \sum_{\substack{0\le j<m \\ m>0, \, n \ne 0}} \gamma \left[e^{-V e^{2(11-d)\phi }\scal{ \frac{|m \tau + ( j + n \taubis)|^2}{\tau_2 \taubis_2}- 2 mn}}\right] \ . 
\ee
Using the property that 
\be 
- (\tau - \bar \tau)^2  \partial_{\tau} \partial_{\bar \tau} e^{-V e^{2(11-d)\phi }\scal{ \frac{|m \tau + ( j + n \taubis)|^2}{\tau_2 \taubis_2}- 2 mn}} =  - (u - \bar u)^2  \partial_{u} \partial_{\bar u} e^{-V e^{2(11-d)\phi }\scal{ \frac{|m \tau + ( j + n \taubis)|^2}{\tau_2 \taubis_2}- 2 mn}} \ , 
\ee
one obtains that modulo a boundary term 
\bea 
&& - (u - \bar u)^2  \partial_{u} \partial_{\bar u}  \int_0^\infty  \frac{dV}{V^{5-d-k}} \int_{\mathds{C}^+}\frac{d\tau_1 d\tau_2}{\tau_2^{\; 2}}  \Phi^\ord{s}_\gra{p}{q}(\tau,\bar\tau) \sum_{\substack{0\le j<m \\ m>0, \, n \ne 0}}  e^{-V e^{2(11-d)\phi }\scal{ \frac{|m \tau + ( j + n \taubis)|^2}{\tau_2 \taubis_2}- 2 mn}} \CR
&\sim& s(s-1) \int_0^\infty  \frac{dV}{V^{5-d-k}} \int_{\mathds{C}^+}\frac{d\tau_1 d\tau_2}{\tau_2^{\; 2}}  \Phi^\ord{s}_\gra{p}{q}(\tau,\bar\tau) \sum_{\substack{0\le j<m \\ m>0, \, n \ne 0}}  e^{-V e^{2(11-d)\phi }\scal{ \frac{|m \tau + ( j + n \taubis)|^2}{\tau_2 \taubis_2}- 2 mn}}  \ . 
\eea
All $E_{d(d)}$ Casimir operators acting on this function decomposes into polynomials in the derivative with respect to $\phi$ and the Laplace operator on $u$, such that these functions are by construction eigenfunctions of all Casimir operators, up to a source term associated to a boundary integral. To determine the eigenvalues, we can by construction choose to compute this integral for an arbitrary solution to the Laplace equation $ \Phi^\ord{s}_\gra{p}{q}(\tau,\bar\tau)$, which we choose to be  $\tau_2^{\; s}$. One computes  with $k=2p+3q-2$ that
\begin{align}
\label{2loopGene}
&=2 \pi^{5-d}   \int_0^\infty  \frac{dV}{V^{5-d-k}} \int_{\mathds{C}^+}\frac{d\tau_1 d\tau_2}{\tau_2^{\; 2-s}} \sum_{\substack{0\le j<m \\ m>0, \, n \ne 0}} e^{-V e^{2(11-d)\phi }\scal{ \frac{|m \tau + ( j + n \taubis)|^2}{\tau_2 \taubis_2}- 2 mn}}&\CR
&=  2\pi^{\frac{11}{2}-d}  e^{-(11-d) \phi} \sqrt{ \taubis_2} \sum_{\substack{0\le j<m \\ m>0, \, n \ne 0}}  \frac{1}{m} \int_0^\infty   \frac{dV}{V^{\frac{11}{2}-d-k}} \int_0^\infty \frac{d\tau_2}{\tau_2{}^{\frac{3}{2}-s}}e^{-e^{2(11-d)\phi} \scal{ \frac{m^2}{\taubis_2} V \tau_2 + n^2 \taubis_2 \frac{V}{\tau_2}}} &\CR
&=  \pi^{\frac{11}{2}-d} e^{-(11-d) \phi} \sqrt{ \taubis_2} \sum_{\substack{0\le j<m \\ m>0, \, n \ne 0}} \frac{1}{m} \int_0^\infty  \hspace{-3mm}dx \, x^{\frac{d+k-s-6}{2}} \int_0^\infty dy\,  y^{\frac{d+k+s-7}{2}} e^{-e^{2(11-d)\phi} \scal{ \frac{m^2}{\taubis_2} y  + n^2 \taubis_2 x}} &\CR
&= 2\pi^{1+k}  \xi(d+k+s-5)\xi(d+k-s-4)  e^{-2(11-d)(d-4+k) \phi} u_2^{\; s} \ . &  
\end{align}
This function is by construction an eigenfunction of all Casimirs: it is the character associated to the weight vector  $\lambda =( d-4+k-s)\Lambda_{d-1} + 2 s\Lambda_d-\rho$. This weight defines the Eisenstein series 
\be  
E_{\frac{d-4+k-s}{2} \Lambda_{d-1} + s \Lambda_d} =  \sum_{\gamma \in P_{d-1} \backslash E_{d(d)}}\gamma\left[ e^{-2(11-d)(d-4+k) \phi} E_{[s]}(u,\bar u) \right] \ , 
\label{GenericEis}  
\ee
that describes some of the contributions to the threshold functions $\cE_\gra{p}{q}^{\scriptscriptstyle \mbox{\tiny 2-loop}}$ for $k=2p+3q-2$. One checks that this series is consistent with the various large radii limits, such that one has for instance
\be 
E_{\frac{d-4+k-s}{2} \Lambda_{d-1} + s \Lambda_d}  = \frac{\xi(d-5+k-s) \xi(d-6+k+s)}{\xi(d-4+k-s) \xi(d-5+k+s)} r^{2\frac{5+k}{9-d}} E_{\frac{d-5+k-s}{2} \Lambda_{d-2} + s \Lambda_{d-1}} + \dots \ , 
\ee
where for $d=3$
\be 
E_{\frac{-1+k-s}{2} \Lambda_{2} + s \Lambda_{3}} = E_{[s]} E_{[0,k-1]} + E_{[k-1]} E_{[\stfrac{-1+k-s}{2},s]} \ ,
\ee
according to the convention that $\Lambda_2$ is associated to the type IIB  limit for the first function and to the type IIA limit for the second. Note that the terms hidden in the dots are not necessarily subleading in the large radius limit.  
 
The string perturbation theory limit of these series for $d=7$ corresponds to corrections at the various loop orders $2,\, \frac{k-s+2}{2},\,  \frac{k+s+1}{2},\, k,\, k-s-3,\, k+s-4,\, \frac{3k-s-5}{2},\, \frac{3k+s-6}{2} ,\, 2k-5$. Consistency with string perturbation theory suggests therefore that $s$ must be an integer, although for generic integral values of $k$ and $s$, none of these contributions vanish and one necessarily have components corresponding to half integer loop order. Producing formally half-integer loop contributions is a generic property of homogenous solutions, and one sees that they cancel eventually in the complete solution. The facts that $s$ ranges over integers follows from the property that $\Phi_\gra{p}{q}(\tau,\bar\tau)$ descends from a homogenous rational function of $\Omega$. We note moreover that the series satisfy the functional relation 
\be 
E_{\frac{d-4+k-s}{2} \Lambda_{d-1} + s \Lambda_d} = \frac{\xi(2s-1)}{ \xi(2s)}  E_{\frac{d-5+k+s}{2} \Lambda_{d-1} + (1-s) \Lambda_d} \ , 
\ee
which exhibits the symmetry $s\rightarrow 1-s$ that is manifest in \eqref{GenericEis}.  For $d=7$ this series is equivalent to the one with the replacement $k\rightarrow 7-k$, such that 
\be 
E_{\mbox{\DEVII00000{\mathnormal{\frac{10\mbox{-}k\mbox{-}s}{2}}}{s}}}  = {\textstyle \frac{\xi(2k-6)\xi(2k-2)\xi(k+s+2)\xi(k+s-1)\xi(k-s+3)\xi(k-s)}{\xi(2k-11)\xi(2k-7)\xi(k+s-9)\xi(k+s-6)\xi(k-s-8)\xi(k-s-5)}} E_{\mbox{\DEVII00000{\mathnormal{\frac{3\mbox{+}k\mbox{-}s}{2}}}{s}}}  \ .  
\ee
We conclude that consistency implies that the relevant homogeneous solutions must be 
\be  
\cE_{\gra{p}{q}\, {\scriptscriptstyle \mbox{\tiny ho}}}^{\scriptscriptstyle \mbox{\tiny 2-loop}}  = \sum_{s\in \mathds{Z}} c_s^{p,q}  \xi(d - 6+ 2 p + 3 q - s)\xi(d - 7+ 2 p + 3 q + s)E_{\frac{d - 6+ 2 p + 3 q - s}{2} \Lambda_{d-1} + s \Lambda_d}\ , 
\ee
where the coefficients $ c_s^{p,q}  $ do not depend on $d$, and are only non-zero for finitely many values of $s$. The range of $s$ can be determined by the Laplace equation, using 
\be 
\Delta E_{\frac{d-4+k-s}{2} \Lambda_{d-1} + s \Lambda_d}  = \Scal{\frac{(4-k-d)(5-11k+(5+k)d)}{9-d}+ s(s-1)}E_{\frac{d-4+k-s}{2} \Lambda_{d-1} + s \Lambda_d} \ . 
\ee
This reproduces indeed the correct $\cE_{\gra{1}{0}}^{\scriptscriptstyle \mbox{\tiny 2-loop}}$ threshold function with $c_s^{1,0}=8\pi \delta_{s,0}$. For $  \cE_{\gra{0}{1}}^{\scriptscriptstyle \mbox{\tiny 2-loop}}$ the Laplace equation implies $s=4$, and using \eqref{E7HomoWeyl} one obtains in four dimensions (with $d=7+2\epsilon$)
\be
c_4^{0,1}  \xi(2\epsilon) \xi(7+2\epsilon) E_{\mbox{\DEVII00000{\mathnormal{\epsilon}}{4}}} =c_4^{0,1} \frac{\xi(7+2\epsilon)\xi(9-2\epsilon)\xi(12-2\epsilon)}{\xi(4-2\epsilon)} E_{\mbox{\DEVII{\mathnormal{(6\mbox{-}\epsilon)}}00000{2\epsilon}}} \ .
\ee
This reproduces the homogeneous solution displayed in \cite{Bossard:2015oxa} for 
\be 
c_4^{0,1}  = \frac{4 \pi^2}{9} \frac{\xi(8)}{\xi(7)} \ . 
\ee
Using this one can also infer which loop orders in string perturbation theory can get contributions from the exceptional field theory $2$-loop amplitude. One concludes for instance that  $\cE_\gra{2}{0}^{\scriptscriptstyle \mbox{\tiny 2-loop}}$ gets contributions at $0,\, 1,\, 2,\, 3,\, 4,\, 5$ loops, and  $\cE_\gra{1}{1}^{\scriptscriptstyle \mbox{\tiny 2-loop}}$ at $0,\, 1,\, 2,\, 3,\, 4,\, 5,\, 6,\, 7$ and $9$ string loops. It is rather striking that  $\cE_\gra{1}{1}^{\scriptscriptstyle \mbox{\tiny 2-loop}}$ does not get any $8$-string loop contribution, which is the order at which $\cE_\gra{1}{1}$ could exhibit a logarithmic term in the string coupling constant that would be associated to a potential 8-loop divergence in supergravity.  
 
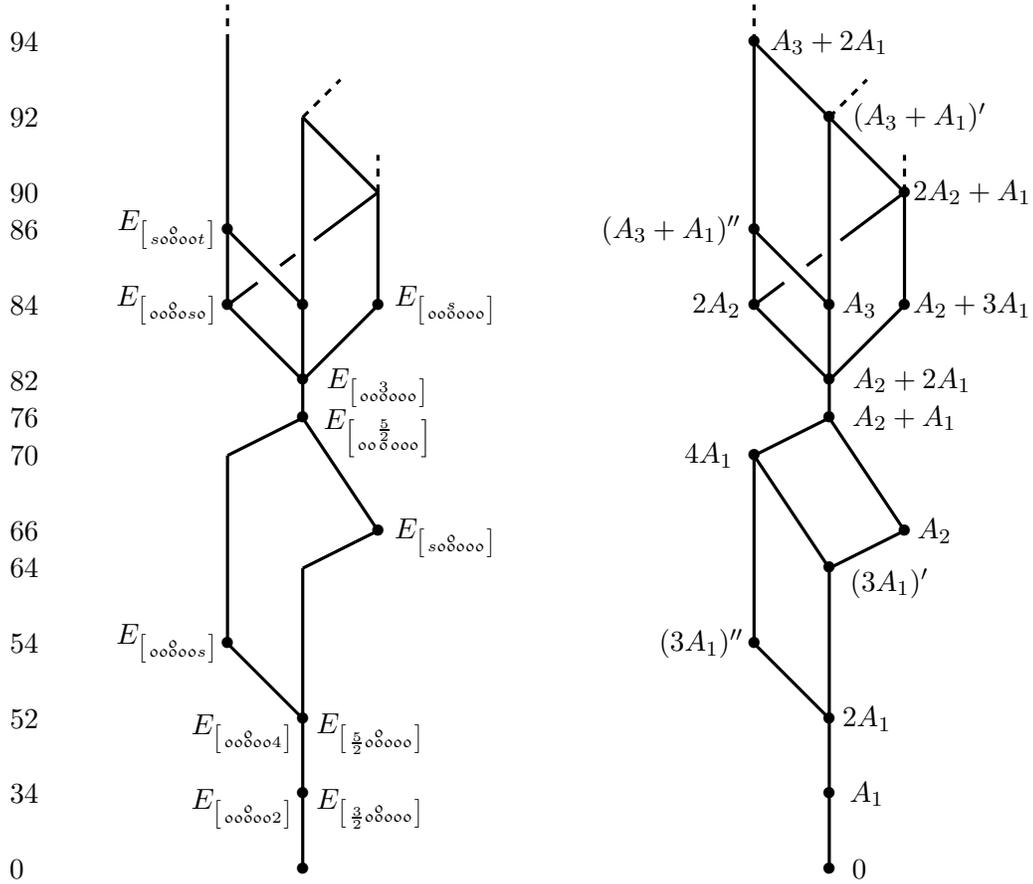
\begin{figure}[htbp]
\def\deltax{7} 
\begin{center}
 \begin{tikzpicture}

\def\DEVIIS#1#2#3#4#5#6#7{{\tiny $ { \left[ \begin{array}{ccccccc}  & &\hspace{0.5mm} \mathfrak{#2} \hspace{-0.9mm}&&&& \vspace{ -0.7mm} \\ \mathfrak{#1}\hspace{-0.7mm} &  \mathfrak{#3} \hspace{-1.2mm}& \mathfrak{#4} \hspace{-1.2mm} & \mathfrak{#5}\hspace{-0.7mm}&\mathfrak{#6}\hspace{-0.7mm}& #7 \hspace{-0.8mm} \end{array}\right] }$}}
\draw (\xmin + 1.9,\ymin + 5.5) node{$E_{\mbox{\DEVII0{\mathnormal{s}}0000{\mathfrak{0}}}}$};
  \draw (\xmin - 1.8,\ymin + 5.5) node{$E_{\mbox{\DEVII00000{\mathnormal{s}}{\mathfrak{0}}}}$};
  \draw (\xmin - 1.8,\ymin + 6.5) node{$E_{\mbox{\DEVII{\mathnormal{s}}00000{t}}}$};
  \draw (\xmin + 0.9,\ymin - 2.2) node{};
  \draw (\xmin - 0.8,\ymin - 1.2) node{$E_{\mbox{\DEVII0000002}}$}; \draw (\xmin + 0.9,\ymin - 1.2) node{$E_{\mbox{\DEVII{\mbox{$\frac{3}{2}$}}00000{\mathfrak{0}}}} $};
 \draw (\xmin - 0.8,\ymin - 0.2) node{$E_{\mbox{\DEVII0000004}}$}; \draw (\xmin + 0.9,\ymin - 0.2) node{$E_{\mbox{\DEVII{\mbox{$\frac{5}{2}$}}00000{\mathfrak{0}}}} $};
  \draw (\xmin - 1.8,\ymin + 1) node{$E_{\mbox{\DEVII000000{{s}}}}$};
   \draw (\xmin + 1.9,\ymin + 2.4) node{$E_{\mbox{\DEVII{\mathnormal{s}}00000{\mathfrak{0}}}}$};

   \draw (\xmin+1.0,\ymin + 3.8) node{$E_{\mbox{\DEVIIS0{\mbox{$\frac{5}{2}$}}0000{\mathfrak{0}}}}$};
\draw (\xmin+1.0,\ymin + 4.4) node{$E_{\mbox{\DEVII0{\mbox{$3$}}0000{\mathfrak{0}}}}$};

 \draw (\xmin,\ymin) node{\textbullet};
 \draw (\xmin,\ymin - 1) node{\textbullet};
  \draw (\xmin,\ymin - 2)  node{\textbullet};
  \draw (\xmin,\ymin + 4)  node{\textbullet};
  \draw (\xmin + 1,\ymin + 2.5)  node{\textbullet};
    \draw (\xmin - 1 ,\ymin + 1) node{\textbullet};
     \draw (\xmin,\ymin + 4.5)  node{\textbullet};
        \draw (\xmin,\ymin + 5.5)  node{\textbullet};  
             \draw (\xmin+1,\ymin + 5.5)  node{\textbullet};  
             \draw (\xmin-1,\ymin + 5.5)  node{\textbullet};  
             \draw (\xmin-1,\ymin + 6.5)  node{\textbullet};  
  \draw[-,draw=black,very thick](\xmin,\ymin) -- (\xmin,\ymin + 2 );
   \draw[-,draw=black,very thick](\xmin - 1,\ymin + 1) -- (\xmin - 1,\ymin + 3.5);
    \draw[-,draw=black,very thick](\xmin,\ymin + 2) -- (\xmin + 1,\ymin + 2.5);
    \draw[-,draw=black,very thick](\xmin + 1,\ymin + 2.5) -- (\xmin,\ymin + 4);
    \draw[-,draw=black,very thick](\xmin - 1,\ymin + 3.5) -- (\xmin,\ymin + 4);
\draw[-,draw=black,very thick](\xmin,\ymin + 4) -- (\xmin,\ymin + 4.5);
  \draw[-,draw=black,very thick](\xmin,\ymin) -- (\xmin - 1,\ymin + 1);
\draw[-,draw=black,very thick] (\xmin,\ymin - 1) -- (\xmin,\ymin);
\draw[-,draw=black,very thick] (\xmin,\ymin - 2) -- (\xmin,\ymin - 1);
 \draw[-,draw=black,very thick](\xmin,\ymin+4.5) -- (\xmin,\ymin+ 5.5 );
  \draw[-,draw=black,very thick](\xmin,\ymin+4.5) -- (\xmin+1,\ymin+ 5.5 );
  \draw[-,draw=black,very thick](\xmin,\ymin+4.5) -- (\xmin-1,\ymin +5.5 );
  \draw[-,draw=black,very thick](\xmin-1,\ymin+5.5) -- (\xmin-1,\ymin +9 );
  \draw[-,draw=black,very thick](\xmin,\ymin+5.5) -- (\xmin-1,\ymin +6.5 );
  \draw[-,draw=black,very thick](\xmin+1,\ymin+5.5) -- (\xmin+1,\ymin +7 );
  \draw[-,draw=black,very thick](\xmin,\ymin+8) -- (\xmin+1,\ymin +7 );
  \draw[-,draw=black,very thick](\xmin,\ymin+5.5) -- (\xmin,\ymin +8 );
\draw[dashed,draw=black,very thick](\xmin,\ymin +8) -- (\xmin+0.5,\ymin + 8.5);
\draw[dashed,draw=black,very thick](\xmin+1,\ymin+7)--(\xmin+1,\ymin+7.5);
     \draw[dashed,draw=black,very thick](\xmin-1,\ymin +9) -- (\xmin-1,\ymin + 9.5);
     \draw[-,draw=black,very thick](\xmin+0.15,\ymin +6.3625) -- (\xmin+1,\ymin + 7);
     \draw[-,draw=black,very thick](\xmin-1,\ymin +5.5) -- (\xmin-0.6,\ymin + 5.8);
     \draw[-,draw=black,very thick](\xmin-0.3,\ymin +6.025) -- (\xmin-0.1,\ymin + 6.175);

 \draw (\xmin+\deltax,\ymin) node{\textbullet};
 \draw (\xmin+\deltax,\ymin - 1) node{\textbullet};
  \draw (\xmin+\deltax,\ymin - 2)  node{\textbullet};
  \draw (\xmin+\deltax,\ymin + 2)  node{\textbullet};
  \draw (\xmin+\deltax,\ymin + 4)  node{\textbullet};
  \draw (\xmin+\deltax + 1,\ymin + 2.5)  node{\textbullet};
   \draw (\xmin+\deltax - 1,\ymin + 3.5)  node{\textbullet};
    \draw (\xmin+\deltax - 1 ,\ymin + 1) node{\textbullet};
     \draw (\xmin+\deltax,\ymin + 4.5)  node{\textbullet};
        \draw (\xmin+\deltax,\ymin + 5.5)  node{\textbullet};  
             \draw (\xmin+\deltax+1,\ymin + 5.5)  node{\textbullet};  
             \draw (\xmin+\deltax-1,\ymin + 5.5)  node{\textbullet};  
             \draw (\xmin+\deltax-1,\ymin + 6.5)  node{\textbullet};  
             \draw (\xmin+\deltax+1,\ymin + 7)  node{\textbullet};  
             \draw (\xmin+\deltax,\ymin + 8)  node{\textbullet};  
             \draw (\xmin+\deltax-1,\ymin + 9)  node{\textbullet};  
  \draw[-,draw=black,very thick](\xmin+\deltax,\ymin) -- (\xmin+\deltax,\ymin + 2 );
   \draw[-,draw=black,very thick](\xmin+\deltax - 1,\ymin + 1) -- (\xmin+\deltax - 1,\ymin + 3.5);
    \draw[-,draw=black,very thick](\xmin+\deltax - 1,\ymin + 3.5) -- (\xmin+\deltax,\ymin + 2);
  \draw[-,draw=black,very thick](\xmin+\deltax,\ymin + 2) -- (\xmin+\deltax + 1,\ymin + 2.5);
    \draw[-,draw=black,very thick](\xmin+\deltax + 1,\ymin + 2.5) -- (\xmin+\deltax,\ymin + 4);
    \draw[-,draw=black,very thick](\xmin+\deltax - 1,\ymin + 3.5) -- (\xmin+\deltax,\ymin + 4);
\draw[-,draw=black,very thick](\xmin+\deltax,\ymin + 4) -- (\xmin+\deltax,\ymin + 4.5);
  \draw[-,draw=black,very thick](\xmin+\deltax,\ymin) -- (\xmin+\deltax - 1,\ymin + 1);
\draw[-,draw=black,very thick] (\xmin+\deltax,\ymin - 1) -- (\xmin+\deltax,\ymin);
\draw[-,draw=black,very thick] (\xmin+\deltax,\ymin - 2) -- (\xmin+\deltax,\ymin - 1);
 \draw[-,draw=black,very thick](\xmin+\deltax,\ymin+4.5) -- (\xmin+\deltax,\ymin+ 5.5 );
  \draw[-,draw=black,very thick](\xmin+\deltax,\ymin+4.5) -- (\xmin+\deltax+1,\ymin+ 5.5 );
  \draw[-,draw=black,very thick](\xmin+\deltax,\ymin+4.5) -- (\xmin+\deltax-1,\ymin +5.5 );
  \draw[-,draw=black,very thick](\xmin+\deltax-1,\ymin+5.5) -- (\xmin+\deltax-1,\ymin +9 );
  \draw[-,draw=black,very thick](\xmin+\deltax,\ymin+5.5) -- (\xmin+\deltax-1,\ymin +6.5 );
  \draw[-,draw=black,very thick](\xmin+\deltax+1,\ymin+5.5) -- (\xmin+\deltax+1,\ymin +7 );
  \draw[-,draw=black,very thick](\xmin+\deltax-1,\ymin+9) -- (\xmin+\deltax+1,\ymin +7 );
  \draw[-,draw=black,very thick](\xmin+\deltax,\ymin+5.5) -- (\xmin+\deltax,\ymin +8 );
\draw[dashed,draw=black,very thick](\xmin+\deltax,\ymin +8) -- (\xmin+\deltax+0.5,\ymin + 8.5);
\draw[dashed,draw=black,very thick](\xmin+\deltax+1,\ymin+7)--(\xmin+\deltax+1,\ymin+7.5);
     \draw[dashed,draw=black,very thick](\xmin+\deltax-1,\ymin +9) -- (\xmin+\deltax-1,\ymin + 9.5);
     \draw[-,draw=black,very thick](\xmin+\deltax+0.15,\ymin +6.3625) -- (\xmin+\deltax+1,\ymin + 7);
     \draw[-,draw=black,very thick](\xmin+\deltax-1,\ymin +5.5) -- (\xmin+\deltax-0.6,\ymin + 5.8);
     \draw[-,draw=black,very thick](\xmin+\deltax-0.3,\ymin +6.025) -- (\xmin+\deltax-0.1,\ymin + 6.175);

\draw (\xmin+\deltax+0.4 ,\ymin - 2) node{$0$};
\draw (\xmin+\deltax+0.5,\ymin - 1) node{$A_1$};
\draw (\xmin+\deltax+0.5,\ymin) node{$2A_1$};
\draw (\xmin+\deltax-1.7,\ymin + 1) node{$(3 A_1)^{\prime\prime}$};
\draw (\xmin+\deltax+0.8,\ymin + 1.8) node{$(3 A_1)^{\prime}$};
\draw (\xmin+\deltax+1.4,\ymin + 2.5) node{$A_2$};
\draw (\xmin+\deltax-1.6,\ymin + 3.5) node{$4A_1$};
\draw (\xmin+\deltax+1.0,\ymin + 4) node{$A_2+A_1$};
\draw (\xmin+\deltax+1.1,\ymin + 4.5) node{$A_2+2A_1$};
\draw (\xmin+\deltax+0.4,\ymin + 5.5) node{$A_3$};
\draw (\xmin+\deltax+1.9,\ymin + 5.5) node{$A_2+3A_1$};
\draw (\xmin+\deltax-1.5,\ymin + 5.5) node{$2A_2$};

\draw (\xmin+\deltax-2.1,\ymin + 6.5) node{$(A_3+A_1)^{\prime\prime}$};
\draw (\xmin+\deltax+1.9,\ymin + 7) node{$2A_2+A_1$};
\draw (\xmin+\deltax+1.2,\ymin + 8) node{$(A_3+A_1)^{\prime}$};
\draw (\xmin+\deltax+0.0,\ymin + 9) node{$A_3+2A_1$};

\draw (\xmin - 3 + 0.2 - 1,\ymin - 2) node{$0$};
\draw (\xmin - 3 + 0.3 - 1,\ymin - 1) node{$34$};
\draw (\xmin - 3 + 0.3 - 1,\ymin) node{$52$};
\draw (\xmin - 3 + 0.3 - 1,\ymin + 1) node{$54$};
\draw (\xmin - 3 + 0.3 - 1,\ymin + 2) node{$64$};
\draw (\xmin - 3 + 0.3 - 1,\ymin + 2.5) node{$66$};
\draw (\xmin - 3 + 0.3 - 1,\ymin + 3.5) node{$70$};
\draw (\xmin - 3 + 0.3 - 1,\ymin + 4) node{$76$};
\draw (\xmin - 3 + 0.3 - 1,\ymin + 4.5) node{$82$};
\draw (\xmin - 3 + 0.3 - 1,\ymin + 5.5) node{$84$};
\draw (\xmin - 3 + 0.3 - 1,\ymin + 6.5) node{$86$};
\draw (\xmin - 3 + 0.3 - 1,\ymin + 7) node{$90$};
\draw (\xmin - 3 + 0.3 - 1,\ymin + 8) node{$92$};
\draw (\xmin - 3 + 0.3 - 1,\ymin + 9) node{$94$};

\end{tikzpicture}
\end{center}
\caption{\small Nilpotent orbits associated to Eisenstein series in the $E_{7(7)}$ closure diagram, where we removed the non-special orbits on the left. (There is another orbit of dimension $94$ that is not depicted as it is not relevant to our discussion.)  The diagram on the left indicates also Eisenstein series associated with the various orbits. If generic parameters $s$ or $t$ are chosen one obtains the orbits shown; for specific values the wavefront set of the Eisenstein series (or its leading residue) can be smaller. The reduction of the wavefront set can be studied by analysing the degenerate Whittaker vectors as in~\cite{Fleig:2013psa,Bossard:2015oxa}.}
\label{ClosureDiag}
\end{figure}

In four dimensions the threshold functions $\cE_\gra{p}{q}^{\scriptscriptstyle \mbox{\tiny 2-loop}}$ are therefore associated to the Eisenstein series $E{\mbox{\DEVII00000{\mathnormal{\frac{3\mbox{+}k\mbox{-}s}{2}}}{s}}} $. Up to a Weyl group transformation, this series is related to $E{\mbox{\DEVII{\mathnormal{\frac{9\mbox{-}k\mbox{+}s}{2}}}00000{(k\mbox{-}1)}}}$, by the functional relation
\be 
E_{\mbox{\DEVII00000{\mathnormal{s}}{t}}} = \frac{\xi(2s-8)\xi(2s-11)}{\xi(2s)\xi(2s-3)} E_{\mbox{\DEVII{\mathnormal{(6\mbox{-}s)}}00000{(2s\mbox{+}t\mbox{-}4)}}} \ . 
\label{E7HomoWeyl} 
\ee
Therefore one expects the wavefront set of such functions to be associated to the dimension $86$ nilpotent orbit associated to the weighted Dynkin diagram \DEVII200000{\mathfrak{2}}. In order to prove this one needs to show that the wavefront set of such a function includes this nilpotent orbit, and does not include the orbit of type $A_2+3A_1$ associated to the weighted Dynkin diagram \DEVII020000{\mathfrak{0}}, as one deduces from the closure diagram \ref{ClosureDiag}.  The fact that the Eisenstein series $E{\mbox{\DEVII00000{\mathnormal{s}}{t}}}$ is attached to the degenerate principal series of dimension $86$ implies also that generically the wavefront set has the $(A_3+A_1)''$ orbit as a maximal component. The harder part is showing the absence of the $A_2+3A_1$ orbit.\footnote{For the labelling of the orbits with Bala--Carter labels we are following the convention of~\cite{CollingwoodMcGovern}.}

In order to argue that this is indeed the case, we will now discuss the wavefront set associated to the product of Eisenstein series in the fundamental that sources the $2$-loop functions respective Laplace equations. Let us consider first the M-theory limit decomposition 
\be 
\mathfrak{e}_{7(7)} \cong{\bf 7}^\ord{-4}  \oplus \overline{\bf 35}^\ord{-2}\oplus \scal{ \mathfrak{gl}_1\oplus \mathfrak{sl}_7}^\ord{0} \oplus {\bf 35}^\ord{2}\oplus \overline{\bf 7}^\ord{4} \ , 
\label{A6Grad}
\ee
in order to show that the wavefront set does not include the orbit  of type $A_2+3A_1$. 
In this decomposition, the Fourier modes of the series  $E{\mbox{\DEVII000000s}}$ satisfy the constraint 
\be 
\varepsilon^{nrstuvw} q_{rst}q_{uv[p} q_{qm]w} = 0 \ ,  \qquad \varepsilon^{mnqrstu} q_{pqr} q_{stu} p^p = 0 \ , \label{AdjointSL7} 
\ee 
that gives $27=21+6$ linearly independent solutions, with typical representative
\be  
\frac{1}{6} q_{mnp} dy^m \wedge dy^n \wedge dy^p = dy^1 \wedge \scal{ q_1 dy^2 \wedge dy^3 + q_2 dy^4 \wedge dy^5 + q_3 dy^6 \wedge dy^7}  \ . 
\label{Rank3} 
\ee
Because all components of this representative include the first direction $dy^1$, it follows that the $n^{\rm th}$ power of $q_{mnp}$ decomposes into irreducible representations of $SL(7)$ that contain at least $n$ columns, whereas all the others vanish according to the constraint \eqref{AdjointSL7}. Therefore the $n^{\rm th}$ power of $q_{mnp}$ is only non-zero in the following irreducible representations 
\bea 
q &\in& [0,0,1,0,0,0] \CR
q^2 &\in& [0,0,2,0,0,0]\oplus [1,0,0,0,1,0]\CR
q^3 &\in& [0,0,3,0,0,0]\oplus [1,0,1,0,1,0]\oplus [2,0,0,0,0,0]\CR
q^4 &\in& [0,0,4,0,0,0]\oplus [1,0,2,0,1,0] \oplus [2,0,0,0,2,0]\CR
q^{n} &\in& \bigoplus_{n_1+2n_2+3n_3=n}[n_2+2n_3,0,n_1,0,n_2,0] 
\label{TensorCubic3Form}
\eea
A generic antisymmetric rank 3 tensor $q_{mnp}$ admits a non-zero septic invariant 
\be 
I_7(q) = \varepsilon^{{\fontsize{5.pt}{5.5pt}\selectfont  \mbox{$n_1n_2n_3n_4n_5n_6n_7$} \fontsize{12pt}{14.5pt}\selectfont }}
\varepsilon^{{\fontsize{5.pt}{5.5pt}\selectfont  \mbox{$p_1p_2p_3p_4p_5p_6p_7$} \fontsize{12pt}{14.5pt}\selectfont }}
\varepsilon^{{\fontsize{5.pt}{5.5pt}\selectfont  \mbox{$q_1q_2q_3q_4q_5q_6q_7$} \fontsize{12pt}{14.5pt}\selectfont }} q_{n_1n_2n_3} q_{p_1p_2p_3} q_{q_1q_2q_3} q_{ n_4n_5p_4} q_{p_5q_4q_5} q_{n_6p_6p_7}q_{n_7q_6q_7} \ , 
\ee
whereas there is no singlet in the tensor product of the representations involved in the $n^{\rm th}$ product of a charge satisfying the constraint \eqref{Rank3} and its $(7-n)^{\rm th}$ tensor product, as one can easily conclude using \eqref{TensorCubic3Form}. The sum of two 3-form satisfying \eqref{Rank3} admits a vanishing septic invariant, and is therefore not generic. We conclude that the product function  $E{\mbox{\DEVII000000s}}E{\mbox{\DEVII000000t}}$ wavefront set does not include the nilpotent orbit {\DEVII020000{\mathfrak{0}}} of type $A_2+3A_1$.

However, the tensor product of two $[1,0,1,0,1,0]$ includes the $[0,0,0,1,0,0]$ such that the derivative of the septic invariant does not vanish. Therefore the wavefront set does include the nilpotent orbit of type $A_2+2A_1$. 

The orbit of type $(A_3+A_1)^{\prime\prime}$ is instead associated to the graded decomposition 
\bea 
\mathfrak{e}_{7(7)} &\cong&{\bf 1}^\ord{-6}  \oplus {\bf 16}^\ord{-2} \oplus \scal{ {\bf 10}\oplus \overline{\bf 16}}^\ord{-2}\oplus \scal{ \mathfrak{gl}_1\oplus  \mathfrak{gl}_1\oplus \mathfrak{so}(5,5)}^\ord{0}  \oplus \scal{ {\bf 10}\oplus {\bf 16}}^\ord{2} \oplus\overline{\bf 16}^\ord{4} \oplus {\bf 1}^\ord{6} \ ,  \CR
{\bf 56} &\cong& {\bf 1}^\ord{-5}  \oplus {\bf 10}^\ord{-3} \oplus \scal{ {\bf 1}\oplus \overline{\bf 16}}^\ord{-1}\oplus  \scal{ {\bf 1}\oplus {\bf 16}}^\ord{1} \oplus {\bf 10}^\ord{3} \oplus {\bf 1}^\ord{5}\ . 
\label{D5Grad}
\eea
A representative of the nilpotent orbit is obtained as a generic doublet of a vector $q^a$ and a Majorana--Weyl spinor $\zeta_\alpha$ in the grad two component, that satisfy 
\be 
\eta_{ab} q^a q^b \ne 0 \ , \qquad   \Gamma_a{}^{\alpha\beta} q^a \zeta_\alpha \zeta_\beta \ne 0 \ , 
\ee
with stabilizer $Spin(3,4)\subset Spin(5,5)$. An element of the $3A_1$ nilpotent orbit satisfies instead 
\be 
q_a =  c \, \Gamma_a{}^{\alpha\beta}  \zeta_\alpha \zeta_\beta \ , 
\ee
for some $c$, which implies that $q^a$ is a null vector and that $\zeta_\alpha$ satisfies  the corresponding Dirac equation. An element of the minimal nilpotent orbit can be realised by a vanishing vector and a pure spinor, or a vanishing spinor and a null vector. If we add to $q^a$ an arbitrary null vector $p^a$, it is clear that 
\be 
\eta_{ab} ( q^a +p^a) (q^b+p^b) = 2 c \,  \Gamma_a{}^{\alpha\beta} p^a  \zeta_\alpha \zeta_\beta \ne 0 \ , \qquad \Gamma_a{}^{\alpha\beta}( q^a+p^a)  \zeta_\alpha \zeta_\beta =  \Gamma_a{}^{\alpha\beta} p^a  \zeta_\alpha \zeta_\beta  \ne 0 \  , 
\ee
for an appropriate $p^a$. It follows that the generic sum of an element of type $A_1$ and an element of type $3A_1$ defines a generic element of type $(A_3+A_1)^{\prime\prime}$. We conclude that the product function  $E{\mbox{\DEVII000000s}}E{\mbox{\DEVII000000t}}$ wavefront set does include the nilpotent orbit {\DEVII200000{\mathfrak{2}}}. Note that because one element can be in the minimal nilpotent orbit, this applies also to the degenerate case  $E{\mbox{\DEVII0000002}}E{\mbox{\DEVII0000003}}$ relevant for the $\nabla^8 R^4$ 2-loop threshold function.

\section{Concluding comments}
\label{sec:Specs}

In this concluding section we collect various comments on our analysis.

\subsection{Massive states as BPS solitons in supergravity}

The sum over the rank one charges $\Gamma$ was interpreted here as a sum over $\tfrac12$-BPS states in M-theory. One can also interpret these corrections directly in supergravity in $D=11-d$ dimensions. One understands then this sum as a sum over $\tfrac12$-BPS solitons, that are the $\tfrac12$-BPS black holes in $11-d$ dimensions. According to \cite{Papageorgakis:2014dma}, the contribution of zero size solitons is not necessarily exponentially suppressed, and cannot be disregarded in perturbation theory. By construction, both $\tfrac12$-BPS and $\tfrac14$-BPS black holes have a vanishing horizon area as classical (two derivative) solutions, and as such can be considered as zero size solitons. In four dimensions there are moreover $\tfrac18$-BPS black holes of vanishing horizon area. It was argued in \cite{Bianchi:2009wj} that these solitons should not be considered in supergravity (in four dimensions) because they define singular solutions for which the horizon is replaced by an infinitely red-shifted naked singularity. However, singularities of classical solutions are standard in quantum field theory (\eg the static electron in quantum electrodynamics), and we shall argue instead that the vanishing horizon solitons cannot consistently be removed from the spectrum in perturbation theory, whereas the finite horizon black hole soliton contributions should be exponentially suppressed and considered as non-perturbative corrections in quantum field theory.

The effective theory describing these solitons involves by construction additional massive states with the BPS mass equal to the largest eigenvalue of the central charge matrix $Z(\Gamma)$. Semi-classical Dirac--Schwinger--Zwanziger  quantisation of the charges $\Gamma$ implies that this effective theory should be invariant with respect to the arithmetic subgroup $E_{d(d)}(\mathds{Z})$ of the classical Cremmer--Julia symmetry. Although this effective theory would be extremely complicated to derive from first principles, it seems reasonable to conjecture that the effective theory describing $\tfrac12$-BPS solitons is uniquely determined by symmetry to be the exceptional field theory on $\mathds{R}^{1,10-d}\times T^d$ studied in this paper. The fact that the effective theory takes almost the form of a higher-dimensional local field theory naturally follows from the property that $\tfrac12$-BPS massive irreducible representations of the maximally extended super-Poincar\'e group are homomorphic to massless irreducible representations of the maximally extended super-Poincar\'e group in one dimension higher \cite{Ferrara:1980ra}. 

We have exhibited in this paper that the ultra-violet behaviour of the theory was improved by the consideration of such $\tfrac12$-BPS states, through the cancelation of the 1-loop divergence in eight dimensions, and the cancelation of the 2-loop divergence in seven dimensions. These cancelations suggest that the theory including all zero-size solitons could possibly be free of ultra-violet divergences, and would define a consistent perturbation theory free of the usual ambiguities inherent to non-renormalisable field theories. One may wonder then if the appropriate quantisation of supergravity including all relevant soliton contributions could reproduce the complete M-theory low-energy effective action. This paper exhibits that supergravity does include the exact corrections to the BPS protected threshold functions in string theory, providing a first hint that this proposal is not completely wrong. However, these BPS couplings are known to be entirely determined by the symmetries of the theory up to an overall factor, and at this level we have to admit that the evidence for the proposal is not very strong. Our proposal does relate to some extent to the conjecture that the complete $(2,0)$ non-abelian conformal field theory in six dimensions on a circle is described by the supersymmetric Yang--Mills theory in five dimensions including `instanton--like' solitons \cite{Douglas:2010iu,Lambert:2010iw}. In this later framework it was proposed that the 6-loop logarithmic divergence \cite{Bern:2012di} could be canceled by soliton contributions \cite{Papageorgakis:2014foa}. This kind of cancelation between massless ultraviolet divergences and soliton contributions is advocated by the cancelations we exhibited  in exceptional field theory. 

On the contrary to the $\tfrac12$-BPS massive multiplets, the $\tfrac14$-BPS massive multiplets do not correspond to massless multiplets in higher dimensions. In four dimensions, one has for example the  lowest spin massive multiplet \cite{Ferrara:1980ra}
\be \begin{tabular}{l|c|c|c|c|c|c|c}
spin & $0$ & $\frac{1}{2}$ & $1$ & $\frac{3}{2}$ & $2$ & $\frac{5}{2}$ & $3$ \\\hline
$Sp(6)$ & ${\bf 429}_{6} $& ${\bf 572}$& ${\bf 429}_{4} $& ${\bf 208}$& ${\bf 65}$& ${\bf 12}$& ${\bf 1}$ 
\end{tabular} \ee
where the symplectic traceless antisymmetric  rank $6-2j$ tensor representations of $Sp(6)$ split into irreducible representations of the $Sp(2)\times SU(4) \subset Sp(2)\times Sp(4) \subset Sp(6)$ subgroup of automorphisms of the supersymmetry algebra with a $\tfrac14$-BPS central charge. It is rather clear form the structure of this multiplet that the effective theory describing such massive states does not naturally admit a higher dimensional realisation, and cannot be described within a framework similar to exceptional field theory. 

\subsection{1/4-BPS states and renormalisation}

Because the $R^4$ type supersymmetric invariant can only be written as a superspace integral over sixteen fermionic coordinates in the linearised approximation \cite{Howe:1981xy}, it follows that $R^4$ is subject to a non-renormalisation theorem, such that only  $\tfrac12$-BPS states can contribute to it \cite{Green:1997tv}. More precisely, it is known that the Fourier support of the $R^4$ threshold function in string theory only includes $\tfrac12$-BPS instanton charges \cite{Obers:1999um,Green:2011vz}. This is a consequence of supersymmetry, through the differential equation that the threshold function must satisfy in order for this coupling to extend to a supersymmetry invariant \cite{Bossard:2014lra}. It is therefore expected that  only  $\tfrac12$-BPS solitons can contribute to the threshold function $\cE_\gra{0}{0}$, explaining that the exceptional field theory amplitude we computed at 1-loop provides the exact result. On the contrary, the $\nabla^4 R^4$ threshold function is known to admit $\tfrac14$-BPS states contributions \cite{Green:2011vz}, and one does not expect the $\tfrac14$-BPS soliton contributions to cancel. Nonetheless, it appears that the 1-loop and the 2-loop amplitudes we have computed contain the exact $\nabla^4 R^4$ and $\nabla^6 R^4$ threshold functions in string theory, while the second also receives $\tfrac18$-BPS states contributions. This, however, might be a consequence of the strong constraints implied by supersymmetry on these couplings, in general one does expect $\tfrac14$-BPS and $\tfrac18$-BPS to be included in exceptional field theory for a consistent and gauge invariant formulation as discussed in section~\ref{sec:FT}.

The fact that the one-loop exceptional field theory $R^4$ coupling is finite is a consequence of the property that it includes all M-theory state contributions, and as such should be free of any divergence. On the other-hand, the one-loop contributions to higher derivative threshold functions does not include all the M-theory states in the loop, and one finds indeed that the corresponding contribution diverge in various dimensions. In particular, the $\nabla^4 R^4$ threshold function then diverges in seven and in six dimensions. It is expected that the contribution from the  $\tfrac14$-BPS states in M-theory should compensate precisely this divergence in an appropriate regularisation scheme. However, we verified in appendix \ref{1/4BPSEpstein} that the naive formula one could write for a candidate function obtained as a lattice sum over $\tfrac14$-BPS charges admit singularities at finite moduli, and it is yet unclear how such a sum could cancel the divergent Eisenstein series contribution we obtained in this paper.  We shall nonetheless assume that this happens through some mechanism  that remains to be clarified.

Because one expects such a contribution to arise independently in all dimensions, it appears that if one has to renormalise the exceptional field theory in order to obtain finite amplitudes, one should do it consistently in all dimensions. We therefore propose the following renormalisation prescription for the theory: If a given threshold function diverges in dimensional regularisation at a given loop order for some value $3 \le d\le 7$, one should add a local counterterm in $D$ dimensions to remove this contribution in all dimensions. 

This is indeed the renormalisation prescription we have adopted in this paper, such that one removes completely the 1-loop contribution to the $\nabla^4 R^4$ threshold function by adding the relevant counterterm. Then the 2-loop contribution to this threshold function turns out to provide the exact string theory function, and the full amplitude is finite. Note that the $\nabla^4 R^4$ 1-loop counterterm should be taken into account at 2-loop, but one can check that it only contributes to $\nabla^{8+2k} R^4$ type threshold functions~\cite{Basu:2014hsa}.

Although we have obtained that the sum of the 1-loop and the 2-loop $\nabla^6 R^4$ threshold functions define the exact string theory threshold function, it is divergent in dimensional regularisation and should be removed by adding the corresponding counterterm within our prescription. We expect the contribution of the 3-loop Mercedes diagram in exceptional field theory to define again the exact string theory  $\nabla^6 R^4$ threshold function, in such a way that its pole in dimensional regularisation would cancel precisely the 3-loop divergences in supergravity, \ie the 3-loop 4-graviton divergence in six dimensions, and the $R^4$ and $\nabla^4 R^4$ sub-divergences in five and four dimensions. 

Our analysis might have some bearing on the role of the strong section constraint for the construction of higher derivative effective actions in exceptional field theory. As we already pointed out in section~\ref{sec:EFTamps}, it is possible to construct Feynman diagrams already at tree-level with external charges $\Gamma$ that do not all satisfy the strong section constraint in a pairwise manner, see \eg~\eqref{4-pointMIIB}, and it is also possible to embed such processes in loop diagrams, as for example 
 \def\xshift{5}
  \def\xmin{1}
 \def\ymin{-2}
 \def\yshift{1.6}
\be
 \begin{tikzpicture}
  \draw (\xmin,\ymin) node{$(0,0,n_3,0)$};
\draw[-,draw=black,very thick](\xmin+1,\ymin) -- (\xmin+1.5,\ymin);
\draw[-,draw=black,very thick](\xmin+1.5,\ymin) -- (\xmin+2,\ymin+0.5);
 \draw[-,draw=black,very thick](\xmin+1.5,\ymin) -- (\xmin+1.5,\ymin-0.5);
  \draw (\xmin+3.2,\ymin+0.6) node{$(0,0,0,n^{12})$};
  \draw (\xmin+1.5,\ymin-0.8) node{$(0,0,n_3,n^{12})$};
\draw[-,draw=black,very thick](\xmin-1,\ymin) -- (\xmin-1.5,\ymin);
\draw[-,draw=black,very thick](\xmin-1.5,\ymin) -- (\xmin-2,\ymin+0.5);
 \draw[-,draw=black,very thick](\xmin-1.5,\ymin) -- (\xmin-1.5,\ymin-0.5);
  \draw (\xmin-3.1,\ymin+0.6) node{$(n_1,0,0,0)$};
  \draw (\xmin-1.5,\ymin-0.8) node{$(-n_1,0,n_3,0)$};
      \draw (\xmin,\ymin-\yshift) node{$(0,0,n_3,0)$};
\draw[-,draw=black,very thick](\xmin+1,\ymin-\yshift) -- (\xmin+1.5,\ymin-\yshift);
\draw[-,draw=black,very thick](\xmin+1.5,\ymin-\yshift) -- (\xmin+2,\ymin-\yshift-0.5);
 \draw[-,draw=black,very thick](\xmin+1.5,\ymin-\yshift) -- (\xmin+1.5,\ymin-\yshift+0.5);
  \draw (\xmin+3.2,\ymin-\yshift-0.6) node{$(0,0,0,-n^{12})$};
\draw[-,draw=black,very thick](\xmin-1,\ymin-\yshift) -- (\xmin-1.5,\ymin-\yshift);
\draw[-,draw=black,very thick](\xmin-1.5,\ymin-\yshift) -- (\xmin-2,\ymin-\yshift-0.5);
 \draw[-,draw=black,very thick](\xmin-1.5,\ymin-\yshift) -- (\xmin-1.5,\ymin-\yshift+0.5);
  \draw (\xmin-3.1,\ymin-\yshift-0.6) node{$(-n_1,0,0,0)$};
\end{tikzpicture}
\label{4-pointMIIB1-loop}
\ee
where we use the same notation as in \eqref{4-pointMIIB}, and the sum over the momentum $n_3$ in the loop should be extended to the sum over all momenta consistent with the strong section constraint with $(n_1,0,0,0)$ and $(0,0,0,n^{12})$. Although we have not computed this diagram, we expect that it might carry a divergence in dimensional regularisation for the \textit{massive} $\nabla^4 R^4$ threshold function in seven and six dimensions, as for the massless case. Note that the low energy expansion of such amplitudes only makes sense if the masses of the external states are much lower than $1$ in units of $\ell^{-1}$, but one can easily check that this can be achieved for some specific choice of moduli near a specific (cusp) boundary. Divergences in such diagrams would have to be removed by counterterms for massive modes in exceptional field theory, which will be local in $D$ dimensions. However, they will be defined by construction as products of fields that do not satisfy the strong section constraint, and that we expect to be non-local in the extended space (\ie not with a polynomial dependence in the integral external momenta). In fact, we suspect a close connection between the notion of locality and the strong section constraint for such counterterms in exceptional field theory.

\subsection{Extension to the Kac--Moody case $E_{d(d)}$ with $d>8$}
\label{sec:KMext}

We can also formally consider the case $d>8$ where the hidden symmetry group $E_{d(d)}$ becomes of Kac--Moody type~\cite{Julia:1980gr,Nicolai:1987kz,Nicolai:1988jb,Hull:1994ys,West:2001as,Damour:2002cu}. Expression for the correction functions $\cE_\gra{0}{0}$ and $\cE_\gra{1}{0}$ have been conjectured in~\cite{Fleig:2012xa} and passed further consistency tests in~\cite{Fleig:2013psa}. Exceptional field theory for Kac--Moody groups has been discussed in~\cite{West:2003fc,Tumanov:2015iea}.

For the $R^4$ correction term $\cE_\gra{0}{0}$ the result from the one-loop calculation in exceptional field theory takes the form (cf.~\eqref{oneloopsplit}):
\begin{align}
\mathcal{E}_\gra{1}{0}^{\scriptscriptstyle \mbox{\tiny 1-loop}} = 4\pi \xi(d-3) E_{\alpha_d,\tfrac{d-3}{2}} = 2\zeta(3) E_{\alpha_1,\stfrac32},
\end{align}
after formally using the functional relation~\eqref{eq:FR} applied with elements of the Kac--Moody Weyl group for $E_{d(d)}$ with $9\leq d\leq 11$.\footnote{For the affine $E_{9(9)}$ case, the functional relation was proven in~\cite{Garland1}.}
This function agrees with the `minimal' series proposed for $R^4$ in~\cite{Fleig:2012xa}. The fact that it could follow from a truncated $E_{11(11)}$ one-loop calculation with $\tfrac12$-BPS states circulating in the loop was later proposed in \cite{West:2012qm}. We also note that our expression~\eqref{BT1} provides a conjectural alternative definition of the minimal Eisenstein series in the Kac--Moody case.

Interestingly, the $\nabla^4 R^4$ contribution from the non-degenerate orbit at two loops given in~\eqref{twoloopind} can also be converted into a more standard form using a suitable Kac--Moody Weyl group element. One can map the Eisenstein series on node $\alpha_{d-1}$ to a series on node $\alpha_1$ for $9\leq d \leq 11$. The resulting two-loop threshold function is 
\begin{align}
\mathcal{E}_\gra{1}{0}^{\scriptscriptstyle \mbox{\tiny 2-loop, n.d.}} =  \zeta(5)   E_{\alpha_1,\stfrac52}
\end{align}
in all cases, corresponding to the correctly normalised Eisenstein series discussed in~\cite{Fleig:2012xa} for the $\nabla^4 R^4$ correction. Just as the `minimal' series above, this `next-to-minimal' series has the special property that it only possesses a finite number of constant terms~\cite{Fleig:2012xa} and very simple (degenerate) Whittaker vectors~\cite{Fleig:2013psa}.

\subsection*{Acknowledgements}

We would like to thank N.~Boulanger, M.~Cederwall, G.~Dall'Agata, J.-H. Park, B.~Pioline, S.-J. Rey, J.~Russo, H.~Samtleben, E.~Sezgin, P.~Sundell and P.~West  for useful discussions. We are grateful to the organisers of the 2015 Benasque String Theory workshop for providing a stimulating environment for presenting and finalising the results of this paper. AK would like to acknowledge CERN and the Mainz Institute for Theoretical Physics (MITP) for its hospitality and support during part of this work.
  
\appendix  

\section{Eisenstein series \`a la Langlands}
\label{app:ES}
\def\DEVII#1#2#3#4#5#6#7{{\tiny $ { \left[ \begin{array}{ccccccc}  & & \mathfrak{#2} \hspace{-0.7mm}&&&& \vspace{ -1mm} \\ \mathfrak{#1}\hspace{-0.7mm} &  \mathfrak{#3} \hspace{-0.7mm}& \mathfrak{#4} \hspace{-0.7mm} & \mathfrak{#5}\hspace{-0.7mm}&\mathfrak{#6}\hspace{-0.7mm}& #7 \hspace{-0.8mm} \end{array}\right] }$}}

In this appendix, we summarise briefly the pertinent definitions from Langlands~\cite{LanglandsFE} for Eisenstein series on split real hidden symmetry groups $E_{d(d)}$ with invariance under $E_{d(d)}(\mathds{Z})$. The material here is discussed also in previous publications~\cite{Green:2010kv,Fleig:2012xa,Pioline:2015yea,Bossard:2015oxa}, we follow in particular~\cite{Bossard:2015oxa}.

An Eisenstein series can be defined for almost all weights $\lambda$ of the Lie algebra $\mf{e}_{d(d)}$ of $E_{d(d)}$. Let us denote the fundamental weights by $\Lambda_i$ for $i=1,\ldots,d$. They are dual to the simple roots $\alpha_j$ in the Cartan--Killing metric $\lp\alpha_i | \Lambda_j\rp = \delta_{ij}$. Let $g \in E_{d(d)}$ be a representative of the coset $E_{d(d)}/K_d$ with $K_d$ the maximal compact subgroup of $E_{d(d)}$. Using the Iwasawa decomposition $g=nak$ with $a=e^h$ in a fixed maximal split torus (and $h$ in the Cartan subalgebra), $k\in K_d$ and $n$ a unipotent element one can define the map 
\begin{align}
H(\lambda,\cdot) \,:\,  G \to \mathds{C}^\times,\quad H(\lambda,g)= H(\lambda,a) = (\lambda+\rho)(h)
\end{align}
using the pairing between the Cartan subalgebra and the space of weights. $\rho=\sum_{i=1}^d \Lambda_i$ is the Weyl vector. Depending on the weight $\lambda$, the function $H(\lambda,\cdot)$ has a stabiliser $P\subset E_{d(d)}(\mathds{Z})$ that always contains the Borel subgroup $B(\mathds{Z})$ and therefore is contained in a parabolic subgroup. We will parametrise the weight $\lambda$ as $\lambda = 2\sum_{i=1}^d s_i \Lambda_i -\rho$ and write it as a labelled Dynkin diagram. For example for $E_{7(7)}$ we would have in general
\begin{align}
\lambda= 2\sum_{i=1}^7 s_i \Lambda_i -\rho \qquad \longleftrightarrow\qquad 
\mbox{\DEVII{\mathnormal{s_1}}{\mathnormal{s_2}}{\mathnormal{s_3}}{\mathnormal{s_4}}{\mathnormal{s_5}}{\mathnormal{s_6}}{\mathnormal{s_7}}}.
\end{align}
The stabiliser $P$ is the parabolic subgroup determined by the vanishing $s_i$; if only one $s_i$ is non-zero $P$ will be a maximal parabolic subgroup. 

The Eisenstein series defined by Langlands then is given by averaging the exponential of the function $H(\lambda,\cdot)$ over the action of the discrete subgroup $E_{d(d)}(\mathds{Z})$:
\begin{align}
\label{ESdef}
E(\lambda,g) = \sum_{\gamma\in P\backslash E_{d(d)}} e^{H(\lambda, \gamma g)}.
\end{align}
Here, it is understood that the sum is over elements of the discrete subgroup $E_{d(d)}(\mathds{Z})$. For compactness we also denote the Eisenstein series with a labelled Dynkin diagram and suppress the dependence on $g$, \eg
\begin{align}
E_{\mbox{\DEVII{\mathnormal{s_1}}{\mathnormal{s_2}}{\mathnormal{s_3}}{\mathnormal{s_4}}{\mathnormal{s_5}}{\mathnormal{s_6}}{\mathnormal{s_7}}}}.
\end{align}

The action of $\gamma\in E_{d(d)}$ transforms the coordinates on the symmetric space $E_{d(d)}/K_d$. We will denote this action by $\gamma\left[\cdot\right]$, where $\cdot$ stands for any function of the coordinates on the symmetric space. For example, for $d=1$ and $E_{1(1)} \cong SL(2,\mathds{R})$ (in the case of type IIB)\footnote{For type IIA in $D=10$, the corresponding group would be just the scaling $\mathds{R}^*_+$.} we can choose standard coordinates $\tau=\tau_1+ i \tau_2$ on the upper half plane such that the fraction linear action $\tau\to \frac{a\tau+b}{c\tau+d}$ under $\gamma=\left(\begin{smallmatrix}a&b\\c&d\end{smallmatrix}\right)\in SL(2,\mathds{R})$ leads to
\begin{align}
\gamma \left[ \tau_2 \right] = \frac{\tau_2}{|c\tau+d|^2}.
\end{align}
The Eisenstein series~\eqref{ESdef} can also be written as
\begin{align}
E(\lambda,g) = \sum_{\gamma\in P\backslash E_{d(d)}} \gamma\left[ e^{H(\lambda,g)}\right].
\end{align}
For the standard non-holomorphic Eisenstein series on $SL(2,\mathds{R})$ we would simply get
\begin{align}
E_{[s]} = \sum_{\gamma\in B(\mathds{Z})\backslash SL(2,\mathds{Z})} \gamma\left[ \tau_2^s \right],
\end{align}  
with $B(\mathds{Z})$ the upper triangular Borel subgroup of $SL(2,\mathds{Z})$.

The definition of Eisenstein series~\eqref{ESdef} is only absolutely convergent when $\textrm{Re} \, s_i>1$ for all $i=1,\ldots,d$. The definition can be analytically continued by using Langlands's functional relation
\begin{align}
\label{eq:FR}
E(\lambda,g) = M(w,\lambda) E(w\lambda,g),
\end{align}
where $w$ belongs to the Weyl group of $E_{d(d)}$. The coefficient appearing in this relation is
\begin{align}
M(w,\lambda) = \prod_{\substack{\alpha>0 \\ w\alpha<0}} \frac{\xi((\lambda|\alpha))}{\xi(1+(\lambda|\alpha))}
\end{align}
in terms of the completed Riemann zeta function $\xi(s) = \pi^{-s/2} \Gamma(s/2) \zeta(s)$ that satisfies $\xi(s)=\xi(1-s)$ and has simple poles at $s=0$ and $s=1$ on the real axis. Eisenstein series can be expanded into constant terms and Fourier modes and general formulas for this can be found in~\cite{LanglandsFE,MoeglinWaldspurger,Green:2010kv,Fleig:2012xa}.

\section{$SO(n,n)$ Eisenstein series at special values}
\label{MinimalOrtho}

In this appendix we want to demonstrate that the minimal unitary representation $SO(n,n)$ Eisenstein series are determined exactly by their approximation to the two summands of highest degree in the graded decomposition with respect to the maximal parabolic $P_1$ associated to the first root. This was claimed below~\eqref{BT1}. 

For orthogonal groups, the minimal unitary representation series can be realised both in the Weyl spinor $S_+$ and the vector representation $V$, owing to the Langlands functional identity
\be 
\xi(n-2) E^{D_{n}}_{V,\frac{n-2}{2}} = \xi(2) E^{D_{n}}_{S_+,1} \ .\label{LanglandsDn}  
\ee
We will analyse this equation in the following for $SO(n+1,n+1)$ rather than $SO(n,n)$ since it makes some formulas more compact. The starting point is the parabolic decomposition 
\bea 
\mathfrak{so}(n+1,n+1)\cong {\bf V}^\ord{-2} \oplus \scal{ \mathfrak{gl}_1 \oplus   \mathfrak{so}(n,n)}^\ord{0} \oplus {\bf V}^\ord{2} \ , \CR
{\bf V}\cong {\bf 1}^\ord{-2} \oplus {\bf V}^\ord{0} \oplus {\bf 1}^\ord{2} \ , \qquad 
{\bf S}_+ \cong {\bf S}_-^\ord{-1} \oplus {\bf S}_+^\ord{1} \ . \label{SOa1} 
\eea
Let us first consider the spinor series, which is a sum over the non-zero Weyl spinor $P\in {\bf S}_+$ subjected to the constraint that $P \gamma_{n-3} P = 0 $, where $\gamma_k$ is the antisymmetric product of $k$ gamma matrices of $SO(n+1,n+1)$.\footnote{Strictly speaking the pure spinor constraint gives  $P \gamma_{n+1-4k} P = 0 $ for all $k\ge1$ such that $4k\le n+1$, but the first equation for $k=1$ implies the others.} Using this decomposition one obtains 
\bea 
2 \xi(2s)  E^{D_{n+1}}_{S_+,s} &=&  \pi^{-s} \Gamma(s) \sum_{\substack{P \in {\bf S}_+^*\\ P \gamma_{n-3} P = 0 }} |Z(P)|^{-2s} \ \CR
&=& 2 \xi(2s) e^{-2s\phi}  E^{D_{n}}_{S_+,s} +\hspace{-0.5mm}\sum_{\substack{q \in {\bf S}_-^*\\ q \gamma_{n-4} q = 0 }}\sum_{\substack{p \in {\bf S}_+\\ q \gamma_{n-3} p = 0 }}  \int \frac{dt}{t^{1+s}} e^{ - \frac{\pi}{t} \scal{ e^{2\phi} |Z(p+ \baS a q)|^2 + e^{-2\phi} |Z(q)|^2} } \, .  \hspace{5mm} 
\eea
To be able to Poisson resume the sum over $p$s, we use the fact that the spinor $q$ can always be rotated to a relative integer in the highest grad component of the parabolic decomposition 
\bea
{\bf V} &\cong&\hspace{10.0mm} {\bf n}^\ord{-2}\hspace{2mm} \oplus\hspace{2mm} \overline{\bf n}^\ord{2} \hspace{2mm}\ , \CR
 \mathfrak{so}(n,n)&\cong& {\bf \frac{n(n-1)}{2}}^\ord{-4} \hspace{-1.2mm} \oplus \scal{ \mathfrak{gl}_1 \oplus   \mathfrak{sl}(n)}^\ord{0} \oplus \overline{\bf \frac{n(n-1)}{2}}^\ord{4} \ , \CR
\Lambda^{n-4}{\bf V} &\cong& \qquad \dots \qquad  \oplus {\bf \frac{n(n-1)(n-2)(n-4)}{24}}^\ord{2n-8}  \ , \CR
\Lambda^{n-3}{\bf V} &\cong& \qquad \dots \qquad  \oplus {\bf \frac{n(n-1)(n-2)}{6}}^\ord{2n-6}  \  , \CR
{\bf S}_- &\cong& \qquad \dots \qquad  \oplus {\bf \frac{n(n-1)}{2}}^\ord{n-4} \oplus {\bf 1}^\ord{n} \ , \CR
{\bf S}_+&\cong& \qquad \dots \qquad  \oplus {\bf \frac{n(n-1)(n-2)}{6}}^\ord{n-6} \oplus {\bf n}^\ord{n-2} \ . 
\eea
such that the sum over $p$s reduces to the unconstrained sum over $\mathds{Z}^n$ in the grad $n-2$ component. Using this one obtains 
\bea && \sum_{\substack{q \in {\bf S}_-^*\\ q \gamma_{n-2} q = 0 }}\sum_{\substack{p \in {\bf S}_+\\ q \gamma_{n-3} p = 0 }}  \int \frac{dt}{t^{1+s}} e^{ - \frac{\pi}{t} \scal{ e^{2\phi} |Z(p+ \baS a q)|^2 + e^{-2\phi} |Z(q)|^2} }   \\
&=& \sum_{\gamma \in P_n\backslash SO(n,n)} \sum_{q\in \mathds{Z}_*} \sum_{p\in\mathds{Z}^n} \int \frac{dt}{t^{1+s}} \gamma \biggl[ e^{ - \frac{\pi}{t} \scal{ e^{2\phi+2(n-2) \upsilon } M(p+a q) +e^{-2\phi+2n \upsilon} q^2  }} \biggr] \CR
&=&2 \xi(2s-n) e^{2(s-n)\phi} E^{D_n}_{S_-,s-1} \CR
&& \hspace{-3mm}+ 2  e^{-n\phi}  \hspace{-5mm}\sum_{\gamma \in P_n\backslash SO(n,n)} \sum_{q\in \mathds{Z}_*} \sum_{\tilde{p}\in\mathds{Z}^n_*} \gamma \biggl[ \frac{ e^{-2n (s-1)\upsilon} \scal{  e^{4\upsilon} q^2 M(\tilde{p})}^{\frac{s-\frac{n}{2}}{2}} }{q^{2s-n}} K_{s-\frac{n}{2}}\scal{ 2\pi e^{-2\phi+2\upsilon} \sqrt{q^2 M(\tilde{p})}} e^{2\pi i \langle q \tilde{p},a\rangle } \biggr] \nonumber 
\eea
Now we note that $\tilde{ p}$ only appears in the sum through the vector $Q \equiv ( \tilde p \gamma_1 q ) $, which satisfies by construction  
\be  
\baa Q q = 0 \ , \qquad \langle Q , Q \rangle = 0 \ , \qquad {\rm gcd}(q) | {\rm gcd}( Q) \  . 
\ee
and the sum over $q$ and $\tilde{p}$ reduces to the sum over all pure spinor $q$ and integral vectors $Q$ satisfying these constraints. It is convenient to analyse this second sum in terms of the graded decomposition \eqref{SOa1}, such that 
\bea
&&  \sum_{\gamma \in P_n\backslash SO(n,n)} \sum_{q\in \mathds{Z}_*} \sum_{\tilde{p}\in\mathds{Z}^n_*} \gamma \biggl[ \frac{ e^{-2n (s-1)\upsilon} \scal{  e^{4\upsilon} q^2 M(\tilde{p})}^{\frac{s-\frac{n}{2}}{2}} }{q^{2s-n}} K_{s-\frac{n}{2}}\scal{ 2\pi e^{-2\phi+2\upsilon} \sqrt{q^2 M(\tilde{p})}} e^{2\pi i \langle q \tilde{p},a\rangle }  \biggr] \CR
&=&  \sum_{\gamma \in P_1\backslash SO(n,n)} \sum_{Q\in \mathds{Z}_*} \sum_{\substack{q \in {\bf S}^*_-\\ {\rm gcd}(q)| Q  }} \gamma \biggl[ \frac{ e^{-2 (s-1)\upsilon_1} |M(q)|^{-2(s-1)} \scal{  e^{2\upsilon_1}Q}^{s-\frac{n}{2}} }{{\rm gcd}(q)^{2-n}} K_{s-\frac{n}{2}}\scal{ 2\pi e^{-2\phi+2\upsilon_1} Q} e^{2\pi i \langle Q ,a\rangle }  \biggr]\CR 
&=&  \ \sum_{\gamma \in P_1\backslash SO(n,n)} \sum_{Q\in \mathds{Z}_*} \sum_{\substack{q \in {\bf S}^*_-\\ {\rm gcd}(q)| Q  }} \gamma \biggl[ \frac{{\rm gcd}(q)^{n-2}   }{|M(q)|^{2(s-1)}}  \frac{  Q^{s-1} }{\scal{ e^{2\upsilon_1}Q}^{\frac{n-2}{2}}} K_{s-\frac{n}{2}}\scal{ 2\pi e^{-2\phi+2\upsilon_1} Q} e^{2\pi i \langle Q ,a\rangle }  \biggr]
\  . 
\eea
One can therefore rewrite the series as
\bea 2 \xi(2s)  E^{D_{n+1}}_{S_+,s} &=& 2 \xi(2s) e^{-2s\phi}  E^{D_{n}}_{S_+,s}  + 2 \xi(2s-n) e^{2(s-n)\phi} E^{D_n}_{S_-,s-1} \\
&& + 4e^{-n\phi} \hspace{-3mm}\sum_{\substack{Q \in \mathds{Z}^{2n}_*\\ \langle Q,Q\rangle = 0 }} \Bigl( \sum_{r|Q}r^{n-2s}\Bigr) \frac{{\rm gcd}(Q)^{s-1}}{|Z(Q)|^{\frac{n-2}{2}}}  E^{D_{n-1}}_{S_-,s-1}(v_Q) K_{s-\frac{n}{2}}(2\pi e^{-2\phi} |Z(Q)|) e^{2\pi i \langle Q,a\rangle }\ ,  \nn \eea
where $v_Q$ is the $SO(n-1,n-1)$ coset representative in the stabilizer of $Q$. 

This series evaluated at $s=1$ gives 
\bea 2 \xi(2)  E^{D_{n+1}}_{S_+,1} &=& 2 \xi(n-1) e^{-2(n-1)\phi}  + 2 \xi(2) e^{-2\phi}  E^{D_{n}}_{S_+,1}  \CR
&& \quad + 4 \hspace{-2mm}\sum_{\substack{Q \in \mathds{Z}^{2n}_*\\ \langle Q,Q\rangle = 0 }} \Bigl( \sum_{r|Q} r^{n-2}\Bigr) \frac{e^{-n\phi}}{|Z(Q)|^{\frac{n-2}{2}}}  K_{\frac{n-2}{2}}(2\pi e^{-2\phi} |Z(Q)|) e^{2\pi i \langle Q,a\rangle }\  .  \hspace{8mm} \label{SpinorSeries} \eea
If we now evaluate the vector series according to the same decomposition, one gets equivalently
\bea 2 \xi(2s)  E^{D_{n+1}}_{V,s} &=& 2 \xi(2s) e^{-4s\phi} + 2 \xi(2s-1) e^{-2\phi} E^{D_n}_{V,s-\frac{1}{2}} \CR
&&  + 4  \hspace{-3mm}\sum_{\substack{Q \in \mathds{Z}^{2n}_*\\ \langle Q,Q\rangle = 0 }} \Bigl( \sum_{r|Q} r^{2s-1}\Bigr) \frac{e^{-(2s+1)\phi}}{|Z(Q)|^{s-\frac{1}{2}}}  K_{s-\frac{1}{2}}(2\pi e^{-2\phi} |Z(Q)|) e^{2\pi i \langle Q,a \rangle } \CR
&& + \sum_{m\in \mathds{Z}_*} \sum_{\substack{Q \in \mathds{Z}^{2n}\\ 2m|\langle Q,Q\rangle  }}\int \frac{dt}{t^{s+1}} e^{ - \frac{\pi}{t} \scal{ e^{4\phi} \bigl( \frac{\langle Q+m a,Q+m a\rangle}{2m}\bigr)^2 +  |Z(Q+m a)|^2 + e^{-4\phi} m^2} }\ . \qquad   \eea
Using the Langlands functional relation \eqref{LanglandsDn} and comparing with \eqref{SpinorSeries}, one concludes that the last line evaluated at $s=\frac{n-1}{2}$ vanishes. Strictly speaking this series is not absolutely convergent, so the cancelation holds for the analytic continuation of the series in $s$ evaluated at $s=\frac{n-1}{2}$, \ie 
\be  
\sum_{m\in \mathds{Z}_*}\sum_{\substack{Q \in \mathds{Z}^{2n}\\ 2m|\langle Q,Q\rangle  }}\int \frac{dt}{t^{s+1}} e^{ - \frac{\pi}{t} \scal{ e^{4\phi}\bigl( \frac{\langle Q+m a,Q+m a\rangle}{2m}\bigr)^2 +  |Z(Q+m a)|^2 + e^{-4\phi} m^2} } = \mathcal{O}\scal{ s-\tfrac{n-1}{2}} \ . 
\ee
This is the vanishing of the extra term as claimed below~\eqref{BT2}.

\section{1/4 BPS-Epstein series}\label{1/4BPSEpstein}
\label{Epstein1/4}

We would like to consider the possibility that an Eisenstein series is produced by $\tfrac{1}{4}$ BPS solitons through a lattice sum over $\tfrac{1}{4}$-BPS charges. A $\tfrac14$-BPS charge defines two quantities invariant under the maximal compact subgroup $K_d\subset E_{d(d)}$. In particular in seven dimensions, the matrix central charge $Z_{ab}(\Gamma)$ for an arbitrary charge $\Gamma$ in the $\overline{\bf 10}$ of SL(5) is an antisymmetric tensor $Z_{ab}$ of $SO(5)$, that satisfies 
\be \cD_{ab} Z^{cd} =\delta^{[c}_{(a} Z_{b)}{}^{d]} - \frac{1}{5} \delta_{ab} Z^{cd} \ . \ee
One computes using this equation that  for a general function $\cE(Z_2,Z_4)$ depending on
\be 
Z_2 = Z_{ab} Z^{ab} \ , \qquad Z_4 = Z_{ab} Z^{bc} Z_{cd} Z^{de} \ , 
\ee
that
\be 
\cD_{ab} \cE(Z_2,Z_4) = \left[ 2 \scal{ Z_a{}^c Z_{bc} - \tfrac{1}{5} \delta_{ab} Z_2 } \partial_2 + 4 \scal{  Z_a{}^c Z_{cd} Z^{de} Z_{eb} - \tfrac{1}{5} \delta_{ab} Z_4 } \partial_4 \right] \cE(Z_2,Z_4) 
\ee
and 
\bea 
&& \cD_{a}{}^c  \cD_{bc} \, \cE(Z_2,Z_4) \nn\\
&=&  \biggl[\Scal{ \frac{9}{10} Z_{a}{}^c Z_{bc}  + \frac{33}{50} \delta_{ab} Z_2 }\partial_2   + \Scal{  -\frac{7}{5}  Z_a{}^c Z_{cd} Z^{de} Z_{eb}  + 2 Z_2 Z_{a}{}^c Z_{bc} + \frac{41}{25} \delta_{ab} Z_4 }\partial_4\CR
&&\quad  + 4 \Scal{  Z_a{}^c Z_{cd} Z^{de} Z_{eb} - \frac{2}{5} Z_a{}^c Z_{bc} Z_2 + \frac{1}{25} \delta_{ab} Z_2^{\; 2}} \partial_2^{\; 2} \\
&& \quad + \Scal{ \frac{24}{5} Z_2  Z_a{}^c Z_{cd} Z^{de} Z_{eb}  +\scal{ \frac{4}{5} Z_4-2 Z_2^{\; 2} }  Z_{a}{}^c Z_{bc}  + \frac{16}{25} \delta_{ab} Z_2 Z_4} \partial_2 \partial_4\CR
&& \quad  + \Scal{ \scal{ 2 Z_2^{\; 2} - \frac{12}{5} Z_4 }  Z_a{}^c Z_{cd} Z^{de} Z_{eb}  + ( 2 Z_2 Z_4 - Z_2^{\; 3} Z_2 ) Z_{a}{}^c Z_{bc} + \frac{16}{25} \delta_{ab} Z_4^{\; 2}} \partial_4^{\; 2}\biggr]  \cE(Z_2,Z_4)  \ .  \nn 
\eea
Using these equations it becomes straightforward to solve the differential equation \cite{Bossard:2014aea}
\be 
\cD_{a}{}^c \cD_{bc} \cE_s  = -\frac{4s-5}{20} \cD_{ab} \cE_s + \frac{3(2s-5)}{25} \delta_{ab} \cE_s \ , 
\ee
that is also solved by the Eisenstein series $E_{[0,0,s,0]}$. One obtains the solution 
\be 
\cE_\pm(Z_2,Z_4) =\frac{2}{\sqrt{ 4 Z_4 - Z_2^{\; 2}}}  \left( \frac{ Z_2 \pm \sqrt{ 4 Z_4 - Z_2^{\; 2}}}{4}\right)^{-s+1}  = \frac{\hspace{4mm} z_\pm^{-s+1}}{z_+-z_-} \ , 
\ee
with the definition 
\be 
z_\pm =\frac{ Z_2 \pm \sqrt{ 4 Z_4 - Z_2^{\; 2}}}{4} \ , 
\ee
such that $z_+\ge z_->0$ define the BPS mass $W =\sqrt{z_+} + \sqrt{z_-}$ of the soliton, and are the two eigenvalues squared of the tensor $Z_{ab}$. $z_-$  is necessarily strictly greater than zero for a rank 4 charge $\Gamma$, but can reach $z_+$ at finite moduli, on a subspace of dimension 10. Therefore $\cE_\pm(Z_2,Z_4)$ admit  singularities at finite moduli. One finds nonetheless that the combination $\cE_+(Z_2,Z_4)-\cE_-(Z_2,Z_4)$ is regular at $z_+=z_-$. The infinite sum\footnote{The notation $\Gamma\times \Gamma=0$ indicates that the product of the charge with itself should not have any component in the ${\bf 5}$ representation appearing in the symmetric product of the two-form $\Gamma\in\overline{\bf 10}$ with itself.}
\be 
\cE^{\scriptscriptstyle 1/4}_{\overline{\bf 10},s} =\sum_{\substack{\Gamma\in \mathds{Z}^{10}\\ \Gamma \times  \Gamma \ne 0}} \frac{\hspace{4mm} z_+(\Gamma)^{-s+1}-z_-(\Gamma)^{-s+1}}{z_+(\Gamma)-z_-(\Gamma)} \ , 
\ee
converges absoutely for $\mbox{Re}(s)>5$ and is regular at finite moduli. By uniqueness of the corresponding automorphic representation, it must therefore be proportional to the Eisenstein series $E_{[0,0,s,0]}$ that satisfies the same differential equation. 

We expect this property to generalise to all dimensions. For example in four dimensions, a $\tfrac14$-BPS central charge satisfies ($i,j,\ldots$ are fundamental $SU(8)$ indices)
\be 
Z_{ik} Z^{kl} Z_{lp} Z^{pq} Z_{qr} Z^{rj} = -\frac{1}{4}(  Z_{kl} Z^{kl} ) Z_{ip} Z^{pq} Z_{qr} Z^{rj}  + \frac{1}{8} \scal{ Z_{kl} Z^{lp} Z_{pq} Z^{qk} - \tfrac{1}{4} (Z_{kl} Z^{kl})^2} Z_{ir} Z^{rj}  \ ,
\ee
and determines two $SU(8)$ invariant functions 
\be 
Z_2 = Z_{ij} Z^{ij} \ , \qquad Z_4 = Z_{ij} Z^{jk} Z_{kl} Z^{li} \ , 
\ee
that determine the BPS mass $W>0$ and the ratio parameter $0\le  x<  1$ such that 
\be 
W = \sqrt{\frac{ Z_2 + \sqrt{ 8 Z_4 - Z_2^{\; 2}}}{8}} \ , \qquad x =  \frac{ Z_2 - \sqrt{ 8 Z_4 - Z_2^{\; 2}}}{Z_2 + \sqrt{ 8 Z_4 - Z_2^{\; 2}}}\ . 
\ee
We expect the $\tfrac14$-BPS differential equations constraining the $\nabla^4 R^4$ threshold function to admit as solution a function of $W$ and $x$ for an arbitrary rank 2 charge $\Gamma$.

\end{document}